\documentclass[fleqn,usenatbib]{mnras}

\usepackage{mathptmx}

\usepackage[T1]{fontenc}
\usepackage{ae,aecompl}
\usepackage{color}
\definecolor{myorange}{rgb}{0.8, 0.3, 0.0}

\definecolor{myred}{rgb}{0.7, 0.05, 0.1}

\definecolor{mygreen}{rgb}{0.0, 0.457, 0.0}

\usepackage{graphicx}	
\usepackage{amsmath}	
\usepackage{amssymb}	

\usepackage{mdwtab}
\usepackage{multirow}
\usepackage{soul}

\newcommand{\lgal}{\texttt{L-Galaxies} }
\newcommand{\gqd}{\texttt{GQd} }

\newcommand{\ptre}{PopIII }
\newcommand{\milltwo}{\texttt{MR-II} }

\title[SMBH seeding in the \lgal SAM]{Multi-flavour SMBH seeding and evolution in cosmological environments}
\author[D. Spinoso et al.]{
D. Spinoso,$^{1,2}$\thanks{E-mail: daniele.spinoso@dipc.org ; dspinoso@cefca.es}
S. Bonoli,$^{1,3}$
R. Valiante$^{4,5}$
R. Schneider$^{6,4,5}$
and D. Izquierdo-Villalba$^{7,8}$
\\
$^{1}$Donostia International Physics Center. Paseo Manuel de Lardizabal, 4, 20018 Donostia-San Sebasti\'an (Gipuzkoa), Spain\\
$^{2}$Centro de Estudios de F\'isica del Cosmos de Arag\'on (CEFCA), plaza San Juan 1, 44001, Teruel, Spain\\
$^{3}$IKERBASQUE, Basque Foundation for Science, E-48013, Bilbao, Spain\\
$^{4}$INAF-Osservatorio Astronomico di Roma, via di Frascati 33,I-00078 Monteporzio Catone, Italy\\
$^{5}$INFN, Sezione di Roma I, P.le Aldo Moro 2, I-00185 Roma, Italy\\
$^{6}$Dipartimento di Fisica, Universit\'a di Roma 'La Sapienza', P.le Aldo Moro 2, I-00185 Roma, Italy\\
$^{7}$Dipartimento di Fisica ``G. Occhialini'', Universit\`{a} degli Studi di Milano-Bicocca, Piazza della Scienza 3, I-20126 Milano, Italy\\
$^{8}$INFN, Sezione di Milano-Bicocca, Piazza della Scienza 3, 20126 Milano, Italy\\
}

\date{Accepted XXX. Received YYY; in original form ZZZ}

\pubyear{2019}

\begin{document}
\label{firstpage}
\pagerange{\pageref{firstpage}--\pageref{lastpage}}
\maketitle

\begin{abstract}
\noindent
We study the genesis and evolution of super-massive black hole (SMBH) seeds through different formation channels, from PopIII remnants to massive seeds, modeled within the \lgal semi-analytic code. We run the model on the \texttt{Millennium-II} simulation (\texttt{MR-II}) merger trees, as their halo-mass resolution ($\rm M_{vir,res}\!\sim\!10^7M_\odot\,h^{-1}$) allows to study in a cosmological volume ($\rm L_{box}=100\,Mpc\,h^{-1}$) the evolution of atomic-cooling halos ($\rm T_{vir}\!\gtrsim\!10^4\,K$) where intermediate-mass and heavy seeds are expected to form. We track the formation of these seeds according to spatial variations of the chemical and radiative feedback of star formation. Not being able to resolve the first  mini-halos ($\rm T_{vir}\!\sim\!10^3\,K$), we inherit evolved PopIII remnants in a sub-grid fashion, using the results of the \texttt{GQd} model. We also include the formation of heavy seeds in gas-rich massive mergers, who are very rare in the \texttt{MR-II} volume. The descendants of light seeds numerically prevail among our SMBHs population at all masses and $z$. Heavier seeds form in dense environments where close neighbors provide the required UV illumination. Overall, our model produces a $z\!=\!0$ SMBHs population whose statistical properties meet current constraints. We find that the BH occupation fraction highly depends on the seeding efficiency and that the scaling relation between BH and stellar mass, in the dwarf-mass regime, is flatter than in the high-mass range. Finally, a fraction of BHs hosted in local dwarf galaxies never grow since they form at $z\!>\!6$.
\end{abstract}




\section{Introduction}
\label{sec:intro}
Recent observations of quasars (QSOs) with bolometric luminosity in excess of $\rm L_{bol}\!>\!10^{46}erg\,s^{-1}$ already at $z\!\gtrsim\!7$ suggest that super-massive black-holes (SMBHs) of $\rm M_{BH}\!\gtrsim\!10^9M_\odot$ formed very early in the Universe history and built-up their mass within few hundreds of Myr from the Big Bang \citep[e.g.,][for recent reviews]{latif_ferrara2016,valiante2017,inayoshi_visbal_haiman2020}. Although several possible pathways have been proposed, current theoretical models still struggle to identify the initial \textit{seeds} of these massive compact objects and to fit their mass growth within less than $\rm1\,Gyr$.

For instance, the evolution of the first metal-free (PopIII) stars is thought to produce \textit{light} seeds at $z\!\gtrsim\!20$, with typical seed masses of $\rm M_{seed}\!\sim\!10^2M_\odot$ \citep[e.g.,][]{schneider2002,yoshida2003,bromm_larson2004}. This process is thought to happen within \textit{mini-halos}, that is: Dark-Matter structures with virial mass $\rm 10^5\!\lesssim\!M_{vir}/M_\odot\!\lesssim\!10^7$ \citep[e.g.,][]{schaerer2002a,schneider2006a,greif2011}. At high-$z$, the gas content of mini-halos is expected to have negligible metallicity (i.e. $\rm Z\!\simeq\!0$) and hence to cool through roto-vibrational transitions of molecular hydrogen \citep[$\rm H_2$,][]{abel_haiman2000,omukai_palla2001}, which ultimately determines the typical mass of PopIII remnants \citep[e.g.,][]{madau_rees2001,schaerer2002b,schneider2006b,haemmerle2018}. Nevertheless, light seeds would need to grow beyond the theoretical Eddington-limit in order to reach $\rm \sim\!10^{\,9\,}M_\odot$ by $z\!\sim\!7$ \citep[][]{volonteri_rees2005,natarajan2011, pezzulli2017,regan2019,haemmerle2021}.

Alternative models assume a more massive origin of SMBHs in order to relax the constraints on their growth-rates. Indeed, \textit{heavy} or \textit{intermediate-mass} BHs (IMBHs), forming with $\rm 10^3\!\lesssim\!M_{seed}/M_\odot\!\lesssim10^5$ already at $z\!\gtrsim\!10$ allow to more easily reach $\rm M_{BH}\!\sim\!10^{\,8-9}M_\odot$ by $z\sim7$  \citep[e.g.,][]{bromm_loeb2003,omukai_schneider_haiman2008,shang_bryan_haiman2010,agarwal2012,volonteri_bellovary2012,regan2014b,visbal_haiman_bryan2014b,latif2015,valiante2016,sassano2021}. On the other hand, these models generally require peculiar physical conditions whose occurrence is yet poorly constrained over cosmological contexts \citep[e.g.,][]{bromm_loeb2003,latif2015,agarwal2016,latif_volonteri_wise2018}. In particular, the formation of massive $\rm M_{seed}\gtrsim10^{\,4-5}M_\odot$ BHs requires that the cooling of pristine, high-$z$ gas clouds is strongly delayed by the suppression of their $\rm H_2$ content. This can be achieved by different heating mechanisms, such as baryon streaming-velocities \citep[e.g.,][]{tanaka_li2014}, dynamical heating \citep[e.g.,][]{yoshida2003,lodato_natarajan2006,fernandez2014,inayoshi_visbal_kashiyama2015,chon2016} or the presence of UV photo-dissociating backgrounds within the specific Lyman-Werner (LW) band\footnote{the Lyman-Werner band is a interval of UV frequencies responsible for the dissociation of $\rm H_2$ molecules, i.e. $\rm \mathit{h}\nu = [11.2 - 13.6]\ eV$.} \citep[][]{dijkstra2008,omukai_schneider_haiman2008,agarwal2012,latif2014a}. Under the latter conditions, pristine gas clouds might directly collapse into a $\rm\!\sim10^5M_\odot$ BH. The required levels of LW flux ($\rm J_{LW}$ hereafter) and the actual occurrence of this direct-collapse BH-seeding scenario (DCBH) are matters of ongoing debates \citep[e.g.,][]{visbal_haiman_bryan2014a,habouzit2016,agarwal2019}.

Several works analyzed an intermediate scenario leading to $\rm M_{seed}\!\sim\!10^{\,3-4}M_\odot$ \citep[e.g.,][]{ebisuzaki2001,omukai_schneider_haiman2008,devecchi_volonteri2009,reinoso2018,regan2020c}. According to this picture, a dense, nuclear stellar cluster might be formed by early star-formation (SF) episodes. Later on, this structure is led to collapse into a single compact object due to runaway stellar mergers \citep[RSM scenario hereafter, see e.g.,][]{ebisuzaki2001,portegieszwart_mcmillan2002,rasio_freitag_gurkan2004,katz_sijacki_haehnelt2015,das2021a}. Assuming that the conditions for RSM-seeds formation can be verified at $z\!\gtrsim\!15$ \citep[as shown by e.g.][]{lupi2014,sassano2021}, their initial mass allow to fit their growth up to few $\rm 10^8M_\odot$ by $z\!\sim\!7$ under typical, Eddington-limited, thin-disk accretion. Finally, further channels based on baryonic physics have been suggested as alternatives to the above scenarios, envisioning the formation of extremely-massive BH seeds (up to $\rm M_{seed}\!\sim\!10^{\,8-9}M_\odot$) through gas-rich galaxy mergers \citep[e.g.,][]{mayer2015} or the collapse of extreme stellar clusters \citep[e.g.,][]{kroupa2020,escala2021}.

This broad ``BH-seeding problem'' has been widely addressed through a multitude of theoretical and numerical approaches, including hydrodynamical simulations \citep[e.g.,][]{agarwal2012,habouzit2016,ardaneh2018,chon_hosokawa_yoshida2018,dunn2018,degraf_sijacki2019,maio2019,latif2021}, semi-analytic and purely-analytic models \citep[e.g.,][]{volonteri_haardt_madau2003,dijkstra2008,volonteri_lodato_natarajan2008,devecchi_volonteri2009,dijkstra2014,valiante2016,volonteri_reines2016}. The first ones can be designed to reach high mass-resolution and directly track the small-scale processes involved in the formation of SMBH seeds. This limits the volume sizes accessible with current computational capabilities, hindering the extrapolation of their results to cosmological scales. At the same time, large-volume hydrodynamical simulations cannot resolve the evolution of high-$z$ mini-halos, hence generally resorting to simplified BH-seeding models \citep[as in e.g.,][]{dubois2014a,dimatteo2017,ni2020}. On the other hand, semi-analytic and analytic codes employ sets of physically-motivated prescriptions to model the evolution of the baryonic component of Dark-Matter (DM) halos \citep[see e.g.,][]{tanaka_haiman2009,valiante2011,lacey2016,lagos2018}. Consequently, these methods are computationally affordable over wide cosmological boxes or large sets of analytically-reconstructed merger trees while employing detailed models for BH-formation \citep[e.g.,][]{lupi_haiman_volonteri2021,piana2021,sassano2021}.

The goal of this work is to embed a comprehensive and physically-motivated model for BH-seeding into a cosmological context and to follow the evolution of the resulting multi-flavour SMBH population down to $z\!=\!0$. To this aim, we use the \lgal semi-analytic code \citep[][]{guo2011, henriques2015}, which was designed to model galaxy formation and evolution on the merger trees of the \texttt{Millennium} \citep[\texttt{MR},][]{mill} and \texttt{Millennium-II} \citep[\texttt{MR-II}][]{boylankolchin2009} cosmological, N-body simulations. In particular, the latter offers a good compromise between halo-mass resolution ($\rm M_{vir,res}\!\sim\!10^7\,M_\odot\,h^{-1}$) and simulated volume ($\rm L_{box}\!=\!100\,Mpc\,h^{-1}$), allowing to study BH-seeding processes in a cosmological environment. Furthermore, differently from purely-analytic codes, \lgal can access to the 3D spatial distribution of structures within its input merger-trees, allowing to track the high-$z$ environments where SMBHs form. Due to the \milltwo mass resolution, the birthplaces of PopIII stars are inaccessible to \texttt{L-Galaxies}, therefore we exploit the results of the \gqd model \citep[][]{valiante2011,valiante2014,valiante2016} where PopIII evolution and the formation of light seeds is followed self-consistently \citep[see also][for recent updates]{sassano2021,trinca2022}. In this way, we implement the formation of light seeds in a sub-grid fashion, while self-consistently tracking the formation of massive BH seeds. This allows us to explore the role of all currently-envisioned seeding channels in the build-up of the global SMBH population over a cosmological volume.

This paper is organized as follows: in Sect. \ref{sec:model:general_sam_description} we detail our upgrades to \texttt{L-Galaxies}, its interplay with \texttt{GQd} and our BH-seeding prescription. We present our results in Sect. \ref{sec:results} and summarize them in Sect. \ref{sec:conclusions}. Throughout this paper, we use \texttt{PLANCK15} cosmological parameters: $\rm h\!=\!0.673$, $\rm\Omega_{m,0}\!=\!0.315$ and $\rm\Omega_\Lambda\!=\!1\!-\!\Omega_{m,0}\!=\!0.685$ \citep[][]{ade2016}.\vspace{-0.5cm}

\section{Black Holes formation model}
\label{sec:model:general_sam_description}
We embed our BH-seeding prescriptions into the \lgal semi-analytical model \citep[SAM, see e.g.,][]{henriques2015}, which is designed to be applied on the merger trees of either the \texttt{MR} or \texttt{MR-II} simulations. For this work we employ the latter, as it offers the best compromise between cosmological volume ($\rm L_{box}=100\,Mpc\,h^{-1}$) and  halo-mass resolution ($\rm M_{vir,res}\!=\!3.84\times10^7\,M_\odot\,h^{-1}$ for \texttt{PLANCK15} parameters). 
 
\begin{figure}
  \centering
  \includegraphics[width=0.97\columnwidth]{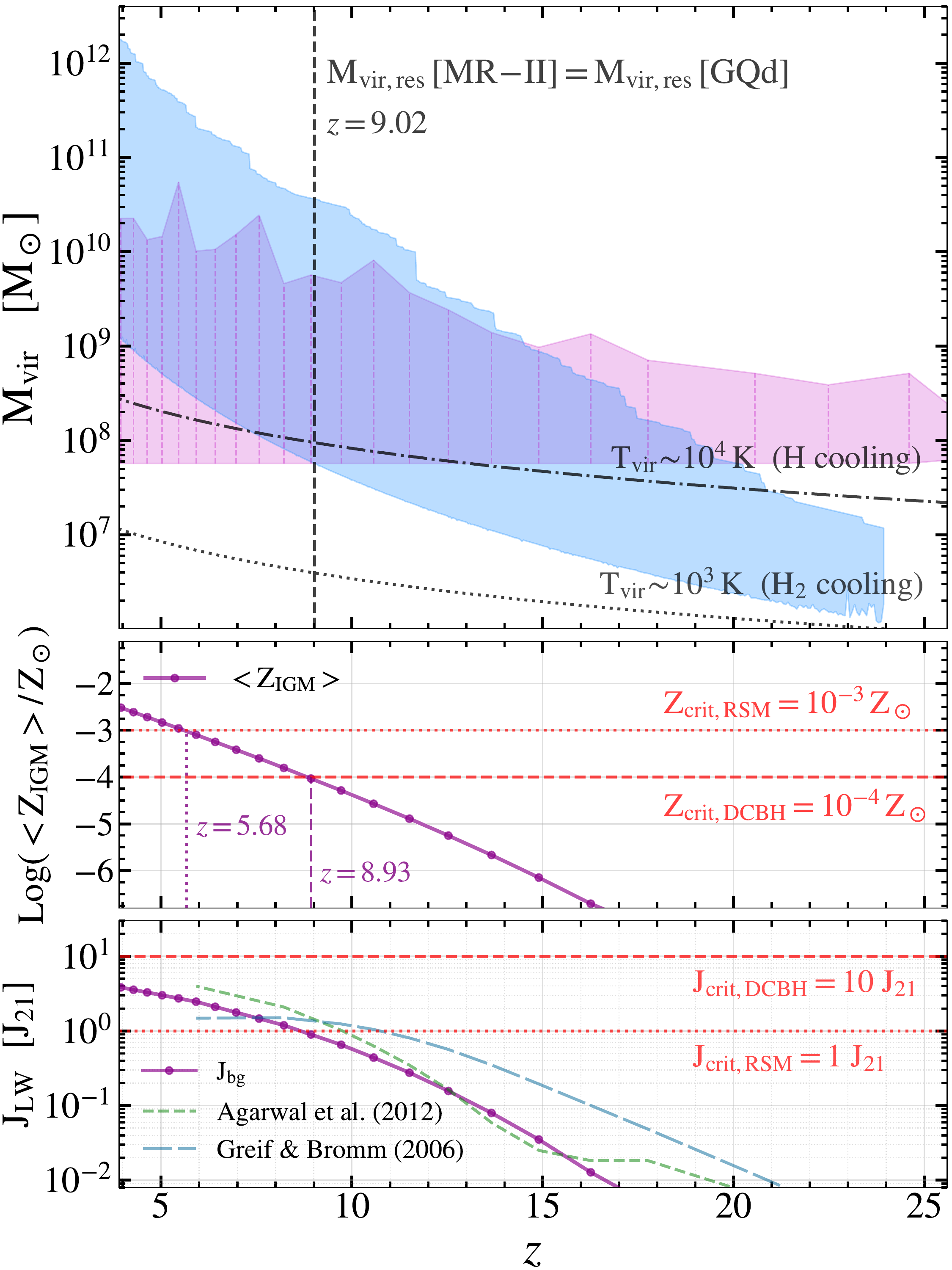}
  \caption{\textit{Upper panel}: comparison between the dynamical ranges of newly-resolved \milltwo halos (pink shaded area) and all the \gqd halos (cyan shaded area), showing the higher mass resolution of \gqd at $z\gtrsim9$. Therefore, \gqd outputs can provide information about the unresolved evolution of structures in \lgal at these very high-$z$. \textit{Middle panel}: Evolution of the average IGM metallicity in \lgal (purple line and dots), tracked over the cosmological volume of the \milltwo ($\rm V_{ol}=(100)^3\,[Mpc\,h^{-1}]^3$). $\rm\langle Z_{IGM}\rangle$ overcomes the critical values of $10^{-4}Z_\odot$ and  $10^{-3}Z_\odot$ respectively at $z\!\sim\!9$ and $z\!\sim\!5.7$. \textit{Bottom panel}: Background level of $\rm J_{LW}$ over the whole \milltwo box, in units of $\rm J_{21}\!=\!10^{\,-21}erg\,cm^{-2}\,s^{-1}\,Hz^{-1} sr^{-1}$ (purple line and dots; see Sect. \ref{sec:model:diffusion_of_metals_and_LW_photons} for details). Our computation provides comparable values to those of \protect\cite{greif_bromm2006} and \protect\cite{agarwal2012}, respectively shown as dashed green and cyan lines.}
  \label{fig:dynaranges}
\end{figure}

This mass-limit implies that all \milltwo halos have $\rm T_{vir}\!\gtrsim\!10^{\,4}K$ at all redshifts above $z\!\gtrsim\!13$ (see Fig. \ref{fig:dynaranges}, upper panel), hence their gas counterparts cool via hydrogen atomic transitions, which defines them as \textit{atomic-cooling} halos \citep[see e.g.,][]{barkana_loeb2001,bodenheimer2011}. In other words, the  \milltwo mass-resolution is not high enough to model the formation of PopIII stars and their evolution into light BH-seeds within $z\!\gtrsim\!15$ mini-halos \citep[e.g.,][]{schaerer2002a,schneider2002,yoshida2003,bromm_larson2004}. We tackle this issue by tracking the formation of light seeds in a sub-grid fashion, feeding the results of the \gqd model \citep[][]{valiante2014,valiante2016,valiante2021} as inputs to \texttt{L-Galaxies}. 

\subsection{The \lgal SAM}
\label{sec:model:lgal_properties_overview}
This work is based on the \lgal model presented by \cite{henriques2015}, with the modifications for the treatment of extremely-minor mergers and the evolution of massive BHs included by \cite{izquierdo-villalba2019,izquierdo-villalba2020}. For further technical details about the physical models summarized below, we refer the reader to these works.

\lgal follows its input merger-trees from high-$z$ down to $z\!=\!0$, associating baryonic counterparts to newly-resolved DM halos and evolving the former in sub-steps ($\rm dt_{step}$) between adjacent input snapshots. Sub-steps are used to track physical processes over typical intervals of $\rm dt_{step}\sim3,10$ and $\rm 20\,Myr$, respectively at $z\!>\!5$, $z\!\sim\!2$ and $z\!<\!1$. The evolution of baryonic structures starts when newly-resolved DM halos at a given initialization redshift $z_{\rm init}$ accrete primordial gas from the IGM, up to a fixed baryonic fraction ($\rm f_b\!=\!0.155$) of their initial mass ($\rm M_{vir,init}$). By following \cite{white_rees1978} and \cite{white_frenk1991}, the newly accreted gas settles in a hot atmosphere with pristine chemical composition (i.e. $\rm75\%$ H and $\rm25\%$ He). Its cooling is regulated by the functions of \cite{sutherland_dopita1993}, which account for both pristine and chemically-enriched plasmas. The mass of cooled gas ($\rm M_{cGas}$) is modelled as a disc-like structure acting as a reservoir for most of the implemented baryonic processes. In the \lgal version of \cite{henriques2015}, the primordial infall has zero-metallicity at all $z$, strongly impacting the potential formation of BH seeds (see Sect. \ref{sec:model:average_igm_metallicity_and_jlw}).

SF can set-in at any sub-step if $\rm M_{cGas}$ overcomes a critical threshold. A fixed fraction ($\rm R\!=\!0.43$) of newly-formed stellar mass is instantaneously recycled into $\rm M_{cGas}$, due to the explosion of massive, short-lived stars. The associated SNe feedback chemically enriches $\rm M_{cGas}$ and re-heats a fraction of it into the hot-phase gas. If the feedback episode is energetic enough, part of the hot gas is then ejected beyond the halo virial radius $\rm R_{vir}$, producing chemically enriched SNe ejecta which are stored in a dedicated reservoir ($\rm M_{ej}$). These mass transfers are therefore responsible for the mixing of the chemical yield of exploding SNe throughout $\rm M_{cGas}$, hot gas and $\rm M_{ej}$ which are consequently enriched self-consistently during the galaxies evolution. In detail, the metallic yield of SNe is computed according to a Chabrier IMF and the cold-gas metallicity at the time of the last SF episode. The newly-produced metals are immediately transferred back into the cold gas component (via ``instantaneous recycling'') and propagated to the hot gas and $\rm M_{ej}$ through the appropriate mass transfers \citep[see][for details]{guo2011,yates2013,delucia2014,henriques2015}.

Galaxy mergers follow the hierarchical assembly of their DM hosts and produce merger-induced SF bursts within their remnants. We use the treatment for \textit{smooth-accretion mergers} of \cite{izquierdo-villalba2019} which introduced significant improvements to the predicted morphology of galaxies. $\rm M_{cGas}$ fuels the growth of central BHs either as a consequence of galaxy mergers or disk-instabilities driven by the secular evolution of galaxies. According to the model introduced by \cite{izquierdo-villalba2020}, the mass growth of BHs proceeds smoothly in time, via the consumption of a thin accretion disk \citep[e.g.][]{shakura_sunyaev1973} or through an advection-dominated accretion flow \citep[ADAF, e.g.,][]{rees1982}. In particular, the thin-disk mass-growth is based on a 2-phase accretion-model, where the first phase proceeds at the Eddington-rate ($\rm\lambda_{Edd}\!=\!1$), in a quasar (QSO)-like fashion. On the other hand, the second phase mimics a quieter AGN-mode with lower, time-declining $\rm\lambda_{Edd}$.Accreting BHs pass from the first phase to the following one as soon as they consume 70\% of the mass available for their growth, which is stored in a dedicated baryonic reservoir \citep[see][for details]{marulli2008a,bonoli2009,izquierdo-villalba2020}. The accretion of $\rm M_{cGas}$ is complemented by a growth-mode fuelled by hot gas, whose activation is independent from the BH-host mergers \citep[e.g.,][]{croton2006}. In our model, BHs coalesce as soon as their host galaxies merge, hence we neglect the implementation of BH-BH merger delays introduced by \cite{izquierdo-villalba2020}. We further neglect the spin-evolution, recoil-velocity, and gravitational waves (GWs) emission described in \cite{izquierdo-villalba2020,izquierdo-villalba2022a}, as we plan to explore their complex interactions with BH-seeds formation in a future work.

\subsection{IGM metals and LW photons in \lgal}
\label{sec:model:diffusion_of_metals_and_LW_photons}
\lgal was designed and calibrated to reproduce the stellar mass ($\rm M_*$) function and the global properties of galaxies at $z\lesssim4$ \cite[][]{guo2011,henriques2015}. Consequently, the high-$z$ chemical and radiative feedback of SF processes on the IGM was not modelled in detail. To provide a physical base to our BH-seeding prescriptions, we introduce a treatment for the average IGM metallicity and the LW background (respectively $\rm\langle Z_{IGM}\rangle$ and $\rm J_{bg}$), as well as a model for their spatial variations \citep[see e.g.,][]{agarwal2014,regan2014b,habouzit2016,maio2019,visbal_bryan_haiman2020}. More in detail, we use the \cite{henriques2015} version of \lgal to compute uniform backgrounds of $\rm Z_{IGM}$ and $\rm J_{LW}$ (Sect. \ref{sec:model:average_igm_metallicity_and_jlw}), on top of which we model local variations (Sect. \ref{sec:model:local_metallicity} and \ref{sec:model:local_LyWer}).

\subsubsection{Uniform $\rm Z_{IGM}$ and  $\rm J_{LW}$ backgrounds}
\label{sec:model:average_igm_metallicity_and_jlw}
In our modified version of \texttt{L-Galaxies}, the metallicity of the primordial infall in each newly-initialized structure at $z_{\rm init}$ is fixed to the average IGM metallicity,  $\rm\langle Z_{IGM}\rangle(\mathit{z}_{init})$. We obtain the latter from the ratio between the total mass of heavy elements ejected in the IGM by all halos, and the total amount of baryonic mass not yet collapsed into halos, i.e. $\rm M_{IGM,gas}(\textit{z})$. Namely:
\begin{equation}
\rm\left\langle Z_{IGM}\right\rangle (\mathit{z}) = \frac{\sum_{\,i}\,M_{ej,met}^i(\mathit{z})}{M_{IGM,gas}(\mathit{z})} = \frac{\sum_{\,i}\,M_{ej,met}^i(\mathit{z})}{\left[M_{DM,box}-\sum_{\,i} M_{vir}^{i}(\mathit{z})\right]\cdot f_b}\,,
\label{eq:zigm}
\end{equation}
where the summations include all the structures in the \milltwo box at a given $z$. In particular, $\rm M_{ej,met}^i(\mathit{z})$ is the mass of metals in the $\rm M_{ej}$ reservoir of the i-th galaxy (see Sect. \ref{sec:model:lgal_properties_overview}). $\rm M_{IGM,gas}(\mathit{z})$ is the total mass of diffuse gas in the IGM, obtained by multiplying the baryon fraction $\rm f_b$ for the total amount of diffuse DM. We obtain the latter from the difference between the total DM mass in the \milltwo box, $\rm M_{DM,box}$, and the mass collapsed in halos, $\rm\sum_i M_{vir}^{\,i}$, with $\rm M_{DM,box}\!=\! N_p\cdot m_p$ where $\rm N_p$ and $\rm m_p$ are the particles number and mass of \texttt{MR-II}. The middle panel of Fig. \ref{fig:dynaranges} shows the $\rm\langle Z_{IGM}\rangle(\mathit{z})$ we obtain. The latter increases towards low redshift as a result of SF activity and SNe feedback in the whole \milltwo volume, reaching $10^{-4} Z_{\odot}$ by $z\!\sim\!9$ and $10^{-3} Z_{\odot}$ by $z\!\sim\!6$.

Similarly, we also compute a uniform $\rm J_{bg}$ over the \milltwo box, in order to account for the large mean free-paths of LW photons \citep[up to $\rm\sim\!100\,Mpc$ at $z\!\gtrsim\!6$, see][]{ahn2009}. For this, we closely follow the approach of \cite{agarwal2012}:
\begin{equation}
    \rm J_{bg} = f_{esc}\frac{\mathit{h}\,c}{4\pi m_H}\,\eta\,\rho_*\,(1+\mathit{z})^3\,,
\label{eq:jlw_background_agarwal}
\end{equation}
where $h$ is the Planck constant, $\rm c$ is the speed of light and $\rm m_H$ is the hydrogen atom mass. $\rm\eta=4\times10^3$ is the number of $\rm H_2$-dissociating photons produced per stellar baryon \citep[see][]{agarwal2012,wise2014,maio2019}, while $\rm\rho_*$ is the mass density of active stars in the whole \milltwo box. Finally, $\rm f_{esc}$ is the escape fraction of LW photons, which we fix to $\rm f_{esc}=1$ by assuming that the galaxies emitting LW photons have been depleted by $\rm H_2$ due to their recent SF \citep[see the discussion in e.g.][]{ahn2009,agarwal2012}. In the bottom panel of Fig. \ref{fig:dynaranges} we show the redshift evolution of our $\rm J_{bg}$, whose values are close to the ones obtained by \cite{greif_bromm2006} and \cite{agarwal2012}.

\subsubsection{Sources of IGM metals}
\label{sec:model:local_metallicity}
To describe the patchy diffusion of metals in the IGM, we assume that strong SNe feedback episodes power the propagation of metallic winds \citep[e.g.,][]{bertone2005,bertone2007,sharma2014}. We model the latter as spherically-symmetric, pressure-driven shells around galaxies by tracking the time-evolution of their radius $\rm r_s$. We follow the analytic approximation of \cite{dijkstra2014} which agrees with the detailed calculations of \cite{madau_ferrara_rees2001, kim_ostriker_railenau2017, yadav2017,fielding2018}:
\begin{equation}
    \rm r_s\left(M,t\right) = \left( \frac{E_0\,\,\xi M_{*,new}}{m_p\,n} \right)^{1/5}\,t^{2/5}\,,
\label{eq:metalshell}
\end{equation}
where $\rm E_0\!=\!10^{51}erg\!=\!1\, Bethe$ is the energy released by a single SN explosion, $\rm\xi$ is the number of SNe per unit-mass of the newly formed stellar mass $\rm M_{*,new}$ within resolved DM halos (i.e. without considering contributions from un-resolved star formation), and $\rm m_p$ is the proton mass. We fix $\xi$ to the recycled fraction of \texttt{L-Galaxies}, namely $\rm\xi=R=0.43$ (see Sect. \ref{sec:model:lgal_properties_overview}). The quantity $\rm n$ at the denominator of Eq. \ref{eq:metalshell} is the number density of baryons in the IGM:
\begin{equation}
    \rm n=60\,\Omega_{b,0}\,\frac{\rho_{crit,0}}{m_p}\,(1+\textit{z})^3\,.
    \label{eq:igm_density}
\end{equation}
Here the factor 60 represents the typical density-contrast, with respect to the cosmic average, of the medium in which SNe ejecta expand \citep[see][]{dijkstra2014}.

For each new SF event which is strong enough to produce $\rm M_{ej}$ (see Sect. \ref{sec:model:lgal_properties_overview}), we calculate the evolution of $\rm r_s\left(M,t\right)$ using Eq. \ref{eq:metalshell}. Since $\rm M_{*,new}$ regulates the speed of the shell expansion (see Eq. \ref{eq:metalshell}), fronts associated to strong SF events might reach slower shells launched earlier in time. In this case, we continue to follow only the evolution of the faster front.
For simplicity, we neglect energy losses during the fronts interaction as well as during the expansion within the IGM.

\subsubsection{Sources of LW photons}
\label{sec:model:local_LyWer}
UV photons produced by SF events can travel for cosmological distances from their source, even in the high-$z$ neutral IGM. Therefore, computing the $\rm J_{LW}$ received by an observer from a star-forming galaxy (i.e., a LW source) would require a computationally expensive radiative-transfer approach. To avoid this, we follow the approach of \cite{ahn2009,agarwal2012} and \cite{dijkstra2014}, i.e.:
\begin{equation}
    \rm J_{LW}\,(d_L) = \frac{f_{esc}}{4\pi\,\Delta\nu}\frac{Q_{LW}\,\,\xi M_{*,new}}{4\,\pi\, d_L^2}\,f_{mod}\,.
\label{eq:lwlocal}
\end{equation}
Here $\Delta\nu$ is the LW frequency-band and $\rm \xi M_{*,new}$ is the newly-formed stellar mass of short-lived, instantly-recycled massive stars as in Eq. \ref{eq:metalshell} \citep[i.e., we assume that only these stars produce significant $\rm J_{LW}$ at each SF event; see][]{schaerer2002a, agarwal2012,wise2014,maio2019}. $\rm d_L = d_c\,(1+\mathit{z}_{rel})$ is the luminosity distance between the observer (at $z_{\rm obs}$) and the LW source (at $z_{\rm s}$), while $\rm d_c$ is their comoving distance in $\rm Mpc$ and $\rm\mathit{z}_{rel}=(1+\mathit{z}_{\,obs})/(1+\mathit{z}_{\,s})$ is the LW source redshift computed from the observer perspective. The quantity $\rm Q_{LW}$ is the energy rate produced within $\Delta\nu$ by $\rm \xi M_{*,new}$. We fix $\rm Q_{LW}=10^{44}\,erg\,s^{-1}$ for the PopII/PopI stellar populations modelled by \lgal \citep[see, e.g.,][]{greif_bromm2006,agarwal2012}. Finally, $\rm f_{mod}$ takes into account the action of redshift and IGM absorption on the propagation of LW photons. Following the phenomenological prescription of \cite{ahn2009}:
\begin{equation}
    \rm f_{mod}\!=\!\left\{\!\!\!
    \begin{array}{lr}
         \rm 1.7\,\exp\,[-(\,d_c\,/\,116.26\,\alpha\,)^{\,0.68}\,]-0.7 &\rm \!\!d_c\leq97.39\alpha\,,  \\
         \rm 0 &\rm \!\!d_c>97.39\alpha\,,
    \end{array}\right.
    \label{eq:jlw_fmod_factor}
\end{equation}
where
\begin{equation}
    \rm \alpha=\left(\frac{0.7}{h}\right)\,\left(\frac{0.27}{\Omega_{m\,;\,0}}\right)^{1/2}\,\left(\frac{21}{1+\textit{z}_s}\right)^{1/2}\,,
    \label{eq:jlw_alpha_factor}
\end{equation}
and $\rm d_{c}\!=\! 97.39\alpha$ is the maximum $\rm d_c$ at which LW photons can be observed before being redshifted out of the LW band or absorbed by the IGM \citep[see e.g.][]{ahn2009}. Throughout the paper, we use the units of $\rm J_{21}\!=\!10^{-21}erg\,s^{-1}\,cm^{-2}\,Hz^{-1}\,sr^{-1}$ for the Lyman-Werner flux.

\subsubsection{Spatial variations of $\rm Z_{IGM}$ and $\rm J_{LW}$}
\label{sec:model:actual_computation_of_local_metals_and_jlw}
The equations detailed in the previous sections allow us to compute the local values of IGM chemical-enrichment ($\rm Z_{local}$) and LW illumination ($\rm J_{local}$). To avoid computationally-expensive approaches, we only evaluate $\rm Z_{local}(\overrightarrow{x}_{\rm s})$ and $\rm J_{local}(\overrightarrow{x}_{\rm s})$ at the specific positions $\rm \overrightarrow{x}_{\rm s}$ of newly-resolved halos which may host the formation of RSM or DCBH seeds. For both quantities, we compute the superposition of their uniform backgrounds and the total contribution of neighboring galaxies. 
For $\rm J_{local}$ we simply write:
\begin{equation}
\rm J_{local}(\overrightarrow{x}_{\rm s})=J_{bg} + J_{PLC}(\overrightarrow{x}_{\rm s})\,,
\label{eq:jlocal_definition}
\end{equation}
where $\rm J_{PLC}(\overrightarrow{x}_{\rm s})$ is the sum of all the contributions from galaxies in the past light-cone (PLC) of the BH-seeding candidate, computed as in Eq. \ref{eq:lwlocal} (see Appendix \ref{sec:appendix:jplc_computation} for technical details). For $\rm Z_{local}(\overrightarrow{x}_{\rm s})$, we consider the mass of metals and gas determined by $\rm\langle Z_{IGM}\rangle(\mathit{z}_{init})$ through the primordial infall of gas, within the $\rm R_{vir}$ of the newly-initialized halo. To these masses of metals and gas, we separately sum the contributions provided by metallic winds launched by neighboring galaxies. In particular, we only sum a fraction $\rm f_m=(r^{\,i}_s/R_{vir})^3$ of the metals and gas within each shell i with radius $\rm r^{\,i}_s$ (given by Eq. \ref{eq:metalshell}) which can reach $\rm\overrightarrow{x}_{\rm s}$ at $\rm \mathit{z}_{init}$. For simplicity, in the following we drop the notation $\rm\overrightarrow{x}_{\rm s}$ on the spatially varying quantities introduced in this section.

With this method, directly accounting for all the possible sources of $\rm J_{LW}$ or metallic winds for each newly-resolved halo in the whole box is computationally prohibitive. Therefore, in our calculation of $\rm Z_{local}$ and $\rm J_{local}$, we consider as metal and LW sources only galaxies that can significantly influence their surroundings with metallic winds or LW flux. More in detail, we define as \textit{metal sources} all galaxies whose associated metallic-shell extends to at least $10$ times the $\rm R_{vir}$ of their DM host. On the other hand, \textit{LW sources} are star-forming galaxies which produce at least $\rm J_{LW}=10\,J_{21}$ at a distance $\rm d_L$ equal to the $\rm R_{vir}$ of their DM host (i.e. for which $\rm J_{LW}(R_{vir})=10\,J_{21}$). We underline that the definitions of these source lists are a trade-off between the inclusion of the highest possible number of sources in each group and reasonable execution-times of our model. Furthermore, the calculation of $\rm Z_{local}$ and $\rm J_{local}$ is only necessary until the last RSM seed can form, therefore we stop storing metal and LW sources as soon as $\rm Z_{IGM}\!>\!Z_{crit,RSM}$.

Metal and LW sources are stored only once, in a preliminary run of \texttt{L-Galaxies}, and used in subsequent runs. During the latter, for each BH-seeding candidate, we compute $\rm Z_{local}$ and $\rm J_{local}$ by following an inside-out procedure. This separately accounts for the contributions of eventual SF episodes within: i) the newly-resolved halo, ii) all its neighbors belonging to the same FoF group\footnote{Here we refer to a FoF as a gravitationally-bound group of DM halos. During its execution, \lgal serially analyzes single FoFs by tracking the evolution of baryonic matter within them.} and iii) galaxies belonging to different FoFs, included in our metal or LW sources lists defined above. In particular, the computation of $\rm J_{PLC}$ proceeds backwards in time and farther away from the time and position where a new BH-seeding candidate is identified for the first time. To avoid the inefficient accounting of the whole catalog of LW sources for each newly-resolved BH-seeding candidate, we only integrate $\rm J_{PLC}$ until the distance at which, if the brightest LW source was to be found there, it would produce a negligible contribution to $\rm J_{local}$ (i.e. 1/100th of $\rm J_{crit}$, see Sect. \ref{sec:model:BH_seeding:RSM_and_DCBH_seeding_in_LGal} for the $\rm J_{crit}$ definition). This approximation allows us to significantly reduce the execution time of our past-light cone integration while still ensuring the convergence of our $\rm J_{PLC}$ computation. Finally, for each BH seed formed, we store the information of its \textit{metal-/LW-contributors}, i.e. all the galaxies actually providing metals or LW photons.

\subsection{The \gqd model}
\label{sec:model:gqd_properties_overview}
The \gqd code reconstructs the merger-trees of DM halos via a Monte Carlo algorithm \citep[based on the extended Press-Schechter formalism, see][]{lacey_cole1993,volonteri_haardt_madau2003} and follows the evolution of their baryonic counterparts in time. Merger-trees are constructed backwards in redshift, recursively splitting a single initial halo and its progenitors in two parent components \citep[see][]{salvadori_schneider_ferrara2007,valiante2011,valiante2016}. Thanks to this, \gqd can reach higher mass-resolution than the \milltwo at $z\!\gtrsim\!9$ and follow the evolution of high-$z$ mini-halos (see Fig. \ref{fig:dynaranges}, upper panel).

\gqd includes recipes for tracking gas cooling, \ptre and PopII star formation, enrichment of the inter-stellar medium (ISM) by both metals and dust, SNe and AGN feedback, as well as the physics involved in galaxy interactions and SMBHs growth \citep[see][and \citealt{valiante2014} for technical details]{salvadori_ferrara_schneider2008,valiante2011}. Most importantly for our work, SMBHs in \gqd form via  both light and heavy seeding channels \citep[see][]{valiante2016,pezzulli2017}. The former are obtained as $\rm M_{seed}\!\lesssim\!3\times10^2M_\odot$ remnants of PopIII stars of (40-140)$\rm M_\odot$ and (260-300)$\rm M_\odot$, forming in halos with ISM metallicity $\rm Z_{ISM}\!<\!Z_{cr}\!=\!10^{-3.8} Z_\odot$ and $\rm J_{LW}\!<\!300 J_{21}$ \citep[see][]{valiante2016,debennassuti2017}. Heavy seeds of $\rm M_{seed}\!=\!10^5M_\odot$ instead form under $\rm J_{LW}\!>\!300 J_{21}$ in metal-poor ($\rm Z\!<\!Z_{cr}$) atomic-cooling halos.

In this work, we use the \gqd results obtained for 10 different realizations of the merger-tree of a DM halo with a final $\rm M_{vir}=10^{13}M_\odot$, hosting a luminous QSO at $z=2$ \citep[see][]{valiante2021}. Each of these \gqd runs tracks a single merger tree, hence modelling a moderately-biased region of the Universe instead of different cosmological environments simultaneously over large-volumes \citep[but see][for an extension of \gqd to a whole AGN population]{trinca2022}. Consequently, baryonic properties such as stellar mass and SFR obtained by \gqd cannot be assumed as valid initial conditions for \milltwo structures. We note that \cite{sassano2021} recently presented an updated version of \gqd which follows the formation of light, intermediate and heavy seeds. Nevertheless, the \milltwo mass-resolution allows to resolve the halos hosting the formation of RSM and DCBH seeds, hence we do not use their results.

\subsection{SMBH-seeding prescription}
\label{sec:model:BH_seeding_prescriptions}
We check for the occurrence of favourable BH-seeding conditions in every newly-resolved halo. Our model includes the formation of BH seeds over a wide mass range, from $\rm M_{seed}\!\sim\!10^2\,M_\odot$ to $\rm M_{seed}\!\sim\!10^{\,5}\,M_\odot$, in four different BH-seeds flavours: i) light PopIII remnants, ii) intermediate-mass BHs resulting from the RSM scenario, iii) heavy DCBHs originating from the monolithic compression of pristine gas clouds and finally iv) merger-induced direct-collapse BHs (miDCBHs) produced at the center of gas-rich galaxy mergers.
\begin{figure}
  \centering
  \includegraphics[width=0.97\columnwidth]{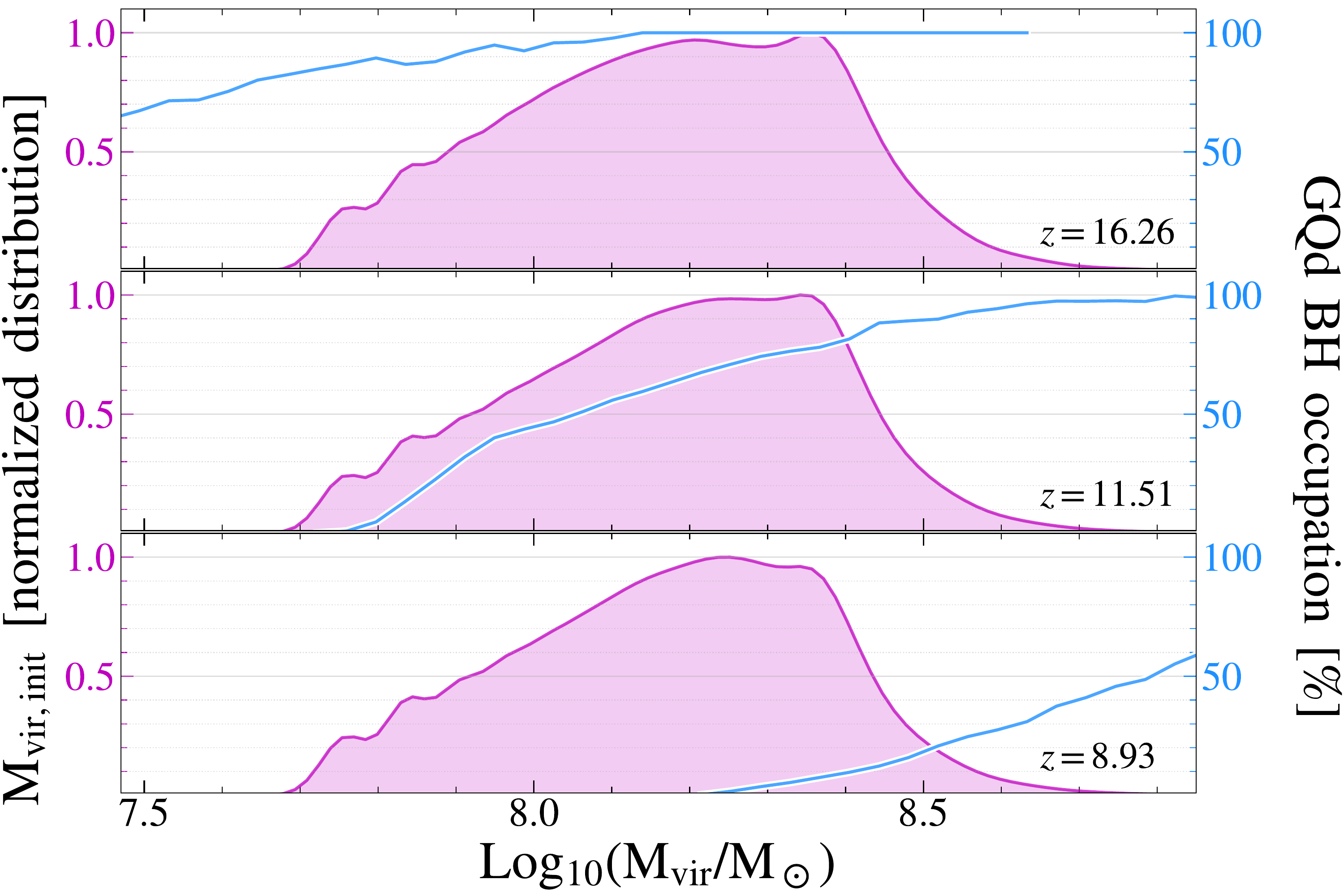}
  \caption{Comparison between the normalized distribution of $\rm M_{vir,init}$ (pink shaded areas) and the BH occupation-fraction within the \gqd halos used for grafting (solid cyan lines) as a function of $\rm M_{vir}$. From top to bottom, we sample the epoch of grafting in its initial, intermediate and final phases. Most of the \lgal halos grafted from \gqd can inherit a BH at $z\gtrsim12$, while an increasingly large fraction of low-mass structures with $\rm M_{vir}\!\lesssim\!5\times10^8\,M_\odot$ at $z\lesssim12$ is associated to \gqd halos without BHs.}
  \label{fig:GQd_BH_occupation}
\end{figure}

\subsubsection{Grafting of light-seed descendants from \gqd}
\label{sec:model:BH_seeding_prescriptions:GQd_Grafting}
The first step of our BH-seeding model is to statistically account for the formation and evolution of light PopIII remnants, as this cannot be resolved by \lgal on the \milltwo merger-trees. For this, we populate newly-initialized structures of the \milltwo with evolved light seeds descendants simulated by the \gqd version presented in \cite{valiante2016} and \cite{valiante2021}. We call this process \textit{grafting} of \gqd information into \lgal structures. This procedure does not depend on $\rm Z_{local}$ nor $\rm J_{local}$, since these cannot be tracked during the unresolved evolution of newly-initialized halos. Rather, we match the halos of the \milltwo and \gqd through their $\rm M_{vir}$ and $z$. In particular, to find a \gqd counterpart of a \milltwo structure newly-resolved with $\rm M_{vir,init}$ at $z_{\rm init}$, we consider all \gqd halos between two consecutive snapshots of the \texttt{MR-II}, namely: $z_{\rm init}\leq z_{\rm\texttt{GQd}}<z_{\rm prev}$. Here $z_{\rm\texttt{GQd}}$ is the redshift of \gqd outputs and $z_{\rm prev}$ the redshift of the snapshot previous to $z_{\rm init}$. We randomly extract a \gqd halo from this sample, within a $\rm M_{vir}$ bin of 0.5 dex, centered on $\rm M_{vir,init}$. This matching procedure is straightforward at $8\!\lesssim\! z\!\lesssim\!16$, i.e. in the regions where the dynamical ranges of \gqd and the one of newly-resolved \milltwo halos overlap (respectively, cyan and pink shaded areas in Fig. \ref{fig:dynaranges}, upper panel). On the other hand, at $z\!>\!16$, the most massive, newly-resolved structures of the \milltwo show $\rm M_{vir,init}>M^{max}_{vir,\gqd}(\mathit{z}_{init})$, where $\rm M^{max}_{vir,\gqd}(\mathit{z}_{init})$ is the maximum virial mass of \gqd halos at $z_{\rm init}$. In these specific cases, for our grafting procedures we use \gqd halos with: $\rm M_{vir,\gqd}(\mathit{z}_{\texttt{GQd}})\geq Log_{10}[M^{max}_{vir,\gqd}(\mathit{z}_{init})]-0.25$.
\begin{figure}
  \centering
  \includegraphics[width=0.97\columnwidth]{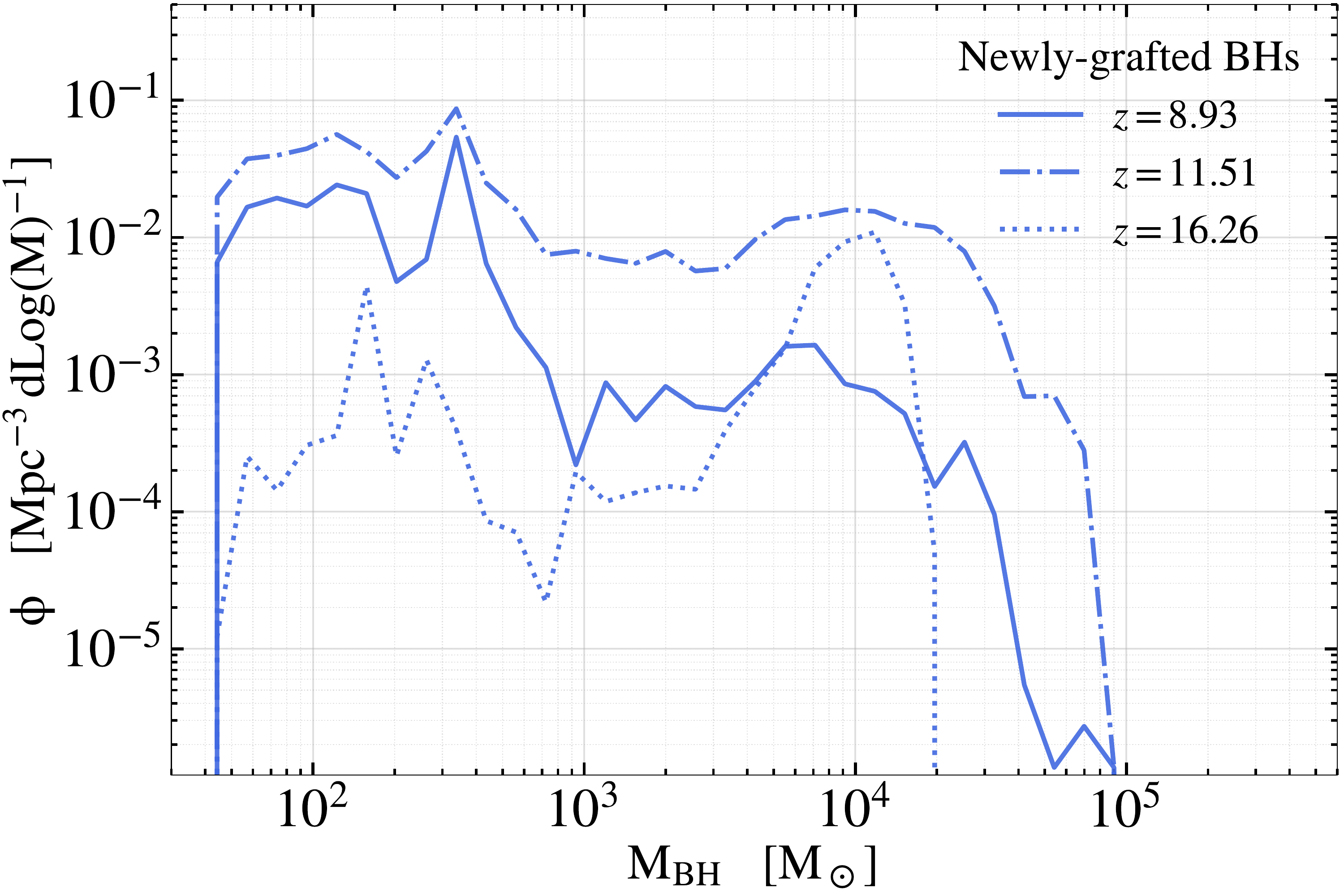}
  \caption{Mass function of evolved light-seed descendants grafted into \lgal structures from \gqd outputs. We show the same redshifts as in Fig. \ref{fig:GQd_BH_occupation} to sample the progress of our grafting procedures at three different moments (respectively: dotted, dashed-dotted and solid light-blue lines). We note that BHs inherited from \gqd descend from light-seeds formed at $z\!>\!20$ in \texttt{GQd}, hence showing an evolved mass distribution spanning $>\!2$ dex already at $z\!>\!16$.}
  \label{fig:BHMF_of_newly_grafted_BHs}
\end{figure}
For a fixed $\rm M_{vir, init}$, it is then likely that the \gqd counterparts of \milltwo halos resolved at $z>16$ are less massive than those of \milltwo structures resolved at $z<16$. We checked that this does not bias our results, since \milltwo halos resolved at $z>16$ outside the \gqd dynamic range only represent the $0.38$\% of all halos which are grafted from \texttt{GQd}.

Finally, we interrupt our grafting procedure as soon as one of the following conditions is verified:
\begin{enumerate}
    \item $\rm \langle Z_{IGM}\rangle\!>\!Z_{\,crit}=10^{\,-4}Z_\odot$,\vspace{1mm}
    \item the mass resolution of \gqd overcomes $\rm M_{vir,res}$.
\end{enumerate}
The first condition generally marks the inhibition of PopIII SF and hence the production of light seeds \citep[see e.g.,][]{bromm_larson2004,schneider2006b,maio2010, valiante2016,pezzulli2017}. The second implies that \gqd no longer tracks the evolution of halos with $\rm M_{vir}\!<\!M_{vir,res}$. In both cases, \gqd cannot provide information about unresolved PopIII remnants. Both conditions are coincidentally verified at $z\!\sim\!9$, as shown in the upper and middle panels of Fig. \ref{fig:dynaranges}. \vspace{1.0mm}\\

\noindent
\textit{Information inherited from \gqd}\vspace{1.0mm}\\ 
\noindent
Once \gqd and \lgal halos are matched, we use the \gqd $\rm M_{BH}$ and seed-type to initialize the central BH in the corresponding \lgal structure. We do not account for other baryonic quantities in order to avoid biases due to the different implementation of physical processes between \gqd and \texttt{L-Galaxies}. In addition, as introduced in Sect. \ref{sec:model:gqd_properties_overview}, \gqd follows the evolution of a relatively biased region of the Universe, hence its outputs cannot account for all the different cosmological environments simulated by the \milltwo. 

The outputs of \gqd we use include both light and heavy seeds (see Sect. \ref{sec:model:gqd_properties_overview}). These two classes can mix within their \gqd evolution, so we consider as heavy seed-types all the \gqd BHs with at least one heavy progenitor, hence joining the ``mixed'' and ``heavy'' classes of \cite{valiante2016}. We ignore the heavy seeds of \gqd since the \milltwo can resolve the atomic-cooling halos in which DCBHs are expected to form. Our grafting procedure naturally reproduces the occupation of \gqd light-seed descendants on the \milltwo halos, leaving a fraction of them without central BHs. We analyze this in Fig. \ref{fig:GQd_BH_occupation} by comparing the normalized distribution of $\rm M_{vir,init}$ of \milltwo halos (pink areas and left y-scale) to the BH occupation in \gqd halos at the corresponding $\rm M_{vir}$ (solid cyan lines and right y-scale). We show  $z\!\sim\!16$, $z\!\sim\!12$ and $z\!\sim\!9$ to uniformly sample the epoch of grafting. The figure shows that an increasing fraction of newly-resolved halos in \lgal does not inherit BHs due to their low occupation in \gqd halos, especially at $\rm M_{vir}\lesssim3\times10^8M_\odot$ and $\rm z\lesssim12$. More in detail, the $\rm M_{vir,init}$ range of \milltwo halos remains relatively constant at any $z\!>\!9$, with the majority of structures being initialized with $\rm M_{vir,init}\!\sim\!10^{8.4}M_\odot$. Taking this value as an example, we note that $100\%$ of the \milltwo halos initialized at $z\!\sim\!16$ inherits a \gqd BH, while this fraction drops to only $10\%$ for the same $\rm M_{vir,init}$ at $z\!\sim\!9$. This wide variation is a consequence of how the $\rm M_{vir,init}$ range samples different kinds of \gqd environments at different times. Indeed, DM halos of $\rm M_{vir}\!\sim\!10^{8.4}M_\odot$ represent a $\gtrsim\!3\sigma$ peak of the DM density field at $z\!\sim\!20$ \citep[see][]{barkana_loeb2001}. These rare and massive overdensities in \gqd are likely to have hosted the formation of light seeds at $z\!\gtrsim\!25$, early mergers with smaller structures and significant evolution already at high-$z$. On the contrary, halos with $\rm M_{vir}\!\sim\!10^{8.4}M_\odot$ at $z\!\sim\!9$ represent more common, $\sim\!1\sigma$ fluctuations, associated to less biased regions. Only a small fraction of their progenitors hosted the formation of light seeds, hence ultimately providing a low BH-occupation to \milltwo halos. These differences are also reflected in the mass distribution of the BHs inherited from \texttt{GQd}, as shown in Fig. \ref{fig:BHMF_of_newly_grafted_BHs}. Indeed, although descending exclusively from light seeds, BHs inherited at $z\!\sim\!16$ are typically massive, with a significantly evolved mass distribution peaked at $\rm M_{BH}\!\sim\!10^4M_\odot$ (dotted line). On the contrary, BHs inherited at $z\!\sim\!9$ typically show $\rm M_{BH}\!\lesssim\!4\times10^2M_\odot$, which is indicative of their quieter evolution in \texttt{GQd}.\vspace{1.0mm}\\

\noindent
\textit{Grafting probability}\vspace{1.0mm}\\
\noindent
Since our grafting procedures do not depend on $\rm Z_{local}$ and $\rm J_{local}$, the inheritance of \gqd BHs is not regulated by the evolution of the IGM gas properties. Therefore, in order to control the abundance of BHs inherited from \texttt{GQd}, we include the possibility to bypass their actual grafting into \lgal structures even if the latter were matched to \gqd halos hosting a BH. In particular, since most of the structures of the \milltwo are resolved with $\rm M_{vir,init}\!\sim\!2\times10^8M_\odot$, we model a \textit{grafting probability} ($\rm\mathcal{P}_{graft}$) as a function of $\rm M_{vir,init}$:
\begin{equation}
    \rm \mathcal{P}_{graft}(M_{vir,init})=G_p \cdot\left(\frac{M_{vir,init}}{M_P} \right)\,.
    \label{eq:grafting_probability}
\end{equation}
$\rm G_p$ is a free parameter ranging from 0 to 1 which controls the magnitude of $\rm\mathcal{P}_{graft}$ at the characteristic mass $\rm M_P=10^8M_\odot$, and it is saturated to 1 when $\rm M_{vir,init}\!>\!M_P/G_p$. In this way, the probability of actually grafting a \gqd BH is reduced linearly with respect to $\rm M_{vir,init}$, allowing to modulate the abundance of grafted BHs within newly-resolved, low-mass halos of the \texttt{MR-II}. The choice of the linear scaling of $\rm\mathcal{P}_{graft}$ with respect to $\rm M_{vir,init}$ is motivated by the simplicity of this empirical prescription, which allows us to easily gain insights about the effect of our grafting procedures at different $\rm M_{vir,init}$ by acting on the $\rm G_p$ parameter.

Unless otherwise stated, we show the results obtained for a run with $\rm G_p\!=\!1$, since this value ensures the best agreement of our results with recent observations of the $z\!=\!0$ BH mass-function (BHMF). In order to bracket the effects of $\rm G_p$, we occasionally compare the results of this fiducial run with those obtained by setting $\rm G_p\!=\!0.01$ and leaving unchanged all the parameters of our model. This choice of $\rm G_p$ is meant to be a representative example providing a significantly lower occupation of \gqd BHs than $\rm G_p\!=\!1$, and illustrate the effect of a global BH-seeding efficiency on our results.

\subsubsection{RSM, DCBH and miDCBH seeds in \lgal}
\label{sec:model:BH_seeding:RSM_and_DCBH_seeding_in_LGal}
After the initialization with \texttt{GQd}, we check if the conditions for the formation of intermediate or heavy seeds in \texttt{L-Galaxies} are verified. By default, we assume that halos hosting the descendants of \gqd PopIII remnants have been chemically enriched beyond the limit for DCBHs formation by unresolved SF episodes. Nevertheless, since mild local pollution allows the formation of RSM seeds \citep[e.g.,][]{omukai_schneider_haiman2008,devecchi_volonteri2009,sassano2021}, we still check the occurrence of the latter within newly-resolved \lgal structures hosting light-seed descendants. We discriminate between different seeding pathways by comparing $\rm Z_{local}$ and $\rm J_{local}$ to specific thresholds, namely: $\rm Z_{crit,RSM}$, $\rm Z_{crit,DCBH}$, $\rm J_{crit,DCBH}$ and  $\rm J_{crit,RSM}$ \citep[e.g.,][]{omukai_schneider_haiman2008,volonteri2010, valiante2017,sassano2021}. miDCBH do not require special conditions for the gas metallicity and LW radiation, so we check the favourable conditions for this channel after every galaxy merger in our box.\\
\begin{table*}
    \centering
    \begin{tabular}{>{\centering}p{2cm} |>{\centering}p{4.cm} | >{\centering}p{3cm} | >{\centering}p{1.5cm} | >{\centering}p{2.5cm} | >{\centering}p{2cm}}
     \hline
     \hline
     \hlx{v[][1.5mm]}
     \textbf{Seed Type}   &     \large{$\rm \mathbf{ Z_{local} }$}    &  \large{$\rm \mathbf{ J_{local} }$}  & \textbf{Grafting}  & \textbf{Merger-driven origin} & \large{$\rm\mathbf{M_{seed}\ [M_\odot]}$}\\
     \hlx{v[][1.5mm]}
     \hline
     \hline
     \hlx{v[][1.5mm]}
     Light  &   -   &   -   &  \checkmark   &   -    &  $\rm 10^2 - 10^5$\\
     \hlx{v[][1.5mm]}
     \hline
     \hlx{v[][1.5mm]}
     DCBH   &  $\rm Z_{local}\leq Z_{crit,DCBH}$    &           $\rm J_{local}\geq J_{crit,DCBH}$           &   -    &   -  &  $\rm 10^5$\\
     \hlx{v[][1.5mm]}
     \hline
     \hlx{v[][1.5mm]}
     RSM   &  $\rm Z_{crit,DCBH} < Z_{local}\leq Z_{crit,RSM}$     &  $\rm J_{local}\geq J_{crit,RSM}$ &  -  &  - &  $\rm 10^3 - 10^4$\\
     \hlx{v[][1.5mm]}
     \hline
     \hlx{v[][1.5mm]}
     miDCBH      &                        -                          &                                -                                 &   -    & \checkmark &  $\rm 8\times10^4$\\
     \hlx{v[][1.5mm]}
     \hline
     \hline
    \end{tabular}
    \caption{Summary of the conditions for the formation of BH seeds in our model.
    From left to right, for each seed-type we specify our requirements for $\rm Z_{local}$ and $\rm J_{local}$ (if any) and highlight whether the BHs are inherited from \gqd or formed during galaxy mergers. The four types detailed in this table can mix through their hosts mergers, producing three \textit{mixed} seed-types which we discuss in the text.}
    \label{tab:BHseeding:prescription_summary}
\end{table*}

\noindent
\textit{Direct-Collapse Black Holes}\vspace{1.5mm}\\
\noindent
As soon as SF sets in, SNe feedback hinders the formation of DCBH seeds \citep[e.g.,][]{ritter2012,agarwal2017,maio2019}. We thus consider as potential DCBH hosts only halos which never hosted SF,
provided a sufficient $\rm J_{LW}$ flux is present to contrast $\rm H_2$-cooling within them. The exact $\rm J_{LW}$ threshold required for this has been widely discussed in the past, ranging between $\rm 0.01\!\lesssim\!J_{crit}\,/\,J_{21}\!\lesssim\!10^5$ \citep[with $\rm J_{21}\!=\!10^{-21}erg\,cm^{-2}\,s^{-1}\,Hz^{-1}\,sr^{-1}$, e.g.,][]{yoshida_omukai_hernquist2007,shang_bryan_haiman2010,regan2014b}. Wide differences are found by considering the emitted spectrum of LW photons \citep[e.g.,][]{latif2015,regan_johansson_wise2016a}, the complete dissociation of $\rm H_2$ and $\rm H^-$ \citep[e.g.,][]{wolcott-green_haiman2011,latif_khochfar2019}, $\rm H_2$ self-shielding \citep[e.g.,][]{hartwig2015c,wolcott-green_haiman2019} or the presence of rotational support \citep[e.g.,][]{latif2014b,latif_volonteri2015}. 

We set $\rm J_{crit,DCBH}=10\,J_{21}$ as critical threshold for DCBH formation, as a compromise among previous approaches in the literature \citep[see the discussions in][]{agarwal2016,regan_downes2018,dunn2018,lupi_haiman_volonteri2021}. Regarding the metallicity of the IGM, we fix the value of $\rm Z_{crit,DCBH}\!=\!10^{-4} Z_\odot$ by following e.g., \cite{bromm_loeb2003,agarwal2012, dijkstra2014} and \cite{ valiante2016}. Finally, we require that newly-resolved DCBH-host candidates posses a large enough reservoir of $\rm M_{cGas}$.
To summarize, we assume that a $\rm M_{seed}\!=\!10^{\,5} M_\odot$ DCBH can form inside a newly-resolved atomic-cooling halo with the following properties: (i) $\rm Z_{local}\!<\!Z_{crit,DCBH}$, (ii) $\rm J_{local}\!\geq\! J_{crit,DCBH}$, (iii) $\rm M_{cGas}\!>\!M_{seed}$ and (iv) $\rm M_*=0$.\\

\noindent
\textit{Formation of IMBHs from runaway stellar mergers}\vspace{1.5mm}\\
\noindent
RSM seeds are thought to form in the core of star-forming atomic-cooling halos under mild chemical enrichment ($\rm 10^{-6}\!\lesssim\!Z/Z_\odot\!\lesssim\!10^{-3}$) and the presence of a $\rm J_{local}$ sufficient to delay $\rm H_2$-cooling in time. This is necessary to increase the critical density for gas fragmentation, so that SF can only proceed in the dense, nuclear regions of the atomic-cooling halo, where the gas numerical density is higher than a metallicity-dependent critical threshold $\rm n_{crit,Z}$ \citep[i.e., $\rm n\!>\!n_{\,crit,Z}$; see][and references therein]{devecchi_volonteri2009}. This ultimately produces a nuclear stellar cluster of mass $\rm M_{cl}\!\sim\!10^{\,5}\,M_\odot$, where efficient stellar collisions eventually lead to the formation of an IMBH seed \citep[see e.g.,][]{omukai_schneider_haiman2008,glebbeek2009,chon_omukai2020,sassano2021}.

We add this seeding channel to \lgal by adapting the purely-analytic prescription of \cite{devecchi_volonteri2009}, which requires $\rm 10^{\,4}\!\lesssim\!T_{vir}/[K]\!\lesssim\!1.8\times10^{\,4}$ to ensure the conditions for the formation of a RSM seed. 
Furthermore, we impose $\rm Z_{\,local}\!\leq\!Z_{\,crit,RSM}\!=\!10^{-3}Z_\odot$ and $\rm J_{local}\!\geq\!J_{crit,RSM}$. Since the formation of RSM seeds does not require the complete dissociation of $\rm H_2$, relatively low levels of $\rm J_{local}$ might be sufficient for this seeding channel \citep[e.g.,][]{omukai2001a,oshea_norman2008}. For this reason, we impose $\rm J_{crit,RSM}\!=\!0.1\,J_{crit,DCBH}$.

Under the above conditions, we compute $\rm n_{crit,Z}$ as in \cite{devecchi_volonteri2009} and follow their approach also to model a two-component density profile for the cold gas within our RSM candidate halos, using their $\rm M_{vir}$, $\rm R_{vir}$ and spin parameters provided by the outputs of the \texttt{MR-II}.
This allows to compute the mass $\rm M_{cl}$ of the nuclear stellar cluster within RSM candidates, which determines the time-scale $\rm t_{cc}$ over which runaway mergers instabilities can produce a central massive object. If $\rm t_{cc}$ is shorter than the typical main-sequence life of massive stars (i.e. $\rm t_{MS}=5\,Myr$) a RSM seed can form with an $\rm M_{seed}$ determined by $\rm M_{cl}$ and $\rm t_{cc}$ \citep[typically $\rm M_{seed}\!\sim\!10^{3-4}M_\odot$; see,][]{portegieszwart1999,devecchi_volonteri2009}.\\
\begin{figure*}
    \centering
    \includegraphics[width=0.92\textwidth]{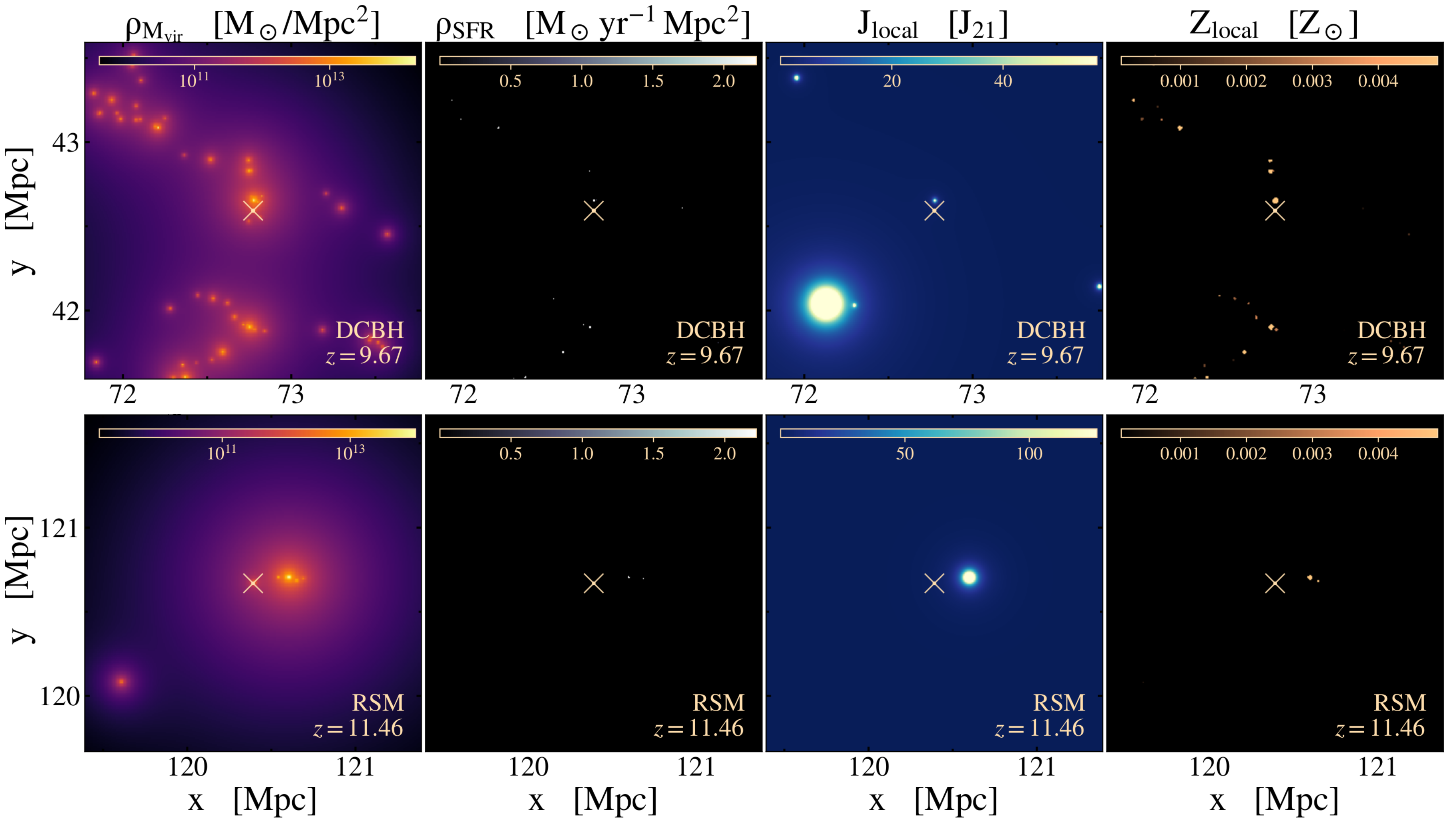}
    \caption{Environment of a newly-formed DCBH (upper row) and RSM seed (lower row), taken as representative examples. The four different spatial-maps on each row are obtained by considering all halos within a $\rm (x,y,z)=(2,2,0.5)\,Mpc$ slice. The position of the newly-formed seeds is marked with a white cross at the center of each panel. From left to right: the projected-density of $\rm M_{vir}$ and SFR (first and second column), the local intensity of LW flux (as received by the BH seed, third column) and the local level of chemical enrichment.}
    \label{fig:fancy_maps}
\end{figure*}

\noindent
\textit{Merger-induced Direct-Collapse Black Holes}\vspace{1.5mm}\\
\noindent
The formation of massive DCBHs might also be triggered within high-$z$, gas-rich major mergers of disc-dominated galaxies \citep[see e.g.,][for a recent review]{mayer_bonoli2019}. Indeed, up to $\rm M_{gas}\!\sim\!10^{7-8}M_\odot$ can be compressed in the nuclear regions of merger remnants by extreme gas-inflow rates, hence providing the conditions to grow a central compact object of $\rm M_{seed}\!\sim\!10^{4-5}M_\odot$ in few $\rm 10^{\,5} yr$, if a massive black hole is not already present \citep[as shown by recent numerical simulations e.g.,][]{mayer2010, inayoshi_visbal_kashiyama2015}. We implement this seeding channel by closely mirroring the work of \cite{bonoli2014}, which tested it on the \texttt{MR} merger trees.

We model the formation of  miDCBH seeds with $\rm M_{seed}\!=\!8\times10^{\,4}M_\odot$ under the following requirements: (i) the merging galaxies  have a baryonic mass ratio  $\rm m_r\!\geq\!0.3$, (ii) the merger remnant has a minimum halo mass of $\rm M_{vir}=10^{9}M_\odot$, (iii) the total $\rm M_{cGas}$ carried by the two interacting galaxies is higher than the (fixed) mass of the forming BH seed, (iv) the two merging galaxies exhibit a maximum bulge-to-total ratio of $\rm B/T=0.2$ (disk-dominated) and, finally, (iv) the merger remnant does not already host a central BH more massive than $\rm M_{BH}=5\times10^{\,4}M_\odot$. These requirements are similar to the ones of \cite{bonoli2014}, to whom we refer for details.

\section{Results}
\label{sec:results}
Our model follows the formation of BH seeds through multiple channels, from light PopIII remnants to heavy DCBHs. Thanks to the N-body origin of the \milltwo merger-trees, we can track the position of BH-seeds hosts. This allows us to study both the environment in which BH seeds form at high-$z$ and the evolution of the lower-$z$ population of their SMBH descendants. We note that the \milltwo volume ($\rm 10^6\,Mpc^3\,h^{-3}$) does not allow to reach the number density of bright, high-$z$ QSOs ($\rm\sim1\,Gpc^{-3}$), hence we focus our analysis on the build-up of the less-extreme objects which constitute the bulk of the SMBHs population over cosmic history.

\subsection{The high-\texorpdfstring{$z$}{z} formation of BHs}
\label{sec:results:high-z_results}
Here we analyze the formation of BHs at high-$z$ in our model, focusing on the environment hosting the occurrence of different seeding scenarios and the build-up of a multi-flavour SMBHs population. In order to keep memory of their origin, we track both the seeding-mass $\rm M_{seed}$ and formation-scenario of all SMBHs in our model. This allows us to define four different \textit{seed classes}, namely: light seeds, inherited from \gqd outputs (see Sect. \ref{sec:model:BH_seeding_prescriptions:GQd_Grafting}) and \textit{RSM}, \textit{DCBH} or \textit{miDCBH} formed in \lgal according to the prescriptions detailed in Sect. \ref{sec:model:BH_seeding:RSM_and_DCBH_seeding_in_LGal} and summarized in Tab. \ref{tab:BHseeding:prescription_summary}.

\subsubsection{Nurturing environment of BH seeds}
\label{sec:results:seeds_environment}
As detailed in sections \ref{sec:model:actual_computation_of_local_metals_and_jlw} and \ref{sec:model:BH_seeding:RSM_and_DCBH_seeding_in_LGal}, we impose a set of conditions on $\rm Z_{local}$ and $\rm J_{local}$ for identifying galaxies hosting BH-seeding processes. Consequently, it is interesting to investigate how these requirements  influence the birthplaces of the various seed flavours.

In Fig. \ref{fig:fancy_maps} we provide a qualitative example of the typical formation environment of different kinds of BH-seeds predicted by our model. As representative cases, we select a DCBH and a RSM seed (respectively upper and lower row) formed thanks to the $\rm J_{LW}$ provided by LW sources in their past light-cone. We show four different spatial-maps computed at the time of BH seeds formation and centered at their, marked with a white cross at center of each image. We consider slices of $\rm (x,y,z) = (2,2,0.5)\,Mpc$ for each map and compute (from left to right in Fig. \ref{fig:fancy_maps}): the projected density of $\rm M_{vir}$ and SFR, the local intensity of LW flux as seen by the newly-formed BHs and the local level of chemical enrichment.

In particular, since we do not have access to the individual DM particles of \texttt{MR-II}, we computed the 2D $\rm M_{vir}$ map by distributing the viral mass of each DM halo over a 3D, Navarro, Frenk and White density profile \citep[see][]{navarro_frenk_white1997}, using the fitting formulas for halo concentration of \cite{dutton_maccio2014} and then projecting the resulting mass distribution along one spatial dimension. We follow a similar procedure for the SFR map, distributing this property over a spherical, top-hat profile with radius equal to the stellar disk of each galaxy. For the $\rm Z_{IGM}$ maps (rightmost column), we simply represent the metallic shells of galaxies as 2D circles, color-coded according to the metallicity within the shells. We underline that the $\rm M_{vir}$, SFR and $\rm Z_{IGM}$ maps are representations of the BH seeds environment at the BH-formation time. On the other hand, the LW map shows the 2D projection of the spatial variations of $\rm J_{local}$, as seen by the newly-formed BH seeds, i.e. produced by LW sources within their past light-cones at the time of the LW photons emission. This implies that LW sources can contribute to this map from previous times with respect to the one at which BH-seeds actually form. Consequently, some of the sources shown in the LW map might be displaced (or even do not appear) in the $\rm M_{vir}$, $\rm\rho_{SFR}$ and $\rm Z_{local}$ maps.

A qualitative comparison between our spatial maps and those presented in recent works underlines interesting similarities, such as the presence of a few neighboring sources illuminating the BH-seed formation site with $\rm J_{LW}$ \citep[as in][]{agarwal2014} or the irregular coverage of the IGM by metal-enriched shells \citep[e.g.,][]{visbal_bryan_haiman2020}. These two works, in particular, focused on relatively small scales. The first one analyzed the formation environments of six DCBHs with a high-resolution hydrodynamic simulation ($\rm L_{box}\!=\!4Mpc$). The second one, analyzed the high-$z$ formation of PopIII stars by applying a self-consistent SAM for early SNe feedback to a high-resolution N-body simulation ($\rm L_{box}\!=\!3Mpc$). In both cases, their simulated IGM shows the presence of photo-ionized patches which are still chemically pristine, due to the different progress of LW illumination and chemical enrichment. This is in line with other recent works focusing on the competing action of these two factors on the formation of BHs at high-$z$ \citep[e.g.][]{visbal_haiman_bryan2014b,agarwal2017,agarwal2019,maio2019}. Our work fits in this panorama by allowing us to generalize on a wide, cosmological volume the results obtained on smaller scales by these previous works.

Although being only qualitative, Fig. \ref{fig:fancy_maps} shows different degrees of occupation by neighboring structures in the environment of DCBH and RSM seeds. We note that relatively dense and biased regions at high-$z$ are likely to host actively SF galaxies, hence allowing to find LW-bright halos in the proximity of a pristine, collapsing gas cloud \citep[for an hydrodynamical study, see e.g.,][]{dunn2018}. Therefore, we expect to preferentially observe the formation of intermediate and massive seeds in over-dense regions, as a consequence of the specific $\rm J_{local}$ requirements.
\begin{figure}
    \centering
    \includegraphics[width=0.49\textwidth]{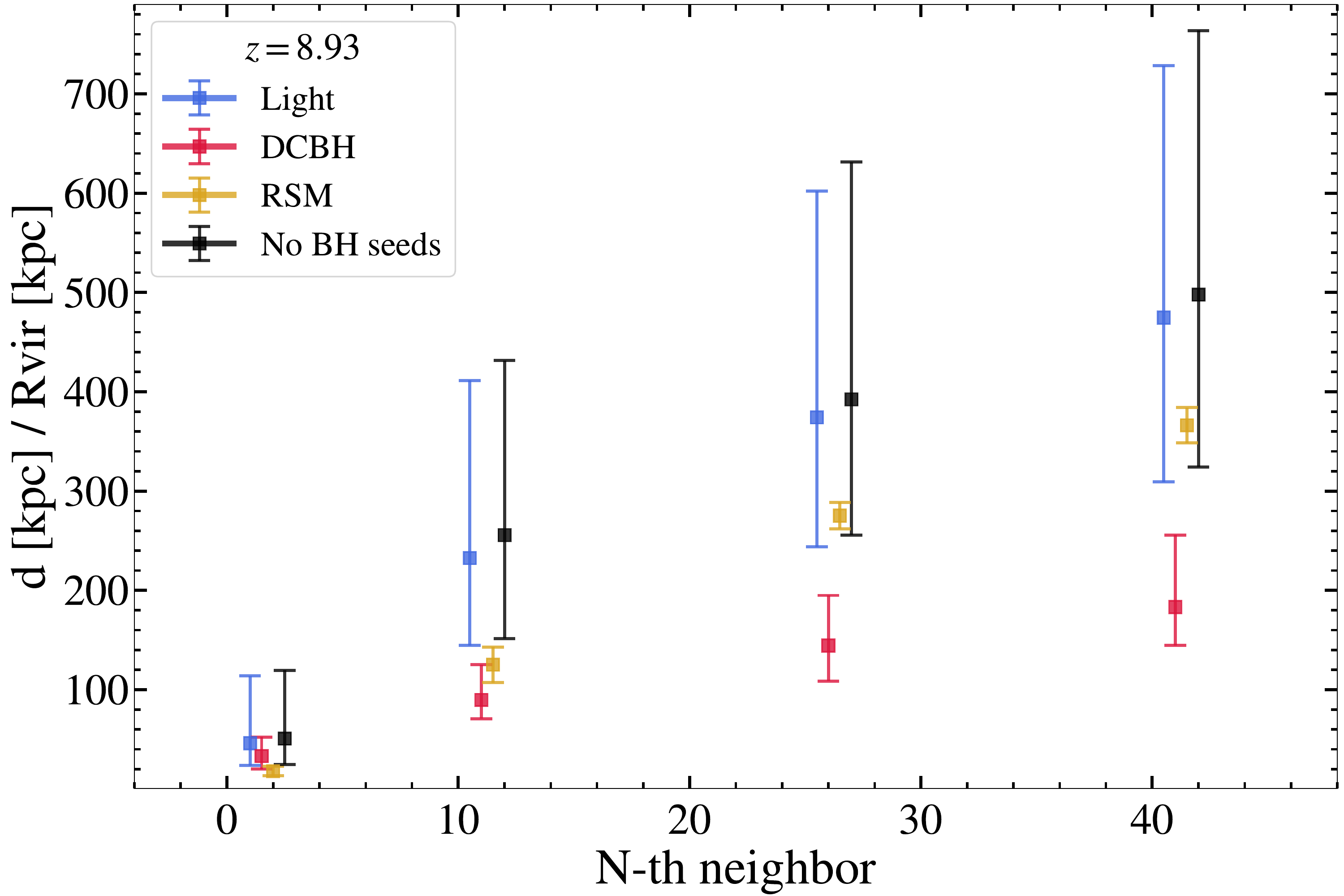}
    \caption{Median distance d at which the 1st, 10th, 25th and 40th neighbor is found in the environment of newly-formed or inherited BH-seeds of different classes (listed in the plot legend). We consider the case of $z\!\sim\!9$ as a representative example. We normalize d to the $\rm R_{vir}$ of BH-seed hosts in order to better compare the results for halos of different sizes and masses. Points are slightly shifted along the x-axis for a better visualization, while the error bars mark the $\rm16th$ and $\rm84th$ percentiles.}
    \label{fig:typical_neighbors_distance_for_seedtypes}
\end{figure}

To statistically quantify the degree of occupation around all newly-formed BH seeds, we calculate the median distance at which their 1st, 10th, 25th and 40th neighbor galaxies are found. This is shown in Fig. \ref{fig:typical_neighbors_distance_for_seedtypes} for BH seeds formed or inherited at $z\!\sim\!9$. To compare our results for halos of different sizes and masses, we normalize the distance between BH-seeds hosts and their neighbors by the $\rm R_{vir}$ of the BH-seed hosts. In this analysis smaller normalized-distances imply denser environments, as in the case of DCBHs hosts (red squares) with respect to light and RSMs hosts (respectively light-blue and yellow squares) or galaxies without BHs (black squares). We use $z\!\sim\!9$ as a representative example because of the larger statistics it provides on the newly-formed RSM and DCBH seeds. We checked that the result of Fig. \ref{fig:typical_neighbors_distance_for_seedtypes} is conserved at $z\!>\!8$.

The difference between the environment of DCBH hosts and the other classes of objects becomes increasingly evident up to the 40th neighbor. Since the formation of DCBHs requires relatively strong $\rm J_{local}$ in our model, it is reasonable to expect that these conditions are more likely to be verified in dense environments rather than under-populated regions. Indeed, the former are likely to host galaxies which formed stars in the recent past of DCBH-seeding candidates and produced the required $\rm J_{PLC}$. This result suggests that the $\rm J_{local}$ and $\rm Z_{local}$ conditions required for DCBH formation might impose a selection effect also on the environment of these BH seeds, at least up to few hundreds times the virial radii of their hosts. Analogously, a fraction of RSM seeds form thanks to the presence of the $\rm J_{PLC}$ provided by neighboring star-forming galaxies (see the lower row of Fig. \ref{fig:fancy_maps} for an example). Consequently, also the formation of RSM seeds is favoured in regions  more biased than the average, although at a lower significance than what we observe for DCBHs.
\begin{figure}
    \centering
    \includegraphics[width=0.49\textwidth]{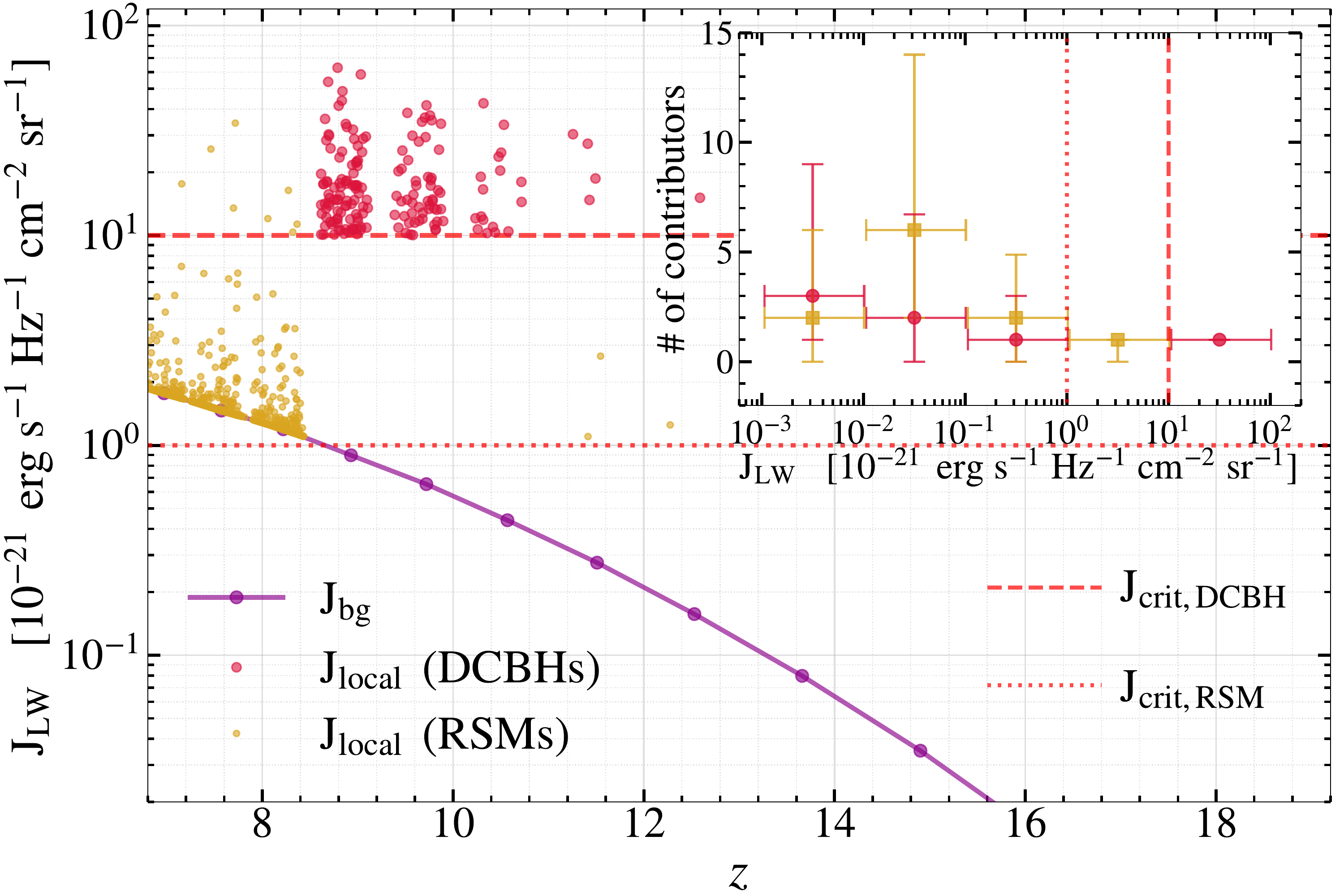}
    \caption{\textit{Main panel}: Comparison between the values of $\rm J_{local}$ received by newly-formed DCBH and RSM seeds (respectively, red and yellow dots) and the background level of $\rm J_{LW}$ (purple line and dots). This shows that the latter is generally lower than $\rm J_{crit,DCBH}$ by at least $\rm1\,dex$. \textit{Inset panel}: median number of LW contributors per decade of $\rm J_{local}$ \textit{received} by newly-formed DCBHs and RSMs (respectively red circles and yellow squares).}
    \label{fig:local_vs_background_Jlw}
\end{figure}

We stress that BHs inherited from \gqd already underwent unresolved evolution at the moment of their grafting in \texttt{L-Galaxies}, as they formed at $z\!>\!20$. Therefore, the result in Fig. \ref{fig:typical_neighbors_distance_for_seedtypes} is not representative of their formation environment in \texttt{GQd}. Rather, \gqd light-seeds hosts trace the average population of newly-initialized halos with $\rm M_{vir}^{init}\!\gtrsim\!10^9\,M_\odot$, since our grafting is independent from environmental conditions.
\begin{figure*}
    \centering
    \includegraphics[width=0.45\textwidth]{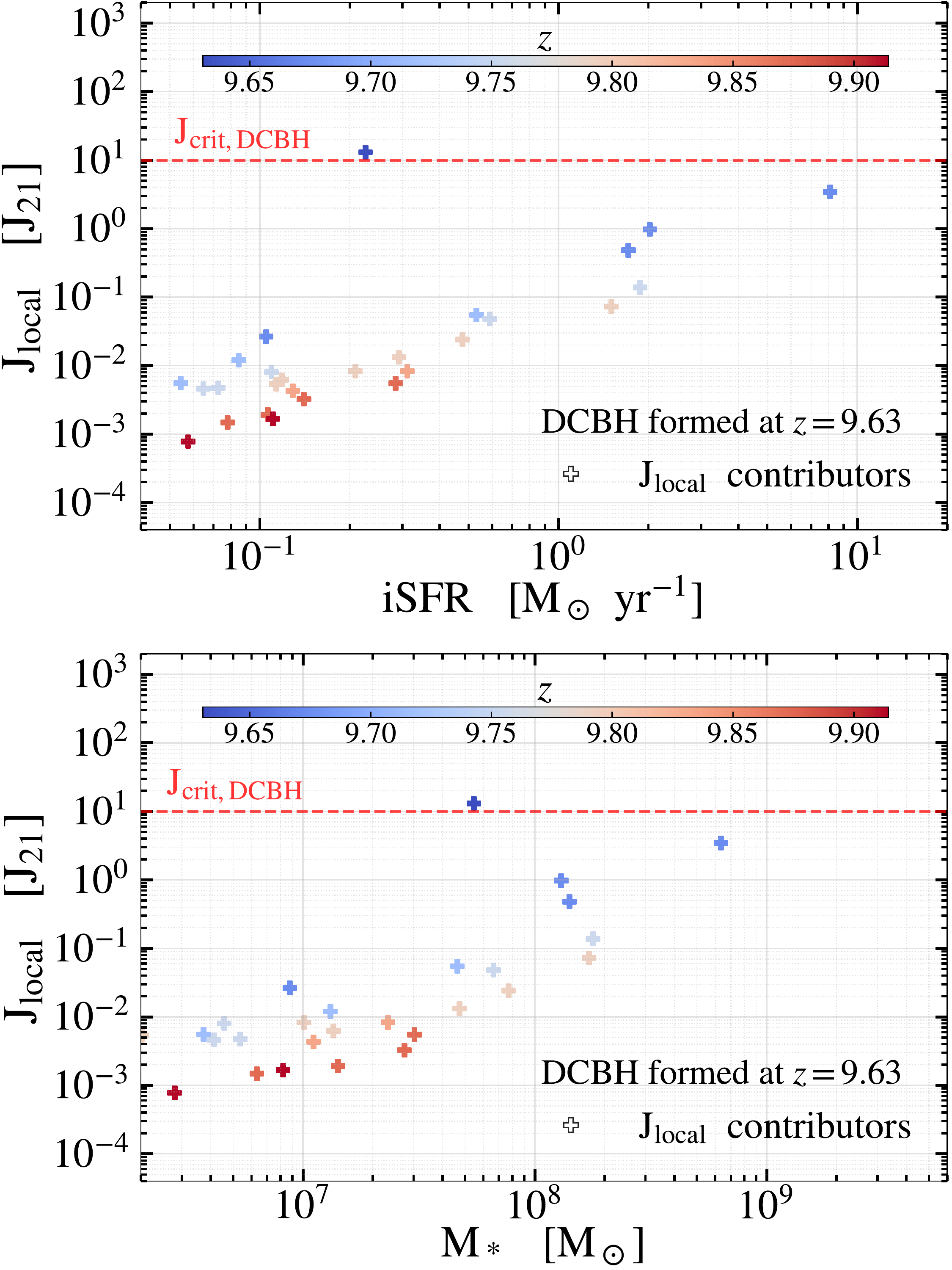}\hspace{5mm}
    \includegraphics[width=0.45\textwidth]{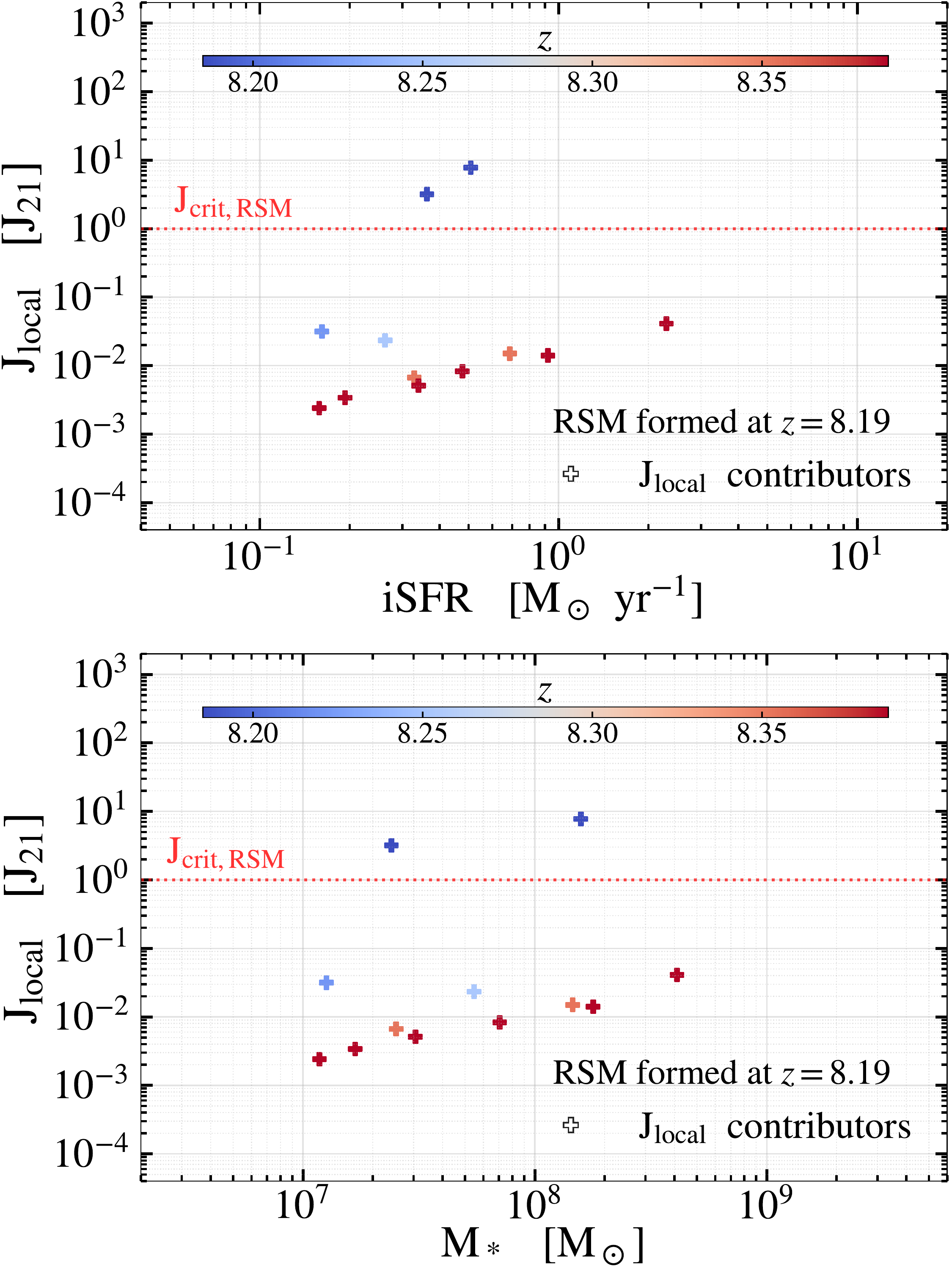}
    \caption{Properties of the $\rm J_{LW}$ contributors for a DCBH seed formed at $z\!\sim\!10$ (left column) and a RSM seed formed at $z\!\sim\!8$ (right column), taken as examples. We show the distributions of $\rm J_{LW}$ received by the forming seeds (i.e. at the time and location of their formation) versus the instantaneous SFR (iSFR) and $\rm M_*$ of the contributors (respectively upper and lower rows). The color code encapsulates the redshift distribution of LW contributors within the past-lightcones of the forming seeds, showing that the luminous LW sources responsible for the overcoming of $\rm J_{crit}$ are also the closest in time to the newly-formed seeds (dark-blue crosses).}
    \label{fig:JLW_contributors_properties}
\end{figure*}

\subsubsection{LW sources contributing to BH seeds formation}
\label{sec:results:seeds_environment:DCBHs}
As discussed in the previous section, DCBH seeds in our model form in relatively dense and active environments, where recent SF episodes provide the required $\rm J_{PLC}$. Indeed, as shown in the bottom panel of Fig. \ref{fig:dynaranges}, $\rm J_{bg}$ never reaches the critical level needed for DCBH formation. On the other hand, the background is sufficient to trigger the formation of RSM seeds when $\rm Z_{crit,DCBH}\!<\!\langle Z_{igm}\rangle\!\leq\!Z_{crit,RSM}$ at $6\!\lesssim\!z\!\lesssim\!8$. This is shown in the main panel of Fig. \ref{fig:local_vs_background_Jlw}, through the comparison between $\rm J_{local}$ at the sites of DCBHs and RSMs formation (respectively red and yellow dots) and the global $\rm J_{bg}$ (purple line and dots). As we can see, strong variations of $\rm J_{LW}$ of $\rm>\!1\,dex$ with respect to $\rm J_{bg}$ are responsible for the formation of DCBHs, hence confirming the key role of $\rm J_{PLC}$ \citep[as in, e.g.,][]{visbal_haiman_bryan2014a,fernandez2014,regan2014b,latif_volonteri2015,regan_downes2018,maio2019}. These local variations are preferentially due to a single, luminous neighbor, in line with recent works suggesting that DCBHs preferably form in close halo-pairs \citep[see e.g.,][]{dijkstra2008,visbal_haiman_bryan2014b,yue2014,agarwal2017,regan2017,agarwal2019}. Our results supports this picture statistically over a wide, cosmological box \citep[as in e.g.,][]{lupi_haiman_volonteri2021}.

Similarly, a few RSM seeds forming at $z\!<\!8$ are also subject to strong $\rm J_{PLC}$ (as shown in Fig. \ref{fig:local_vs_background_Jlw}), in excess by $\sim\!1$ dex with respect to $\rm J_{bg}$. Therefore, our results suggest that the close halo-pairs scenario might hold also for a fraction of RSM seeds. Indeed, the $26\%$ of all the RSM seeds formed in our model at any $z$ receive $\rm J_{PLC}\!>\!J_{crit,RSM}$, while $33\%$ of them require the presence of $\rm J_{bg}$ to compensate $\rm J_{PLC}$ and overcome $\rm J_{crit,RSM}$. The remaining $41\%$ form only thanks to the presence of $\rm J_{bg}$, without any direct LW contributor. Among the $26\%$ of RSM forming with $\rm J_{PLC}\!>\!J_{crit,RSM}$, $92\%$ of the objects (i.e. $\sim\!24\%$ of all the RSMs) form thanks to one luminous neighbor. For comparison, none of our DCBHs forms only thanks to $\rm J_{bg}$ as already commented, and only $7\%$ of them needs $\rm J_{bg}$ to overcome $\rm J_{crit,DCBH}$. Among the $93\%$ of DCBHs forming with $\rm J_{PLC}\!>\!J_{crit,DCBH}$, the $\sim\!82\%$ (i.e. $76\%$ of all DCBHs) only need a single, actively star-forming LW contributor.

To further test the idea that BH seeds may form in synchronized halo pairs, we compute the number of LW contributors per dex of $\rm J_{PLC}$ for each newly-formed DCBH and RSM with $\rm J_{PLC}\!>\!0$, i.e. excluding BH seeds which only form thanks to $\rm J_{bg}$. Our results are shown in the inset panel of Fig. \ref{fig:local_vs_background_Jlw} with yellow squares and red circles respectively for RSMs and DCBHs. We mark the $\rm J_{PLC}$ bins-size with horizontal error bars, while vertical errors show the 16th and 84th percentiles of the distribution in each bin. Statistically, $\rm J_{crit,DCBH}$ (vertical dashed red line) is surpassed thanks to only one luminous LW contributor, supporting the close halo-pairs scenario. Interestingly, this is also the case for $\rm J_{crit,RSM}$, although a significant fraction of RSMs does not show LW-contributors in the $\rm 1\!<\!J_{LW}/J_{21}\!<\!10$ bin, as pointed out by the vertical error bar reaching zero. This is an effect of the low value we set for $\rm J_{crit,RSM}$, which allows $\rm J_{bg}$ to actively play a role and compensate for the $\rm J_{PLC}$ provided by faint LW-contributors.
In other words, when considering BH seeds forming under $\rm J_{PLC}\!>\!J_{crit}$, the contribution of a single, luminous neighbor is sufficient in the large majority of the cases, for both DCBH and RSM seeds. These results provide statistical support to the close halo-pair scenario for the formation of DCBHs over a cosmological volume and suggest that a similar picture can be expected for those RSM seeds receiving $\rm J_{PLC}\!>\!0$.

Finally, in Fig. \ref{fig:JLW_contributors_properties} we show an example of the distributions of $\rm J_{LW}$ measured at the position of a newly-formed DCBH (left) and RSM seed (right) versus the instantaneous SFR (iSFR) and $\rm M_*$ of all their LW contributors (respectively upper and lower panel). In both cases, the $\rm J_{LW}$ of the brightest LW contributor is sufficient to overcome $\rm J_{crit}$, being higher by $\rm\gtrsim\!1\,dex$ with respect to those of the other LW contributors. Interestingly, the brightest LW contributors do not show the highest $\rm M_*$ or iSFR among all the contributors.
This can be understood by analyzing the time-distance (and hence spatial separation) between LW contributors and the forming BH-seeds. Indeed, by color-coding the $z$ distribution of LW contributors, we show that the strongest contributors are the closest to the BH formation event, in line with the halo-pair picture. 
This also shows that distant LW sources are unable to produce strong $\rm J_{local}$ variations, independently of their number, $\rm M_*$ or iSFR \citep[see also, e.g.,][]{yue2014}.

The results presented above are based on the definition of our LW sources list (see Sect. \ref{sec:model:actual_computation_of_local_metals_and_jlw}). It can be argued that, by lowering the thresholds for the list definition, a higher number of LW-faint galaxies would be defined as LW-sources, hence potentially increasing the number of faint LW contributors nearby newly-resolved BH seeds candidates. This might compensate for their shallowness and influence our results about the close halo-pair scenario. We explicitly checked this possibility by using two additional sources lists, obtained by lowering the threshold from $\rm J_{LW}(R_{vir})=10\,J_{21}$ (as detailed in Sect. \ref{sec:model:actual_computation_of_local_metals_and_jlw}) to 
$\rm J_{LW}(R_{vir})=5\,J_{21}$ and $\rm J_{LW}(R_{vir})=1\,J_{21}$.

We find negligible differences for the case of DCBHs, hence confirming that intrinsically-faint LW sources never manage to produce strong $\rm J_{local}$ variations, independently of their abundance. A similar conclusion is also valid for RSM seeds, since we recover all the cases where these seeds form thanks to a single, luminous neighbor. In particular, we do not find cases where DCBH or RSM seeds form only thanks to a group of faint, closeby neighbors, suggesting that our results about the halo-pair scenario (presented in Fig. \ref{fig:typical_neighbors_distance_for_seedtypes}, \ref{fig:local_vs_background_Jlw} and \ref{fig:JLW_contributors_properties}) are robust against the definition of our sources list.

On the other hand, for both of the new lists definitions (i.e. $\rm J_{LW}(R_{vir})=5\,J_{21}$ and $\rm J_{LW}(R_{vir})=1\,J_{21}$) 
the larger number of faint LW-sources produces an increase of LW-contributors providing faint $\rm J_{PLC}$ contributions at the birthplaces of RSM seeds. Interestingly, the $\rm J_{PLC}$ of these faint neighbors are generally comparable to $\rm J_{bg}$, suggesting that the role of faint LW sources converges to that of the LW background. We conclude that using a list of LW-bright sources together with a uniform $\rm J_{bg}$ is roughly equivalent to including LW-faint objects in our LW sources list. We chose to employ the former approach due to it is computational convenience.

\subsubsection{Build-up of a multi-flavour SMBHs population}
\label{sec:results:seed_types_definition}
\begin{figure}
    \centering
    \includegraphics[width=0.49\textwidth]{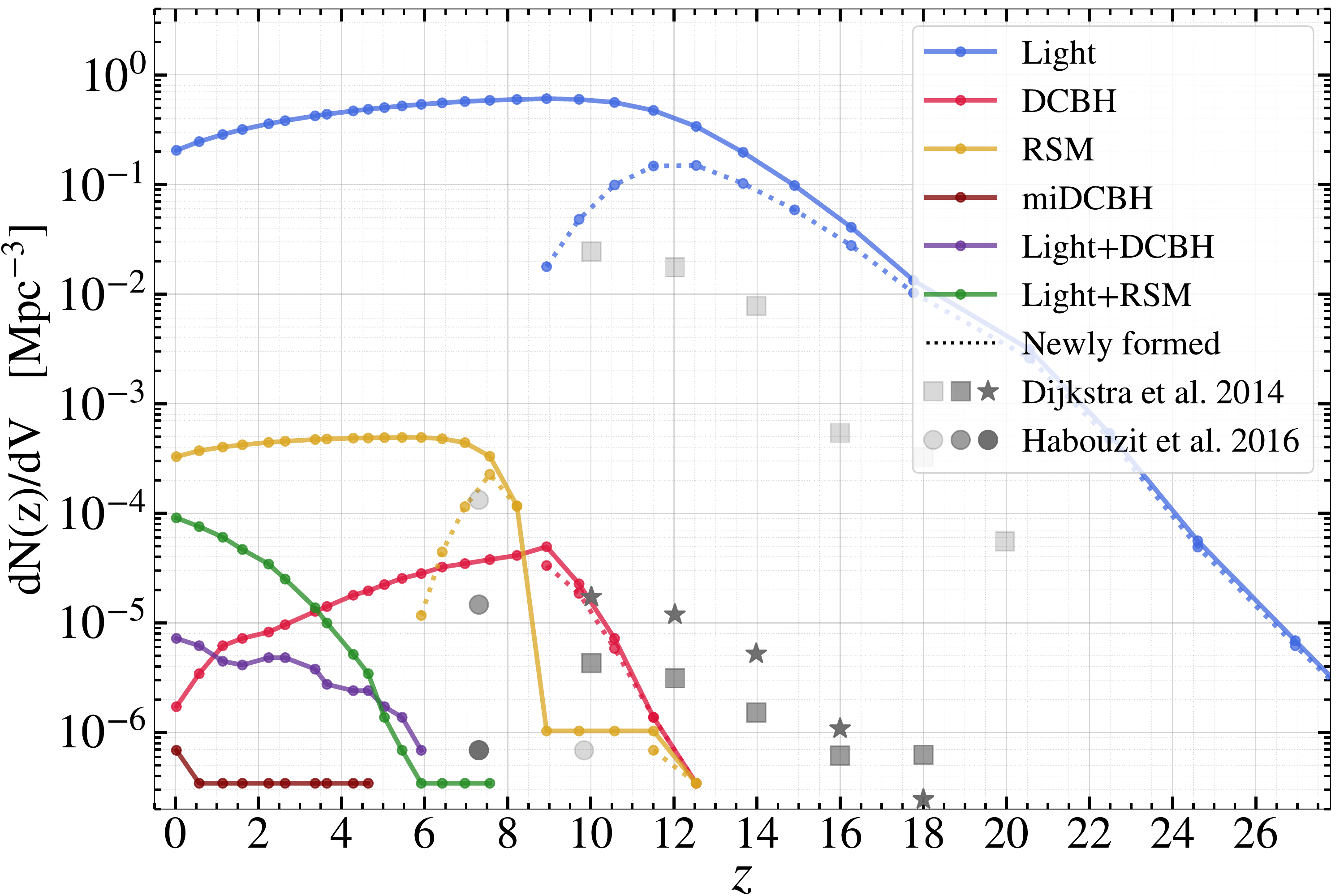}
    \caption{Evolution of the BH number density predicted by our model, split into different seeds-flavours, as shown in the plot legend. The density of newly-formed/inherited BHs is marked with dashed lines, to show the progress of BH seeds formation at high-$z$. The latter stops at $z\!\sim\!8$ for light-descendants inherited from \gqd and DCBHs (respectively light-blue and red lines), while RSM seeds can form down to $z\!\sim\!6$ (yellow lines). The mixing between DCBH, RSM and light seed-descendants produces the increase of mixed seed-flavours, shown as green and purple lines. Only two miDCBH form in the \milltwo box, one at $z\!\sim\!5$ and the other one at $z\!\sim\!0$ (brown lines). Finally, we show the predicted DCBH number density of \protect\cite{dijkstra2014} as grey squares and stars and of \protect\cite{habouzit2016} as grey circles, in order to compare our results with those of previous models.}
    \label{fig:seeding_evolution}
\end{figure}

In our model, each BH-seeding channel naturally stops when the conditions for its occurrence are no longer verified (see Table \ref{tab:BHseeding:prescription_summary}). After this, our population of SMBHs only evolves via gas accretion or through mergers of already-existing objects, which represents a key aspect of our model. Indeed, any low-$z$ property of our SMBH population is a consequence of the self-consistent evolution of BHs formed at high-$z$ through physically-motivated seeding prescriptions. Figure \ref{fig:seeding_evolution} presents the evolution of the BH number-density 
split in the different seed-types classes, showing the build-up and gradual mixing of the multi-flavour population of BH-seeds descendants. We distinguish the total number-density of each seed type (solid lines) from that of newly-formed (or newly-grafted) BHs (dashed lines), to highlight the progress of BH formation at high-$z$. In particular, the grafting of light seeds (light-blue lines) stops under the conditions discussed in Sect. \ref{sec:model:BH_seeding_prescriptions:GQd_Grafting}, while DCBHs (red lines) can only form down to $z\!\sim\!8$, after which $\rm Z_{IGM}\!>\!Z_{crit,DCBH}$ and newly-resolved halos are not chemically pristine. Analogously, the chemical enrichment of the IGM interrupts the formation of RSMs at $z\!\sim\!6$ (yellow lines). As a consequence, the total number density of light-descendants, RSM and DCBH seeds in the \milltwo box decreases in time after reaching a maximum. This is an effect of the hierarchical mergers of SMBHs hosts, which also produce the rare miDCBHs (brown lines) and the mixed light+RSM and light+DCBH (respectively, green and violet lines at $z\!\lesssim\!6$).

We note that light seeds-descendants are numerically dominant over other classes by several orders of magnitude, at all $z$. Their grafting reaches a maximum at $z\!\sim\!12$ as an effect of the interplay between \gqd and \lgal dynamical ranges (see Sect. \ref{sec:model:BH_seeding_prescriptions:GQd_Grafting}). Indeed, at $z\!\lesssim\!12$ we sample the low-end of the \gqd halo-mass range, where the occupation of \gqd BHs is low (as shown in Fig. \ref{fig:GQd_BH_occupation}). On the other hand, DCBHs and RSMs seed-types begin to form at $z\!\lesssim\!13$ as the gradual progress of SF in the \milltwo box provides the necessary $\rm J_{LW}$ and IGM chemical enrichment. Indeed, the mass-resolution of \milltwo does not allow to track the build-up of $\rm H_2$-depleted regions at $z\!\gtrsim\!13$ produced by the energetic output of early SF episodes. This leads us to under-estimate the amount of photo-dissociated regions at high-$z$, hence to predict a lower number-density of BH seeds formed at $z\!\gtrsim\!13$ with respect to previous works \citep*[e.g.,][but also \citealt{holzbauer_furlanetto2012,valiante2021} for a discussion on the effects of mass resolution on  $\rm J_{bg}$ and high-$z$ BH formation]{agarwal2012,yue2014,valiante2016,visbal_bryan_haiman2020,lupi_haiman_volonteri2021,sassano2021}.

On the other hand, once DCBHs start forming in our model, we are able to recover the results obtained by previous works which were able to trace the early build-up of $\rm J_{LW}$ variations. This shows that the permissive thresholds $\rm J_{crit,DCBH}$ and $\rm J_{crit,RSM}$ we impose tend to balance the limitations imposed by the \milltwo mass-resolution. Nevertheless, we underline that our $\rm J_{crit,DCBH}$ is on the lower-end of the typical range explored in the literature (i.e. $\rm0.01\!\lesssim\!J_{crit}/J_{21}\!\lesssim\!10^5$, as commented in Sect. \ref{sec:model:BH_seeding:RSM_and_DCBH_seeding_in_LGal}). Consequently, the amount of DCBHs formed in our model should be intended as an upper bound. As a comparison, we show the number density of DCBHs predicted by \cite{habouzit2016} with grey circles of different shades in Fig. \ref{fig:seeding_evolution}. This work used a suite of hydrodynamic simulations to explore different SNe feedback models, $\rm J_{crit}$ thresholds, box-sizes and mass resolutions. Their results show that, depending on the details of the feedback model employed and on the required $\rm J_{crit}$, the predicted number density of DCBHs at $z\!\sim\!8$ can vary by more than 2 orders of magnitude. Interestingly, they find that the rare DCBHs formed by requiring high $\rm J_{crit}$ offer a valid formation channel for the massive BHs powering high-$z$ QSOs, while the more common population of SMBHs observed at lower redshifts can be accounted for by requiring $\rm J_{crit}\!\sim\!30J_{21}$, comparably to our case. Similarly, \cite{dijkstra2014} analyzed the impact of $\rm f_{esc}$, $\rm J_{crit}$ and the presence of stellar winds, finding extreme variations ($\gtrsim\!6$ dex) of the predicted number density of DCBHs. We mark their results obtained for different $\rm J_{crit}$ thresholds as grey squares of different shades, while grey stars show their results for a model without chemically-enriched, SNe-powered winds. Overall, the comparison between our results and those of these works shows that our model is able to provide a reasonable compromise between the different approaches followed in the past.

Within our global SMBH population, objects formed as DCBHs always play a minor role, with only few of them reaching $z\!\sim\!0$ without mixing with other seed-types. On the contrary, the number of newly-formed RSM seeds shows a striking discontinuity at $z\!\sim\!8$, which rapidly brings the RSM class to be the second more abundant in the \milltwo box. 
The sudden increase of RMS seeds formation is the result of several concurring factors. Firstly, our grafting procedures stop at $z\!\sim\!8$, leaving all the newly-initialized halos potentially available for either DCBH or RSM formation in \texttt{L-Galaxies}. At similar times (i.e. at $z\!\sim\!8.9$), $\rm \langle Z_{IGM}\rangle\!>\!Z_{crit,DCBH}$ so that DCBHs formation is inhibited. Finally, also at $z\!\sim\!8$, the  $\rm J_{bg}\!>\!J_{crit,RSM}$, hence fostering the formation of RSM seeds. The presence of this discontinuity is therefore an effect of both the details of our model for the IGM evolution and the thresholds we impose for RSM formation on $\rm J_{local}$ and $\rm Z_{local}$. Nevertheless, this does not bias significantly the build-up of our global population of SMBHs, since light-seed descendants strongly prevail on all other seed classes. Indeed, although setting $\rm J_{crit,RSM}\!=\!J_{crit,DCBH}$ \citep[as in e.g.,][]{sassano2021} would drastically reduce the number of RSM formed in our model (see the main panel of Fig. \ref{fig:local_vs_background_Jlw}), this would only impact the $\rm dN(\mathit{z})/dV$ of BHs at $z\!<\!8$ by less than $0.1\%$, as it can be appreciated by comparing the evolution of light, RSM and light+RSM seed descendants in Fig. \ref{fig:seeding_evolution}. Similarly, the effects of rising $\rm J_{crit,RSM}$ would minimally affect the statistics of our $z\!=\!0$ SMBH population presented in the next sections. 

We note that only few RSM seeds form at $z\!>\!9$ in our model, under moderately low LW illumination ($\rm 0.1\!<\!J_{local}/J_{21}\!<\!1$). The limiting factor for the occurrence of this seeding channel at $z\!>\!9$ is the mild chemical enrichment required by the RSM scenario, which is difficult to attain at very high-$z$ in our model. Indeed, on one side the $\rm \langle Z_{IGM}\rangle$ at $z\!>\!9$ is too low to foster the formation of RSM seeds, and the rare metallic shells produced by active SF galaxies at these $z$ are never able to pollute neighboring collapsing halos. On the other hand, as soon as high-$z$ \lgal structures are able to form stars, they get rapidly polluted by internal SF processes beyond the limit for RSM seeds formation.

Finally, the formation of miDCBHs is particularly scarce in our cosmological box at all $z$. Indeed, gas-rich mergers of massive galaxies with $\rm m_r\!>\!0.3$ as those required to form miDCBHs are extremely rare in the \milltwo simulated volume \citep[see][]{izquierdo-villalba2019}. In detail, the only two miDCBHs formation episodes happen at $z\!=\!4.64$ and $z\!=\!0.35$. Both objects do not form within the most massive structure of their FoF group, but rather within the remnant of the merger between two satellite halos with $\rm M_{vir}\!\sim\!3\times10^9 M_\odot$. Both halos hosting miDCBHs are significantly gas rich ($\rm M_{cGas}/M_*\!>\!100$) and do not host central BHs before their miDCBH. As shown in Fig \ref{fig:seeding_evolution}, the number density of miDCBHs is few $\rm\sim\!10^7\,Mpc^{-3}$ at any $z$, roughly comparable to what predicted by \cite{bonoli2014} on the \texttt{MR} merger trees. One of the two miDCBHs form at $z\!=\!0.35$, hence much later than the typical BH-seeding epoch ($z\!\gtrsim\!8$). Also in this case, our result is in line with the predictions of \cite{bonoli2014}, where $\rm20\%$ of mergers in the local universe could host the formation of miDCBHs.

\subsection{The \texorpdfstring{$z\!=\!0$}{z=0} SMBH population produced by high-\texorpdfstring{$z$}{z} BH-seeding processes}
\label{sec:results:BHMF_BHoccupation_and_MstarVSmBH}
Thanks to the cosmological nature of the \milltwo merger trees, our model can track the mass build-up of BH-seeds descendants through cosmic times, as determined by gas accretion and BH mergers. Here we analyze this evolution and the properties of the resulting SMBH population at $z\!=\!0$.

\subsubsection{Mass growth of SMBH seeds}
\label{sec:results:seeds_evolution}
The progress of SMBH mass-assembly through cosmic time and the associated feedback on their host galaxies is thought to be one of the key evolutionary phenomena in galaxy evolution \citep[see e.g.,][for recent reviews]{alexander_hickox2012,fabian2012,kormendy_ho2013,heckman_best2014,reines_volonteri2015}. We follow the mass evolution of BH-seeds descendants by using the model described in \cite{izquierdo-villalba2020} for the time-prolonged mass-growth of BHs. The parameters of this growth-model were calibrated on the \texttt{MR} merger trees in order to reproduce the $z\!\sim\!0$ BH mass-function (BHMF) and the AGN luminosity function (LF) at $z\!<\!4$. In this section we simply illustrate that we are able to track the accretion history of BH-seeds descendants down to $z\!\sim\!0$, hence we do not perform a re-calibration of the growth-model parameters. Furthermore, we neglect the effects of BH-spin evolution, BH-BH merger-delay and GW-induced recoil after BH-BH mergers analyzed in \cite{izquierdo-villalba2020}. We leave the inclusion of these physical models and their adaptation to the \milltwo to an upcoming work. 

\begin{figure}
    \centering
    \includegraphics[width=0.48\textwidth]{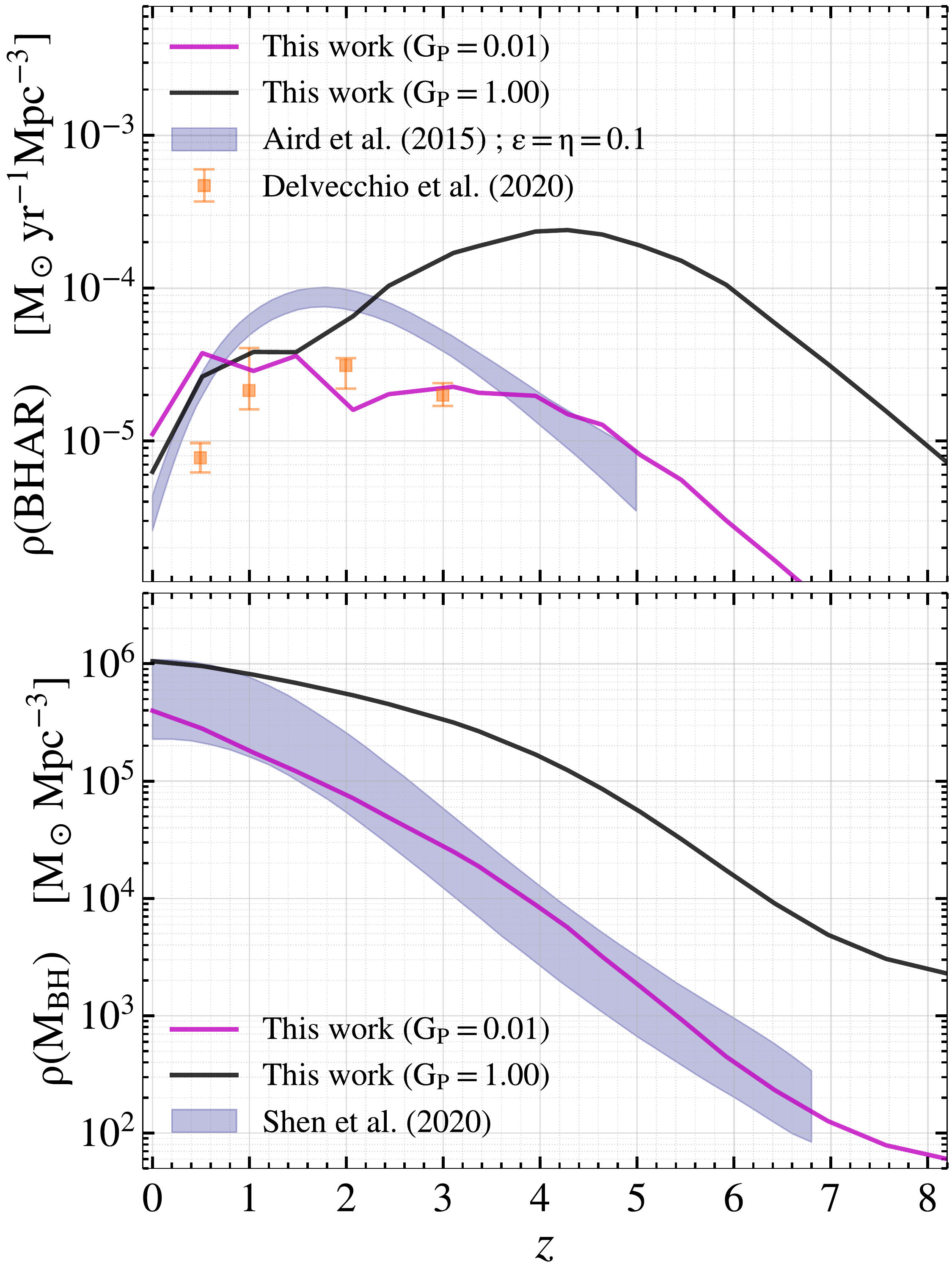}
    \caption{\textit{Upper panel}: evolution of the density of the $\rm M_{BH}$ accretion-rate (BHAR) for two runs of our model with identical parameters except for $\rm G_p$ (see Sect. \protect\ref{sec:model:BH_seeding_prescriptions:GQd_Grafting}). We show the extreme values of $\rm G_p\!=\!1$ and $\rm G_p\!=\!0.01$ (respectively solid black and solid purple lines) to bracket the effect of this parameter on our results. \textit{Lower panel}: evolution of the density of $\rm M_{BH}$ with respect to redshift, for the same runs as in the upper panel. These quantities provide insight about the cosmological mass-growth of SMBHs, showing that BHs in our model grow too rapidly at $z\!>\!2$ with respect to the observational constraints of \protect\cite{aird2015} and \protect\cite{delvecchio2020}, respectively blue shaded area and gold squares (upper panel), and \protect\cite{shen2020} in the lower panel.}
    \label{fig:bhar_and_mbh_densities}
\end{figure}

In Fig. \ref{fig:bhar_and_mbh_densities} we present the evolution of the BHs mass density $\rm\rho\,(M_{BH})$ and BH accretion-rate density $\rm\rho\,(BHAR)$, as predicted by two runs of our model with identical parameters except for $\rm G_p$ (see Sect. \ref{sec:model:BH_seeding_prescriptions:GQd_Grafting}).
In particular, our fiducial run ($\rm G_p\!=\!1$, solid black line) predicts a $\rm\rho\,(M_{BH})$ in excess of $\rm\sim\!1$ dex with respect to current constraints at $z\!\gtrsim\!6$. Similarly, this run strongly over-predicts the evolution of $\rm\rho\,(BHAR)$ with respect to current constraints at $z\!\gtrsim\!3$. We underline that this comparison with observational data is to be considered as illustrative, since it does not account for the selection effects under which observational constraints are obtained. Nevertheless, our fiducial run appears to anticipate in time the activity of SMBHs, by predicting a peak of $\rm\rho\,(BHAR)$ at $z\!\sim\!4$ rather than at $z\!\sim\!2$. On the other hand, the discrepancies we find at high-$z$ are less evident at $z\!<\!2$, where $\rm\rho\,(M_{BH})$ becomes compatible to current constraints as a consequence of the lower mass-growth rates of SMBHs. Indeed, as shown in the lower panel of Fig. \ref{fig:bhar_and_mbh_densities}, $\rm\rho\,(BHAR)$ for the $\rm G_p\!=\!1$ run significantly decreases at $z\!<\!4$, as a combined effect of SF processes, earlier BH growth and AGN feedback, which gradually reduce the amount of $\rm M_{cGas}$ within SMBH hosts. Consequently, the increasingly gas-poor mergers are unable to efficiently fuel the mass-growth of central BHs during their evolution at $z\!<\!4$.
\begin{figure*}
    \centering
    \includegraphics[width=0.848\textwidth]{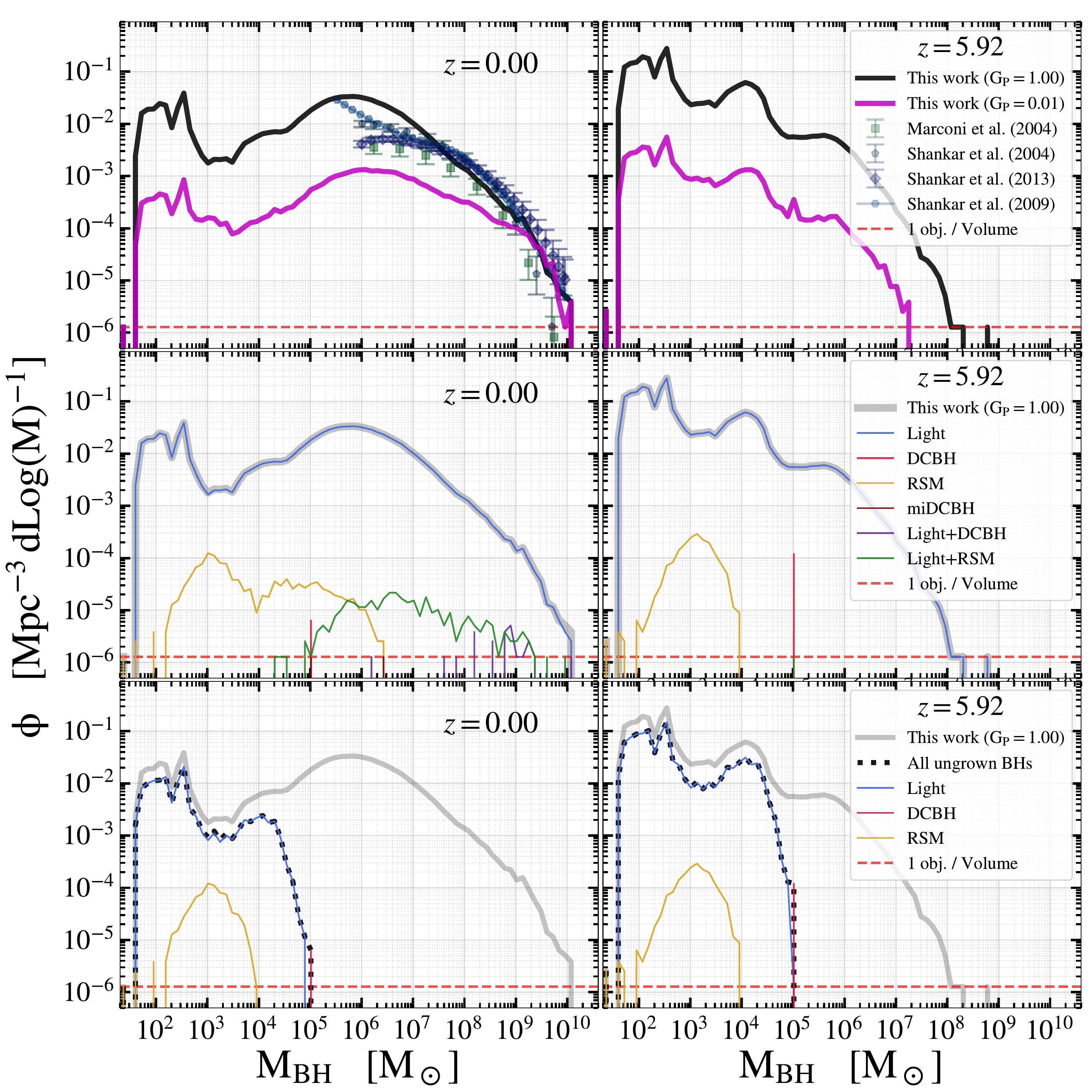}
    \caption{The BHMF produced by our model at the end of BH-formation epoch ($z\!\sim\!6$, right column) and at the endpoint of SMBH evolution ($z\!=\!0$, left column). \textit{Upper row}: comparison between the results obtained with the $\rm G_p\!=\!1$ and $\rm G_p\!=\!0.01$ runs (respectively: solid black and purple line). At $z\!=\!0$ we compare our results with the constraints of \protect\cite{marconi2004}, \protect\cite{shankar2004}, \protect\cite{shankar_weinberg_miralda-escude2009} and \protect\cite{shankar2013} shown as blue and green points and errors (upper left panel). \textit{Middle row:} contribution of each seed-type (color-coded as shown in the legend) to the total BHMF for the $\rm G_p\!=\!1$ run (solid grey line). \textit{Lower row}: contribution of BHs which never underwent any mass-growth since their formation (dotted black line), split into different seed-types (color coded as in the plot legends). The horizontal dashed red line in each panel marks the level at which only one object is present in the whole \milltwo box.}
    \label{fig:BHMFs_wholesome}
\end{figure*}

This picture dramatically changes by reducing the abundance of light-seed descendants inherited from \texttt{GQd}. Indeed, the evolution of $\rm\rho\,(M_{BH})$ predicted by the $\rm G_p\!=\!0.01$ run (solid purple line) shows a better agreement with current determinations at $z\!\gtrsim\!2$. Also in this case, the total $\rm M_{BH}$ density in the \milltwo box at $z\!=\!0$ is comparable with the constraints of \cite{shen2020}, confirming that the BH growth model of \cite{izquierdo-villalba2020} is able to correctly predict the total amount of mass accreted onto BHs throughout cosmic history. At the same time, the BHAR predicted by the $\rm G_p\!=\!0.01$ appear to be in better agreement with current determinations, although the lower number of SMBHs manifests in the noisier evolution of $\rm\rho\,(BHAR)$. Interestingly, the predictions of this run show a rather flat evolution of $\rm\rho\,(BHAR)$ at $1\!<\!z\!<\!4$, rather than showing a clear peak as for the $\rm G_p\!=\!1$ case. This is an effect of the low BH-occupation for this run, which on one side reduces the total amount of mass being accreted onto central BHs at high-$z$ in the whole \milltwo box, and on the other side reduces the action of AGN feedback onto the cold-gas of SMBH hosts. Indeed, these systems retain the $\rm M_{cGas}$ which would have fuelled the growth of their central BH in the $\rm G_p\!=\!1$ run, allowing for a later mass-growth of SMBHs driven by hierarchical mergers. 

In summary, the mass-assembly history of SMBHs predicted by our model favours a low occupation of \gqd light-seed descendants, although the noisy evolution of $\rm\rho\,(BHAR)$ for the $\rm G_p\!=\!0.01$ run and its anticipated peak in redshift for the $\rm G_p\!=\!1$ run suggest that the interplay between our BH formation model and the mass-growth recipes of \cite{izquierdo-villalba2020} require further calibration when the latter are extended to the \milltwo dynamical range.
\subsubsection{Properties of the \texorpdfstring{$z\!=\!0$}{z=0} SMBH population}
\label{sec:results:properties_of_SMBHs_population_at_z0}
Here we focus on the endpoint of the cosmological evolution of our SMBH population, presenting its statistical properties at $z\!=\!0$, in connection to our high-$z$ BH-seeding model. Figure \ref{fig:BHMFs_wholesome} compare our predictions for the BHMF at two representative redshifts, namely: $z\!=\!0$, where our model can be anchored to recent observations (left column), and $z\!\sim\!6$, just at the end of the BH-seeding epoch (right column). 
From top to bottom, we show: i) the differences between the $\rm G_p\!=\!1$ and $\rm G_p\!=\!0.01$ runs (upper row), ii) the contribution of each seed-type class to our BHMF predicted by the  $\rm G_p\!=\!1$ run (middle row) and finally iii) the mass distribution of BHs which never grew since their formation (bottom row).

The top left panel shows that the BHMF predicted by our $\rm G_p\!=\!1$ run is in good agreement with the local determinations of \cite{marconi2004, shankar2004, shankar_weinberg_miralda-escude2009} and \cite{shankar2013}. Therefore, the total integrated growth of SMBHs down to low-$z$ is well-predicted by the \cite{izquierdo-villalba2020} model also on the merger-trees of the \texttt{MR-II}, especially for SMBHs with $\rm M_{BH}\!\gtrsim\!10^7\,M_\odot$. Despite being affected by low-statistics due to the volume limitations of the \milltwo{} (shown as the horizontal, dashed red line), the highest mass bins of our BHMF at $z\sim5-6$ are broadly comparable to the determination of \cite{kelly_shen2013}, which appear to prolong our results beyond $\rm M_{BH}\!\gtrsim\!3\times10^8M_\odot$ (we do not report their data, as they lie outside the number density ranges accessible by our work). On the other hand, the low SMBH occupation of the $\rm G_p\!=\!0.01$ run is not sufficient to recover the BHMF normalization suggested by observational constraints. This contrasts with our predictions for the mass-growth history of SMBHs, which favour a low $\rm G_p$ value (see Sect. \ref{sec:results:seeds_evolution}). This further suggests that the extension of the \cite{izquierdo-villalba2020} model to the \milltwo dynamical range and its interplay with our BH-formation model require accurate adjustments, which we leave for an upcoming work.

The differences induced by varying $\rm G_p$ completely vanish at $\rm M_{BH}\!\gtrsim\!10^9\,M_\odot$ in our $z\!=\!0$ BHMF, showing that the details of our grafting procedure are washed-out by the cosmological growth of SMBHs only at the highest BH masses. This supports the idea that mass-accretion induced by the hierarchical assembly of their hosts primarily drives the growth and evolution of $z\!\sim\!0$ SMBHs \citep[see e.g.,][]{kauffmann_haenelt2000,malbon2007,fanidakis2011}. On the other hand, acting on the grafting of \gqd BHs at high-$z$ produces differences of $\gtrsim1$ dex at $\rm M_{BH}\!\leq\!10^7\,M_\odot$. This suggests that the efficiency of light-seeds formation in our model is imprinted on the abundance of $z\!\sim\!0$ SMBHs with relatively low mass \citep[see also:][]{sesana_volonteri_haardt2007,volonteri_lodato_natarajan2008,valiante2021}. Recently, \cite{degraf_sijacki2019} explored the effect of a varying BH-seeding efficiency on the BHMF computed at different cosmic epochs. Their analysis relies on a post-processing BH-formation and evolution model applied to the Illustris simulation \citep[][]{nelson2015}. Also in their case the efficiency of BH-formation affects the normalization of the BHMF at low redshift, producing a shift at all $\rm M_{BH}$. This difference with our results is expected, since in their model the BH-formation efficiency does not depend on $\rm M_{vir}$, as in our case. Overall, we argue that the common approach of initializing a SMBH in every newly-resolved structure, at any $z$, should be considered with caution when modelling the low-mass end of the BHMF \citep[see e.g.,][]{fanidakis2011,lacey2016,cora2018,trinca2022}.

The middle-row panels of Fig. \ref{fig:BHMFs_wholesome} show the BHMF of our run with $\rm G_p\!=\!1$, split into the contributions of different seed classes. It is evident that the descendants of \gqd BHs are the most numerous class both at $z\!=\!0$ and $z\!\sim\!6$, over the whole $\rm M_{BH}$ distribution. As expected, at $z\!\sim\!6$ (end of the BH-seeding epoch), only light-seed descendants, DCBH and RSM seeds are present, with the latter class showing a mass distribution peaked at $\rm 10^2\!<\!M_{seed}/M_\odot\!\sim\!10^4$, in line with what expected for IMBHs formation \citep[see e.g.,][]{devecchi_volonteri2009,volonteri2010,lupi2014,katz_sijacki_haehnelt2015,sassano2021}. On the other hand, also light+RSM and light+DCBH classes can be found at $z\!=\!0$, although providing minor contributions. Therefore, in our model, the mixing of seed-types does not appear necessary to populate the massive end of the $z\!=\!0$ BHMF. 
Indeed, the highest mass bins of our BHMF are populated by light-seeds descendants whose mass growth has been mainly driven by galaxy mergers, as also testified by the increasing relative abundance of light+RSM and light+DCBH classes at $\rm M_{BH}\!>\!10^{\,6}M_\odot$.

By extending the dynamical range of \texttt{GQd}, the recent work of \cite{trinca2022} showed that a merger-driven model for the growth of light PopIII remnants can bring a significant fraction of the latter in the $\rm10^5\!\lesssim\!M_{BH}/M_\odot\!\lesssim\!10^8$ mass regime already at $z\!\gtrsim\!6$. In light of this, it is not surprising for us to obtain light-seed descendants as massive as $\rm M_{BH}\!\sim\!10^{10}M_\odot$ at $z\!=\!0$, since galaxy mergers able to drive BH-growth are expected to be extremely frequent in the \milltwo \citep[as shown in][]{izquierdo-villalba2019}. Nevertheless, our results extends this scenario over a wide, cosmological volume by employing the prolonged accretion model of \cite{izquierdo-villalba2020}, instead of assuming a fixed mass-accretion timescale following galaxy mergers.

Generally, BH formation models focusing on the origin of the first SMBHs at $z\!>\!6$ find that, if super-Eddington accretion onto light seeds is neglected, the contribution from intermediate or massive seeding channels is necessary in order to reach the masses inferred for the brightest QSOs at $z\!\gtrsim\!7$ \citep[see e.g.,][]{lupi2014,valiante2016,latif_volonteri_wise2018,lupi2021,sassano2021,trinca2022}. Given our simulation volume of $\rm V\!=\!10^{-3}\,h^3\,Gpc^{-3}$, our results cannot be conclusive about the role of intermediate or heavy seeding scenarios in the formation of these extreme objects within the rarest and most biased regions of the Universe. Indeed, the latter might follow a significantly different evolution with respect to the one experienced at the typical halo masses resolved by the \texttt{MR-II} simulation.
Interestingly, our model predicts that a fraction of BHs never grows from their initial $\rm M_{seed}$. In particular, we find that these \textit{ungrown} BHs represent $50\%$ of all BHs hosted in halos with $\rm M_{vir}\!<\!10^9M_\odot$ at $z\!=\!0$, while this fraction rapidly drops to $<\!0.01\%$ at $\rm M_{vir}\!=\!5\times10^{10}M_\odot$. Ungrown BHs are generally hosted in isolated galaxies which never experienced mergers throughout their evolution. Indeed, the driving mechanism for galaxy interactions and BH mass-growth in the \milltwo box is the merger between massive, central galaxies and small satellites, as shown by \cite{izquierdo-villalba2019}. The bottom panels of Fig. \ref{fig:BHMFs_wholesome} show the contribution of ungrown BHs to the total BHMF, highlighting the different seed-types within the class of ungrown BHs. A large fraction of BHs with $\rm M_{BH}\!\lesssim\!10^4M_\odot$ never experienced any growth, both at high-$z$ and low-$z$, with the majority of them being light-seeds descendants \citep[as in the reference model of][]{trinca2022}. Furthermore, by comparing the left panels of the middle and bottom row, we note that most of the RSM seeds found at $z\!=\!0$ with  $\rm M_{BH}\!<\!10^4M_\odot$ never grew since their formation.

Overall, ungrown BHs represent the dominant population at $\rm M_{BH}\!\lesssim\!3\times10^{\,4}M_\odot$ at all $z$ in our model. This suggests that intermediate-mass BHs ($\rm M_{BH}\!\lesssim\!10^{\,4}M_\odot$) at $z\!\sim\!0$ might still carry significant information about early epochs of BH-evolution, in line with recent works \citep[see e.g.,][]{volonteri_lodato_natarajan2008,greene2012,cann2021,mezcua2021,valiante2021}. We underline that our definition of ungrown BHs is only based on the mass-growth we can resolve during the evolution of BHs traced by \lgal on the \milltwo merger trees. However, light-seed descendants are able to efficiently grow before their grafting into \texttt{L-Galaxies}. This early growth (unresolved by the \texttt{MR-II}) typically brings them in the IMBHs mass range (i.e. $\rm 10^3\!<\!M_{BH}/M_\odot\!<\!10^5$), as demonstrated by Fig. \ref{fig:BHMF_of_newly_grafted_BHs}.
Consequently, the ungrown light-seed descendants predicted by our model at $z\!=\!0$ with $\rm M_{BH}\!\gtrsim\!10^3M_\odot$ do not directly represent the ungrown relics of high-$z$ light seeds. 
This is in line with the idea presented in \cite{mezcua2019}, according to which the average population of low-mass SMBHs hosted in dwarf galaxies might not be a good tracer of high-$z$ BH-seeding processes. On the other hand, we also inherit BHs with $\rm M_{BH}\!\lesssim\!10^2M_\odot$, i.e. objects which never grew during their \gqd evolution. As shown in the bottom-left panel of Fig. \ref{fig:BHMFs_wholesome}, $\!\gtrsim50\%$ of these BHs never grow also in \texttt{L-Galaxies}, down to $z\!=\!0$. These direct descendants of light-seeds formed at $z\!>\!20$ in \texttt{GQd} are typically hosted in extremely small \milltwo halos at $z\!=\!0$ ($\rm M_{vir}\!\lesssim5\times10^9M_\odot$, corresponding to $\lesssim\!500$ DM particles). The presence of these ungrown light-seed descendants in our results suggests that at least a fraction of low-mass BHs in local dwarf galaxies might still carry significant information about their formation process \citep[e.g.,][]{mezcua2021}. Nevertheless, this conclusion should be regarded as speculative since the evolution of these low-mass structures might be affected by our mass-resolution limits. Finally, the low-mass end of our BHMF shows few ``wiggles'', especially at $z\!>\!1$ and $\rm M_{BH}\!\lesssim\!10^{\,5}M_\odot$. These are due to our \gqd grafting procedures, from which we inherit an already-evolved population of BHs with an intrinsic mass-distribution, as shown in Fig. \ref{fig:BHMF_of_newly_grafted_BHs} and by the recent work of \cite{trinca2022}.\vspace{1.5mm}\\

\noindent
\textit{BH occupation and $\rm M_*$-$\rm M_{BH}$ scaling relation}\vspace{1.5mm}\\
\noindent
The overall efficiency of BH-seeding processes is expected to affect the fraction of galaxies hosting a central SMBH at $z\!=\!0$ \citep[e.g.,][]{buchner2019}. Indeed, assuming that massive BHs form in the majority of high-$z$ galaxies and that they are retained within their hosts down to low-$z$, it is reasonable to expect the presence of massive BHs within most low-$z$ galaxies \citep[see e.g.,][]{volonteri_lodato_natarajan2008,van-wassenhove2010}. This simplified picture can be modified during the hierarchical assembly of BH hosts if massive BHs are ejected due to gravitational recoils after their merger \citep[e.g.,][]{volonteri2007,volonteri_gultekin_dotti2010,dunn2020,izquierdo-villalba2020,askar_davies_church2021b}. Nevertheless, since we ignore the effect of recoils, we expect to observe an imprint of the efficiency of high-$z$ BH-seeding on the $z\!\sim\!0$ BH occupation, especially for low-mass galaxies with a quiet evolution.
\begin{figure}
    \centering
    \includegraphics[width=0.49\textwidth]{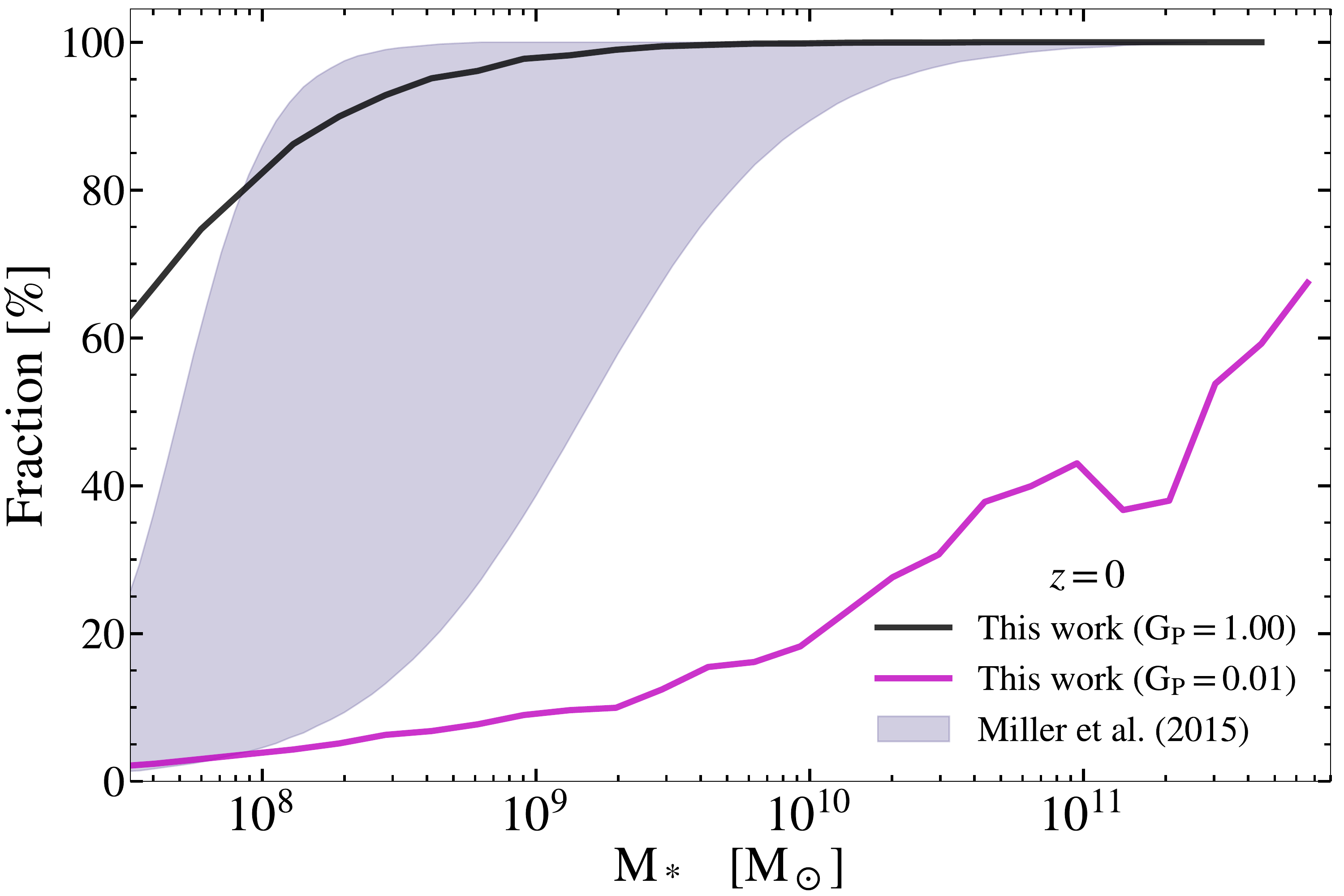}
    \caption{Black Hole occupation fraction as a function of stellar mass, for the two runs with $\rm G_p\!=\!1$ (solid black line) and $\rm G_p\!=\!0.01$ (solid purple line). We compare our results to the local observational constraints of \protect\cite{miller2015}, shown as the blue shaded area. This shows how the efficiency of light-seeds formation at high-$z$ is reflected into the properties of our SMBH population at  $z\!=\!0$.}
    \label{fig:BH_occupation}
\end{figure}

We show this prediction of our model in Fig. \ref{fig:BH_occupation} as a function of the $\rm M_*$ of SMBH hosts, for the two runs with $\rm G_P\!=\!1$ and $\rm G_P\!=\!0.01$. We compare our results to the observations of \cite{miller2015}, which offer relatively solid constraints at $\rm M_*\!\gtrsim\!10^{10}M_\odot$, based on X-ray observations of optically-selected galaxies by the AMUSE survey \citep[][]{gallo2008}. Our fiducial run ($\rm G_P\!=\!1$) predicts that $\rm>\!80\%$ of galaxies with $\rm M_*\!>\!10^8M_\odot$ host a central, massive BH and it recovers the high occupation fraction ($\sim\!100\%$) expected at $\rm M_*\!>\!10^{10}M_\odot$. 
On the other hand, the $\rm G_p\!=\!0.01$ run worsen the agreement of our predictions with observations. Indeed, 
less than $\rm\sim\!50\%$ of galaxies with $\rm M_*\!>\!10^{10}M_\odot$ host a central SMBH, in tension with the high occupation of SMBHs expected in local massive galaxies \citep[see e.g.,][]{volonteri_bellovary2012,kormendy_ho2013,reines_volonteri2015}. This suggests that our model favours a high efficiency of high-$z$ BH-seeding processes in order to recover the statistical properties of SMBHs observed at $z\!\sim\!0$, as for the case of our BHMFs (see Fig. \ref{fig:BHMFs_wholesome}).
\begin{figure}
    \centering
    \includegraphics[width=0.49\textwidth]{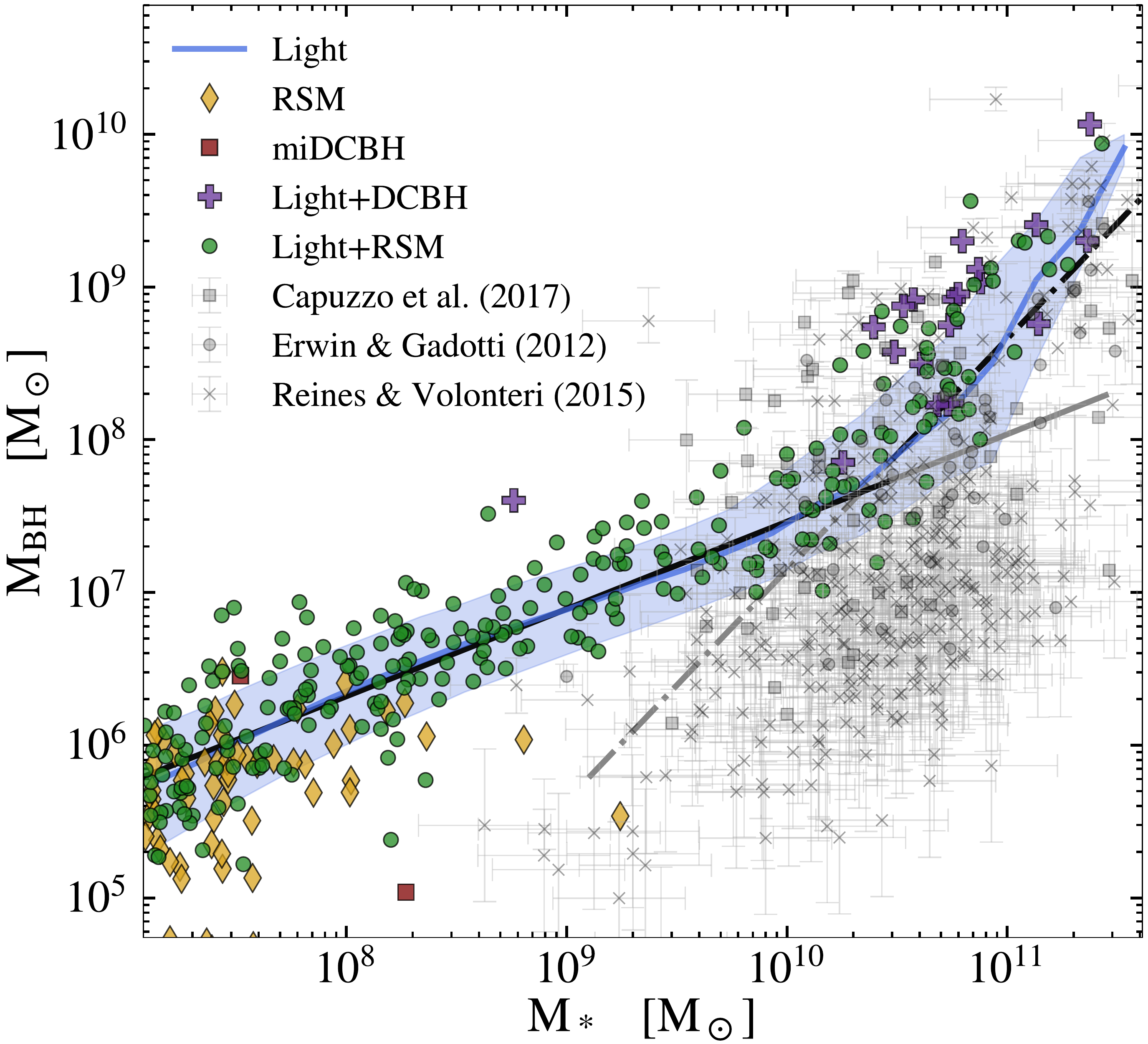}
    \caption{$\rm M_{BH}$-$\rm M_*$ relation for galaxies with $\rm M_{vir}\!>\!3\times10^8M_\odot$, hosting a central SMBH at $z\!=\!0$. We show the median relation and the $\rm16^{th}\!-\!84^{th}$ percentiles of the distribution of light-seed descendants (solid light-blue line and shaded area). The remaining seed-types are instead shown as colored symbols (see the plot legend). The observational results of \protect\cite{erwin_gadotti2012}, \protect\cite{reines_volonteri2015} and \protect\cite{capuzzo-dolcetta_tosta-e-melo2017} are shown as grey dots, crosses and squares. Finally, we split our sample at $\rm M_*\!=\!3\times10^{10}M_\odot$ and perform two linear fits to our predicted relation, in order to quantify its break (black solid and dashed-dotted lines). To highlight their comparison, we also extend the two fits outside their respective ranges (grey solid and dashed-dotted lines).}
    \label{fig:Mstar_vs_MBH_relation}
\end{figure}

We note that the broad constraints of \cite{miller2015} at $\rm M_*\!\lesssim\!3\times10^9M_\odot$ might allow to tune $\rm G_p$ in order to reconcile with current constraints the BH-occupation fraction in dwarf galaxies, our $z\!\sim\!0$ BHMF and the evolution of $\rm\rho\,(BHAR)$ and $\rm\rho\,(M_{BH})$ at the same time. Nevertheless, the GW-induced expulsion of BHs from their host, presented in \cite{izquierdo-villalba2020}, affects the BH occupation in similar ways as our $\rm G_p$ parameter. Indeed, BH-seeds can be efficiently expelled from structures of the \milltwo at high-$z$, due to the shallow gravitational potential of their hosts. Furthermore, GW-kicks are maximized by BH-BH mergers with mass-ratios close to $1$, a condition which is likely verified at $z\!>\!10$ by the newly-grafted BHs covering a relatively narrow mass range ($\rm 10^2\!\lesssim\!M_{BH}/M_\odot\!\lesssim\!10^5$). We plan to address the degeneracy between $\rm G_p$ and GW-kicks in an upcoming work.

Recent works have focused on the population of massive BHs hosted in local dwarf galaxies, trying to pinpoint the imprint of high-$z$ seeding processes on the local scaling relations between massive BHs and their hosts \citep[see e.g.,][]{habouzit_volonteri_dubois2017,mezcua2017,martin-navarro_mezcua2018,kristensen2021}. To explore this, in Fig. \ref{fig:Mstar_vs_MBH_relation} we show the relation between $\rm M_*$ and $\rm M_{BH}$ for all galaxies hosting a central SMBH at $z\!=\!0$. To minimize resolution effects, we only consider structures hosted in DM halos with $\rm M_{vir}\!>\!3\times10^8M_\odot$ (corresponding to halos resolved with at least 25 particles of the \texttt{MR-II}). We present the median $\rm M_*$-to-$\rm M_{BH}$ relation for light-seed descendants (respectively: solid light-blue line and shaded areas). For the less abundant RSM, light+RSM and light+DCBH classes we only show individual points, respectively as yellow diamonds, green circles and purple crosses.

The high-mass end of our $\rm M_*\!-\!M_{BH}$ relation, at $\rm M_*\!>10^{10}M_\odot$ and $\rm M_{BH}\!>10^7M_\odot$, is consistent with the observational constraints of \cite{erwin_gadotti2012}, \cite{capuzzo-dolcetta_tosta-e-melo2017} and \cite{reines_volonteri2015}, although the latter significantly populate also the region between $\rm 10^{10}\!\lesssim\!M_*/M_\odot\!\lesssim\!10^{11}$ and $\rm 10^{6}\!\lesssim\!M_{BH}/M_\odot\!\lesssim\!10^{8}$, differently from our predictions. The origin of this discrepancy can be partially ascribed to our impossibility to replicate the selection effects of \cite{reines_volonteri2015}, which focused on a uniform sample of AGN in low-mass, local galaxies. In addition, is it possible that our model produces galaxies with too little $\rm M_*$ with respect to their central SMBH, especially at $\rm M_*\!\lesssim\!10^{10}M_\odot$. This might be a consequence of our BH-growth model, which may favour the production of massive BHs at the expenses of the $\rm M_*$ of their small-dwarf galaxy hosts. We refrain to further comment on this topic since we plan to thoroughly address it in a follow-up paper. 

Overall, our model predicts a clear break in the $\rm M_*\!-\!M_{BH}$ relation at $\rm M_*\!\sim\!5\times10^{10}M_\odot$, as also shown by recent theoretical works\citep[e.g.,][]{sharma2020,bhowmick2021,habouzit2021}. This result is compatible with the flattening of the relation between $\rm M_{BH}$ and the velocity dispersion of bulge stars ($\sigma_\mathit{v}$) measured by recent observational works at $\rm \sigma_\mathit{v}\!\lesssim\!100\, km\, s^{-1}$ \citep[roughly equivalent to $\rm M_*\!\sim\!3\times10^{10}M_\odot$, see e.g.,][]{mezcua2017,martin-navarro_mezcua2018,mezcua2019,ferre-mateu2021}. Futhremore, we tentatively compare our findings with the results of \cite{reines_volonteri2015}, although this comparison should be considered as illustrative since we cannot replicate the selection effects of AGN observations on our results. More in detail, we perform simple linear fits to our data by using
\begin{equation}
    \rm Log\left(M_{BH}/M_\odot\right) = \alpha + \beta\,Log\left(M_*/[10^{11}\,M_\odot]\right)\,,
    \label{eq:mstar_vs_mbh_relation}
\end{equation}
as in \cite{reines_volonteri2015}, and by splitting our results in two different samples at $\rm M_*\!=\!3\times10^{10}M_\odot$ \citep[following][]{martin-navarro_mezcua2018}. We show the two fits in Fig. \ref{fig:Mstar_vs_MBH_relation} as solid and dashed-dotted black lines and their extension outside their respective fitting ranges as grey lines. For the high-mass range fit, we find $\rm\alpha\!=\!8.66\pm0.03$ and $\rm\beta\!=\!1.51\pm0.09$, which are compatible with the values $\rm\alpha\!=\!8.95\pm0.09$ and $\rm\beta\!=\!1.40\pm0.21$ found by \cite{reines_volonteri2015} for the sample of elliptical galaxies and S0 with classical bulges at $\rm M_*\!\gtrsim\!10^{10}M_\odot$. We note that our high-mass range fit deviates from our $\rm M_*\!-\!M_{BH}$ relation at $\rm M_*\!\gtrsim\!10^{11}M_\odot$, since its slope is mainly driven by the more abundant population of $\rm 10^{10}\!<\!M_*/M_\odot\!<\!10^{11}$ objects. On the other hand, at the low-mass range we find a $\rm\alpha\!=\!8.03\pm0.01$ and $\rm\beta\!=\!0.57\pm0.01$, showing the strong change of slope of our relation. We refrain from further commenting our results for this low-mass regime, due to the current absence of data for the $\rm M_*\!-\!M_{BH}$ relation covering the $\rm M_*\!<\!10^{10}M_\odot$ and $\rm M_{BH}\!<\!10^{7}M_\odot$ ranges.
\begin{figure*}
    \centering
    \includegraphics[width=0.85\textwidth]{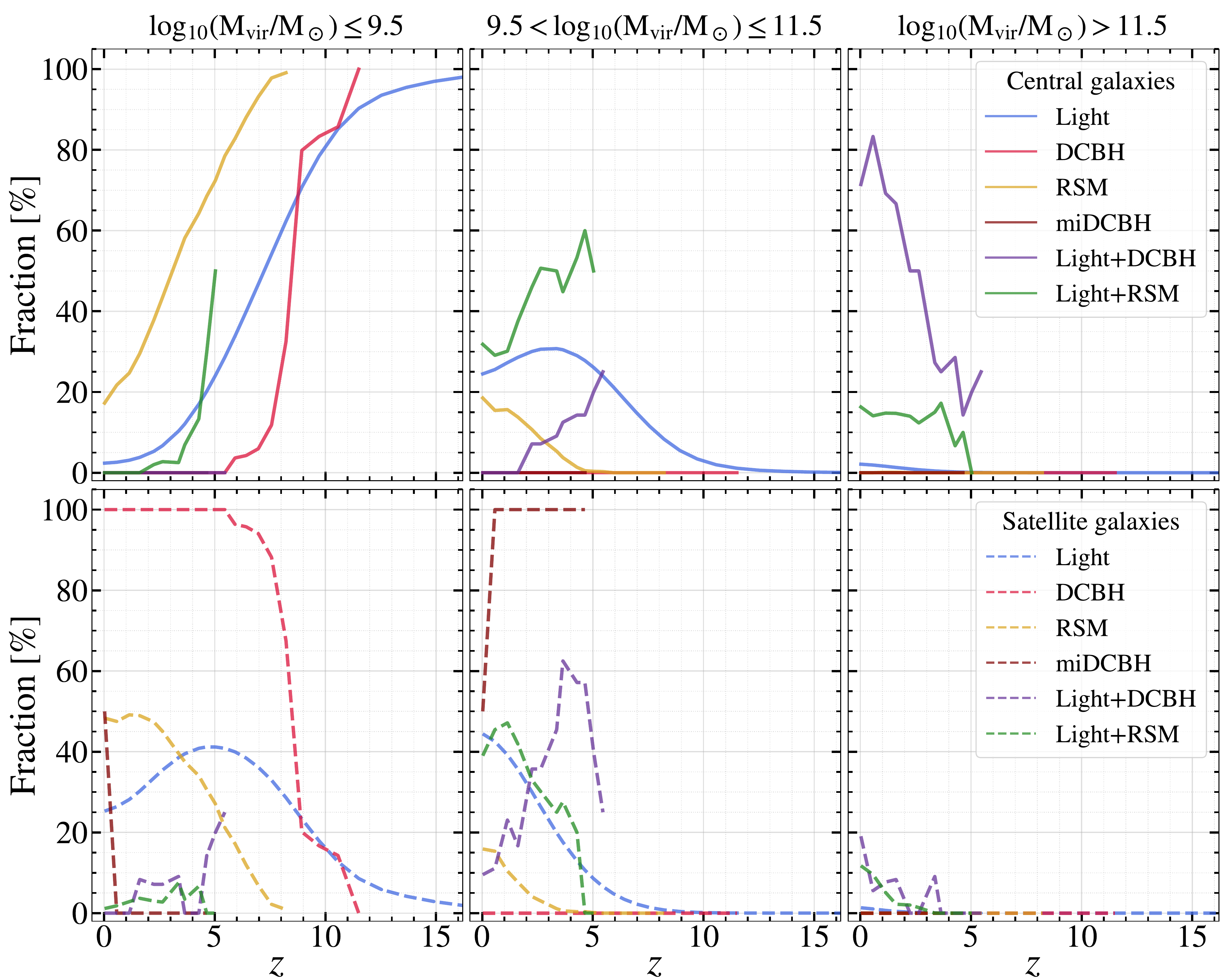}
    \caption{Fractions of BHs hosted in central galaxies (upper row) and satellite galaxies (lower row). Different colored lines represent descendants of different seed-type classes, as shown in the plot legend. For each seed-class and at fixed $z$, fractions add to $100\%$ when considering the six panels together. This shows how BH-seeds descendants mix and distribute over the \milltwo dynamical range.}
    \label{fig:seedtypes_in_centrals_and_satellites}
\end{figure*}

We note that the highest $\rm M_*$ and $\rm M_{BH}$ bins in Fig. \ref{fig:Mstar_vs_MBH_relation} are only populated by light, light+RSM and light+DCBH descendants, showing that DCBH or RSM seeds can only reach $\rm M_{BH}\!>\!10^7M_\odot$ as a consequence of the hierarchical growth of their hosts. Indeed, DCBHs and RSMs descendants at $z\!=\!0$ are only hosted in low-mass galaxies ($\rm M_*\!\lesssim\!10^9M_\odot$, hence supporting the idea that BHs which are able to retain memory of their formation channel (i.e. by experiencing a quiet evolution), can end up in dwarf galaxies by $z\!\sim\!0$ \citep[see][]{van-wassenhove2010,cann2021,mezcua2021}.

These results show that our model is able to draw predictions for the yet poorly explored mass range of local dwarf galaxies, potentially allowing to shed light on the connection between the population of low-mass SMBHs at $z\!\sim\!0$ and high-$z$ BH-seeding processes. Nevertheless, a thorough analysis of the scaling relations between massive BHs and their hosts is beyond the scope of this work, hence we plan to further address this point in a future work.

\subsubsection{Cosmological evolution of BH-seeds hosts}
\label{sec:results:evolution_of_BHseeds_hosts_properties}
The N-body merger trees of the \milltwo allow to focus on a dynamical range encompassing dwarf satellites and Milky Way-like halos, making our model suitable to analyze how the descendants of BHs formed at $z\!\gtrsim\!6$ distribute within the population of $\rm\lesssim\!10^{11} M_*$ galaxies across cosmic history.

As discussed in Sect \ref{sec:results:seeds_environment}, the $\rm J_{local}$ and $\rm Z_{local}$ requirements tend to promote the formation of DCBH and RSM seeds in the vicinity of high-$z$ star-forming galaxies, potentially leaving an imprint on the fraction of BHs hosted into central or satellite galaxies at later epochs. In Fig. \ref{fig:seedtypes_in_centrals_and_satellites} we show the evolution of these fractions in bins of $\rm M_{vir}$ for each seed-flavour. We underline that, for each seed-type and at fixed $z$, the fractions add to $100\%$ when considering the six panels together. In this way, we illustrate how the different seed-types mix in time and distribute over the \milltwo dynamic range. We note that \lgal baryonic structures are always initialized as the only member of their associated DM halos, hence as central galaxies. This explains the high fractions of light, DCBH and RSM seeds hosted in central galaxies at $z\!>\!8$. The two miDCBHs forming in our box represent an exception to this picture, as they form after the merger between $\rm M_{vir}\!\sim\!3\times10^9M_\odot$ satellites of larger halos (see Sect. \ref{sec:results:seed_types_definition}). Thus, they are never hosted by central galaxies, as shown by the brown dashed lines in the bottom panels of Fig. \ref{fig:seedtypes_in_centrals_and_satellites}.

The fraction of seed-types hosted in central galaxies diminishes in time for the lowest $\rm M_{vir}$ bin (upper left panel in Fig. \ref{fig:seedtypes_in_centrals_and_satellites}), as small halos get accreted by larger structures across their evolution. In particular, the fraction of DCBHs hosted in central galaxies drops to zero already by $z\!\sim\!6$, showing that DCBH hosts form promptly become satellites of larger neighbors. This scenario is confirmed by the appearance of the light+DCBH class, also at $z\!\sim\!6$, consequently to the mergers between light and DCBH seeds hosts. Analogously, RSM and light-seed hosts gradually merge producing the mixed light+RSM class (green lines). Nevertheless, RSM seeds can also form in the absence of nearby SF halos due to the sufficient $\rm J_{bg}$ (see Sect. \ref{sec:results:high-z_results}), leading to a milder decrease of the central-galaxies fraction hosting RSM seeds. Indeed, a larger fraction of RSM descendants are found in low-mass centrals at $z\!<\!8$ with respect to any other seed-type (upper left panel). In addition, $\sim\!20\%$ of halos hosting RSM manage to become more massive than $\rm M_{vir}\!=\!10^{9.5}M_\odot$ by $z\!=\!0$. This happens thanks to either: i) the merger with other RSM hosts, ii) the smooth accretion of unresolved DM or iii) the merger with halos initialized after $z\!\sim\!6$ which do not host a central BH.
 
At $z\!\sim\!0$ only $\sim\!30\%$ of light-seed descendants are hosted in central galaxies at any $\rm M_{vir}$. Among these, only $\sim\!3\%$ are hosted in DM halos reaching the highest DM masses of the \milltwo dynamic range (right column). This shows that DM halos with $\rm M_{vir}\!>\!10^{11.5} M_\odot$ are rare in the \milltwo box at any $z$. 
Overall, light-seeds hosts cover a wide dynamic range, being present both as centrals and satellites at all $\rm M_{vir}\!<\!11.5$, especially at $z\!\lesssim\!5$. Indeed, since our grafting procedures are independent from $\rm J_{local}$ and $\rm Z_{local}$, the hosts of light-seed descendants include a combination of central and satellite galaxies with diverse merging histories, hence tracing the average evolution of all the \texttt{MR-II} structures. 
Consequently, also the hosts of light+RSM and light+DCBH are found at all masses, both in satellites and centrals. 
This is opposite to the case of those RSMs and DCBHs which retain memory of their formation channel down to $z\!\sim\!0$. In particular, DCBH hosts are exclusively found at the lowest virial masses at any redshift, showing that in our model DCBHs form in small $\rm M_{vir}\leq10^{9.5}M_\odot$ halos and that a fraction of the latter evolves down to $z\!=\!0$ without experiencing a merger throughout their entire history. 

\section{Conclusions}
\label{sec:conclusions}
We present a complete and detailed semi-analytic model for SMBH-formation, which  accounts  for most of the currently-envisioned channels of BH-seeding, from light PopIII remnants to massive direct-collapse BHs. 

We embed our model in the \lgal SAM, coupling it to the  mass-growth modeling of \cite{izquierdo-villalba2020}. We apply this model to the cosmological box of the N-body simulation \texttt{Millennium-II}, and study the evolution of the multi-flavour population of BHs produced by BH-seeding processes. Due to  the mass-resolution limit of the simulation, the formation and early-evolution of light PopIII remnants is accounted for in a sub-grid fashion, by ``grafting'' the evolved population of PopIII remnants predicted by the \gqd model \citep[][]{valiante2021} into newly-resolved structures of the \texttt{MR-II}. On the other hand, we directly model the formation of intermediate seeds in dense, nuclear star clusters from runaway stellar mergers (RSM) and heavy seeds from the direct collapse of pristine gas clouds (DCBH), as the \texttt{MR-II} allows to resolve atomic-cooling halos above $\rm\sim 10^8 M_\odot$. In our model, the occurrence of these seeding-scenarios depends on the Lyman-Werner (LW) background $\rm J_{bg}$ and on the average IMG metallicity $\rm\langle Z_{IGM}\rangle$, as well as on local variations of the LW flux and IGM metallicity. In addition, we include the formation of merger-induced DCBHs  (miDCBH) within the remnants of gas-rich major-mergers between galaxies hosted in $\rm M_{vir}\!>\!10^9M_\odot$ halos without pre-existing massive BHs.

A large fraction of structures simulated by our model at $z\!>\!12$ inherits a central BH from \texttt{GQd}, hence our results show a strong predominance of light-seeds descendants. As an example, all halos of $\rm M_{vir}\!=\!10^{8}M_\odot$ initialized at $z\!\sim\!16$ inherit a light-seed descendant. This fraction gradually lowers in time ($\rm<\!10\%$ for the same $\rm M_{vir}$ at $z\!\sim\!9$), hence leaving an increasing number of halos devoid of central BHs and eventually prone to the occurrence of other BH-formation scenarios. 

We find that $\rm J_{bg}$ is never strong enough to foster the monolithic collapse of pristine gas clouds. Therefore, DCBHs only form in over-dense regions, within halos close to an actively star-forming companion, which is responsible for strong, local variations of LW flux ($\rm J_{LW}$). This provides additional support to the ``synchronized halo-pairs'' scenario, according to which the optimal birthplaces of DCBHs are the found in the proximity of UV-bright, star-forming galaxies whose chemical feedback did not yet enrich their surroundings \citep[e.g.,][]{dijkstra2008,agarwal2013,visbal_haiman_bryan2014b,regan2017}. Regarding RSM, $\sim74\%$ of them form thanks to presence of $\rm J_{bg}$, which is enough to allow for the formation of dense, nuclear stellar clusters in our model. The remaining $26\%$ of RSMs require the presence of $\rm J_{LW}$ spatial variations, analogously to DCBHs. In these cases, the formation of RSMs also happens in the vicinity of luminous neighbors, extending the ``synchronized halo-pairs'' scenario to the formation of intermediate-mass BH seeds.

The formation of miDCBHs is rare, due to the low frequency of gas-rich major mergers over the dynamical range of the \texttt{MR-II}. In particular, we find only two miDCBHs forming at $z\!=\!4.64$ and  $z\!=\!0.35$ as a consequence of the merger between gas-rich ($\rm M_{cGas}/M_{*}\!>\!100$) galaxies hosted in $\rm M_{vir}\!\sim\!3\times10^9M_\odot$ halos. These numbers and formation-redshifts are compatible with the results presented by \cite{bonoli2014} on the merger tree of the \texttt{MR} simulation.

PopIII remnants inserted in  \lgal from \gqd outputs show an evolved mass distribution covering the interval $\rm10^2\!<\!M_{BH}/M_\odot\!\lesssim\!10^5$, and are able to effectively grow up to $\rm M_{BH}\!>\!10^7M_\odot$ already by $z\!\sim\!7$, hence dominating the number densities of BHs at all masses.  This finding does not exclude that intermediate or massive BH-seeds might be needed in order to reach $\rm M_{BH}\!\gtrsim\!10^9M_\odot$, as inferred for the brightest $z\!\gtrsim\!6$ QSOs \citep[see e.g.,][]{mazzucchelli2017,banados2018a,wang2021}.  In order to be conclusive on this point, our BH-seeding model should be applied to significantly larger volumes than the one of the \milltwo\, in order to probe the spatial density regimes of $\rm\sim\! 1\,Gpc^{-3}$. 

On the other hand, 
when evolved down to the local Universe, our model produces a BH mass function (BHMF) which is in good agreement with current constraints, especially at $\rm M_{BH}\!\gtrsim\!10^7M_\odot$. This shows that the integrated mass-growth of our SMBH population, over its whole cosmological evolution, is well-reproduced by the BH-seeding and the mass-growth models we employ. 
However, while our $z\!\sim\!0$ BHMF is in good agreement with local determinations, the mass-assembly  history of our SMBH population shows  discrepancies    with respect to current constraints, as BHs accrete most of their mass at earlier epochs than observed \citep[i.e., $z\!>\!3$ against $1\!\lesssim\!z\!\lesssim\!2$, see, e.g.,][]{ueda2014,miyaji2015,aird2015,shen2020}.
The discordance between our predictions and current high-z observations moderately improves by acting on the abundance of light seeds descendants, i.e., assuming that the efficiency of the grafting from \gqd is lower than in our default model. Interestingly, lowering initial abundance of light seeds produces $>\!1$ dex differences on the amplitude of the BHMF at $\rm M_{BH}\!<\!10^7M_\odot$, supporting the idea that the low-mass end of the local BHMF carries information about the efficiency of high-$z$ seeding processes \citep[e.g.,][]{sesana_volonteri_haardt2007,volonteri_lodato_natarajan2008,miller2015}. This suggests that further analysis is required in order to disentangle the role of BH-seeding efficiency and mass-assembly history of SMBHs on their low-$z$ observables. We plan to undertake this specific study in an upcoming work, by also exploiting the detailed modelling of BH-spin evolution, GW kicks and recoils as well as BH-BH merger delays presented in \cite{izquierdo-villalba2020,izquierdo-villalba2022a}. 

Furthermore, when studying the population of massive BHs hosted in central and satellite galaxies, we find that DCBHs which are able to retain memory of their formation channel down to $z\!=\!0$ are only found in $\rm M_*\!\lesssim\!10^9M_\odot$ satellites. Similarly, $\sim\!50\%$ of RSMs at $z\!=\!0$ are hosted in low-mass satellites, which testifies the quiet evolution of their hosts. On the other hand, the most massive halos at $z\!=\!0$ only host evolved SMBHs descending from light-seeds and mixed seed-classes, testifying the central role of hierarchical assembly in their evolution. 
We also analyze the population of un-evolved seeds. These are hosted by very low-mass systems ($\rm M_*\!\lesssim\!10^8M_\odot$) at $z\!=\!0$, indicating that small dwarf galaxies could harbour traces of high-$z$ BH-seeding processes. We will also study this in more detail in future works.

Finally, we  analyze the predicted $\rm M_{BH} - \rm M_{*}$ scaling relation for  the evolved SMBH population. In the high-mass range ($\rm M_{BH} > 10^{6}-10^{7} M_{\odot}$), our results are consistent with current observational constraints. In the lower mass regime, still hardly accessed by observational probes, we predict a ``flattening'' of the scaling relation. Intermediate and massive seed remnants that have not merged with light seeds are generally under-massive with respect to the global population. Future  observations on intermediate mass black holes, their relation with the host galaxy and their occupation fraction will provide key tests for seeding and growth models such as the one presented in this work.

\section*{Acknowledgements}
The authors greatly thank the anonymous Referee for their comments, which helped to significantly improve the quality of this manuscript. DS acknowledges the support from grant AYA2015-66211-C2-2 of the Ministerio de Economia, Industria y Competitividad (MINECO/FEDER). SB acknowledges support from the project PGC2018-097585-B-C22, of the  MINECO/FEDER, UE of the Spanish Ministerio de Economia, Industria y Competitividad. 
RS and RV acknowledge support from the Amaldi Research Center funded by the MIUR program ``Dipartimento di Eccellenza'' (CUP:B81I18001170001) and from the INFN TEONGRAV specific initiative.  D.I.V. acknowledges financial support provided under the European Union's H2020 ERC Consolidator Grant ``Binary Massive Black Hole Astrophysic'' (B Massive, Grant Agreement: 818691) and from INFN H45J18000450006.

\section*{DATA AVAILABILITY}

The simulated data underlying this article will be shared on reasonable request to the corresponding author. This work used the 2015 public version of the Munich model of galaxy formation and evolution: \lgal. The source code and a full description of the model are available at http://galformod.mpa-garching.mpg.de/public/LGalaxies/.

{\footnotesize
\bibliographystyle{mnras}
\bibliography{references}

\begin{thebibliography}{}
\makeatletter
\relax
\def\mn@urlcharsother{\let\do\@makeother \do\$\do\&\do\#\do\^\do\_\do\%\do\~}
\def\mn@doi{\begingroup\mn@urlcharsother \@ifnextchar [ {\mn@doi@}
  {\mn@doi@[]}}
\def\mn@doi@[#1]#2{\def\@tempa{#1}\ifx\@tempa\@empty \href
  {http://dx.doi.org/#2} {doi:#2}\else \href {http://dx.doi.org/#2} {#1}\fi
  \endgroup}
\def\mn@eprint#1#2{\mn@eprint@#1:#2::\@nil}
\def\mn@eprint@arXiv#1{\href {http://arxiv.org/abs/#1} {{\tt arXiv:#1}}}
\def\mn@eprint@dblp#1{\href {http://dblp.uni-trier.de/rec/bibtex/#1.xml}
  {dblp:#1}}
\def\mn@eprint@#1:#2:#3:#4\@nil{\def\@tempa {#1}\def\@tempb {#2}\def\@tempc
  {#3}\ifx \@tempc \@empty \let \@tempc \@tempb \let \@tempb \@tempa \fi \ifx
  \@tempb \@empty \def\@tempb {arXiv}\fi \@ifundefined
  {mn@eprint@\@tempb}{\@tempb:\@tempc}{\expandafter \expandafter \csname
  mn@eprint@\@tempb\endcsname \expandafter{\@tempc}}}

\bibitem[\protect\citeauthoryear{{Abel} \& {Haiman}}{{Abel} \&
  {Haiman}}{2000}]{abel_haiman2000}
{Abel} T.,  {Haiman} Z.,  2000, in {Combes} F.,  {Pineau Des Forets} G.,  eds,
  Molecular Hydrogen in Space. p.~237 (\mn@eprint {arXiv} {astro-ph/0002031})

\bibitem[\protect\citeauthoryear{{Agarwal}, {Khochfar}, {Johnson}, {Neistein},
  {Dalla Vecchia}  \& {Livio}}{{Agarwal} et~al.}{2012}]{agarwal2012}
{Agarwal} B.,  {Khochfar} S.,  {Johnson} J.~L.,  {Neistein} E.,  {Dalla
  Vecchia} C.,   {Livio} M.,  2012, \mn@doi [\mnras]
  {10.1111/1365-2966.2012.21651}, \href
  {https://ui.adsabs.harvard.edu/abs/2012MNRAS.425.2854A} {425, 2854}

\bibitem[\protect\citeauthoryear{{Agarwal}, {Davis}, {Khochfar}, {Natarajan}
  \& {Dunlop}}{{Agarwal} et~al.}{2013}]{agarwal2013}
{Agarwal} B.,  {Davis} A.~J.,  {Khochfar} S.,  {Natarajan} P.,   {Dunlop}
  J.~S.,  2013, \mn@doi [\mnras] {10.1093/mnras/stt696}, \href
  {https://ui.adsabs.harvard.edu/abs/2013MNRAS.432.3438A} {432, 3438}

\bibitem[\protect\citeauthoryear{{Agarwal}, {Dalla Vecchia}, {Johnson},
  {Khochfar}  \& {Paardekooper}}{{Agarwal} et~al.}{2014}]{agarwal2014}
{Agarwal} B.,  {Dalla Vecchia} C.,  {Johnson} J.~L.,  {Khochfar} S.,
  {Paardekooper} J.-P.,  2014, \mn@doi [\mnras] {10.1093/mnras/stu1112}, \href
  {http://adsabs.harvard.edu/abs/2014MNRAS.443..648A} {443, 648}

\bibitem[\protect\citeauthoryear{{Agarwal}, {Smith}, {Glover}, {Natarajan}  \&
  {Khochfar}}{{Agarwal} et~al.}{2016}]{agarwal2016}
{Agarwal} B.,  {Smith} B.,  {Glover} S.,  {Natarajan} P.,   {Khochfar} S.,
  2016, \mn@doi [\mnras] {10.1093/mnras/stw929}, \href
  {http://adsabs.harvard.edu/abs/2016MNRAS.459.4209A} {459, 4209}

\bibitem[\protect\citeauthoryear{{Agarwal}, {Regan}, {Klessen}, {Downes}  \&
  {Zackrisson}}{{Agarwal} et~al.}{2017}]{agarwal2017}
{Agarwal} B.,  {Regan} J.,  {Klessen} R.~S.,  {Downes} T.~P.,   {Zackrisson}
  E.,  2017, \mn@doi [\mnras] {10.1093/mnras/stx1528}, \href
  {http://adsabs.harvard.edu/abs/2017MNRAS.470.4034A} {470, 4034}

\bibitem[\protect\citeauthoryear{{Agarwal}, {Cullen}, {Khochfar}, {Ceverino}
  \& {Klessen}}{{Agarwal} et~al.}{2019}]{agarwal2019}
{Agarwal} B.,  {Cullen} F.,  {Khochfar} S.,  {Ceverino} D.,   {Klessen} R.~S.,
  2019, \mn@doi [\mnras] {10.1093/mnras/stz1347}, \href
  {https://ui.adsabs.harvard.edu/abs/2019MNRAS.488.3268A} {488, 3268}

\bibitem[\protect\citeauthoryear{{Ahn}, {Shapiro}, {Iliev}, {Mellema}  \&
  {Pen}}{{Ahn} et~al.}{2009}]{ahn2009}
{Ahn} K.,  {Shapiro} P.~R.,  {Iliev} I.~T.,  {Mellema} G.,   {Pen} U.-L.,
  2009, \mn@doi [\apj] {10.1088/0004-637X/695/2/1430}, \href
  {https://ui.adsabs.harvard.edu/abs/2009ApJ...695.1430A} {695, 1430}

\bibitem[\protect\citeauthoryear{{Aird}, {Coil}, {Georgakakis}, {Nandra},
  {Barro}  \& {P{\'e}rez-Gonz{\'a}lez}}{{Aird} et~al.}{2015}]{aird2015}
{Aird} J.,  {Coil} A.~L.,  {Georgakakis} A.,  {Nandra} K.,  {Barro} G.,
  {P{\'e}rez-Gonz{\'a}lez} P.~G.,  2015, \mn@doi [\mnras]
  {10.1093/mnras/stv1062}, \href
  {https://ui.adsabs.harvard.edu/abs/2015MNRAS.451.1892A} {451, 1892}

\bibitem[\protect\citeauthoryear{{Alexander} \& {Hickox}}{{Alexander} \&
  {Hickox}}{2012}]{alexander_hickox2012}
{Alexander} D.~M.,  {Hickox} R.~C.,  2012, \mn@doi [\nar]
  {10.1016/j.newar.2011.11.003}, \href
  {https://ui.adsabs.harvard.edu/abs/2012NewAR..56...93A} {56, 93}

\bibitem[\protect\citeauthoryear{{Ardaneh}, {Luo}, {Shlosman}, {Nagamine},
  {Wise}  \& {Begelman}}{{Ardaneh} et~al.}{2018}]{ardaneh2018}
{Ardaneh} K.,  {Luo} Y.,  {Shlosman} I.,  {Nagamine} K.,  {Wise} J.~H.,
  {Begelman} M.~C.,  2018, \mn@doi [\mnras] {10.1093/mnras/sty1657}, \href
  {https://ui.adsabs.harvard.edu/abs/2018MNRAS.479.2277A} {479, 2277}

\bibitem[\protect\citeauthoryear{{Askar}, {Davies}  \& {Church}}{{Askar}
  et~al.}{2021}]{askar_davies_church2021b}
{Askar} A.,  {Davies} M.~B.,   {Church} R.~P.,  2021, arXiv e-prints, \href
  {https://ui.adsabs.harvard.edu/abs/2021arXiv210710862A} {p. arXiv:2107.10862}

\bibitem[\protect\citeauthoryear{{Ba{\~n}ados} et~al.,}{{Ba{\~n}ados}
  et~al.}{2018}]{banados2018a}
{Ba{\~n}ados} E.,  et~al., 2018, \mn@doi [\nat] {10.1038/nature25180}, \href
  {https://ui.adsabs.harvard.edu/abs/2018Natur.553..473B} {553, 473}

\bibitem[\protect\citeauthoryear{{Barkana} \& {Loeb}}{{Barkana} \&
  {Loeb}}{2001}]{barkana_loeb2001}
{Barkana} R.,  {Loeb} A.,  2001, \mn@doi [\physrep]
  {10.1016/S0370-1573(01)00019-9}, \href
  {https://ui.adsabs.harvard.edu/abs/2001PhR...349..125B} {349, 125}

\bibitem[\protect\citeauthoryear{{Bertone}, {Stoehr}  \& {White}}{{Bertone}
  et~al.}{2005}]{bertone2005}
{Bertone} S.,  {Stoehr} F.,   {White} S.~D.~M.,  2005, \mn@doi [\mnras]
  {10.1111/j.1365-2966.2005.08772.x}, \href
  {https://ui.adsabs.harvard.edu/abs/2005MNRAS.359.1201B} {359, 1201}

\bibitem[\protect\citeauthoryear{{Bertone}, {De Lucia}  \& {Thomas}}{{Bertone}
  et~al.}{2007}]{bertone2007}
{Bertone} S.,  {De Lucia} G.,   {Thomas} P.~A.,  2007, \mn@doi [\mnras]
  {10.1111/j.1365-2966.2007.11997.x}, \href
  {https://ui.adsabs.harvard.edu/abs/2007MNRAS.379.1143B} {379, 1143}

\bibitem[\protect\citeauthoryear{{Bhowmick} et~al.,}{{Bhowmick}
  et~al.}{2021}]{bhowmick2021}
{Bhowmick} A.~K.,  et~al., 2021, \mn@doi [\mnras] {10.1093/mnras/stab2204},
  \href {https://ui.adsabs.harvard.edu/abs/2021MNRAS.507.2012B} {507, 2012}

\bibitem[\protect\citeauthoryear{{Bodenheimer}}{{Bodenheimer}}{2011}]{bodenheimer2011}
{Bodenheimer} P.~H.,  2011, {Principles of Star Formation}.
Springer, \mn@doi{10.1007/978-3-642-15063-0}

\bibitem[\protect\citeauthoryear{{Bonoli}, {Marulli}, {Springel}, {White},
  {Branchini}  \& {Moscardini}}{{Bonoli} et~al.}{2009}]{bonoli2009}
{Bonoli} S.,  {Marulli} F.,  {Springel} V.,  {White} S. D.~M.,  {Branchini} E.,
    {Moscardini} L.,  2009, \mn@doi [\mnras]
  {10.1111/j.1365-2966.2009.14701.x}, \href
  {https://ui.adsabs.harvard.edu/abs/2009MNRAS.396..423B} {396, 423}

\bibitem[\protect\citeauthoryear{{Bonoli}, {Mayer}  \& {Callegari}}{{Bonoli}
  et~al.}{2014}]{bonoli2014}
{Bonoli} S.,  {Mayer} L.,   {Callegari} S.,  2014, \mn@doi [\mnras]
  {10.1093/mnras/stt1990}, \href
  {http://adsabs.harvard.edu/abs/2014MNRAS.437.1576B} {437, 1576}

\bibitem[\protect\citeauthoryear{{Boylan-Kolchin}, {Springel}, {White},
  {Jenkins}  \& {Lemson}}{{Boylan-Kolchin} et~al.}{2009}]{boylankolchin2009}
{Boylan-Kolchin} M.,  {Springel} V.,  {White} S. D.~M.,  {Jenkins} A.,
  {Lemson} G.,  2009, \mn@doi [\mnras] {10.1111/j.1365-2966.2009.15191.x},
  \href {https://ui.adsabs.harvard.edu/abs/2009MNRAS.398.1150B} {398, 1150}

\bibitem[\protect\citeauthoryear{{Bromm} \& {Larson}}{{Bromm} \&
  {Larson}}{2004}]{bromm_larson2004}
{Bromm} V.,  {Larson} R.~B.,  2004, \mn@doi [\araa]
  {10.1146/annurev.astro.42.053102.134034}, \href
  {https://ui.adsabs.harvard.edu/abs/2004ARA&A..42...79B} {42, 79}

\bibitem[\protect\citeauthoryear{{Bromm} \& {Loeb}}{{Bromm} \&
  {Loeb}}{2003}]{bromm_loeb2003}
{Bromm} V.,  {Loeb} A.,  2003, \mn@doi [\apj] {10.1086/377529}, \href
  {http://adsabs.harvard.edu/abs/2003ApJ...596...34B} {596, 34}

\bibitem[\protect\citeauthoryear{{Buchner}, {Treister}, {Bauer}, {Sartori}  \&
  {Schawinski}}{{Buchner} et~al.}{2019}]{buchner2019}
{Buchner} J.,  {Treister} E.,  {Bauer} F.~E.,  {Sartori} L.~F.,   {Schawinski}
  K.,  2019, \mn@doi [\apj] {10.3847/1538-4357/aafd32}, \href
  {https://ui.adsabs.harvard.edu/abs/2019ApJ...874..117B} {874, 117}

\bibitem[\protect\citeauthoryear{{Cann} et~al.,}{{Cann}
  et~al.}{2021}]{cann2021}
{Cann} J.~M.,  et~al., 2021, \mn@doi [\apjl] {10.3847/2041-8213/abf56d}, \href
  {https://ui.adsabs.harvard.edu/abs/2021ApJ...912L...2C} {912, L2}

\bibitem[\protect\citeauthoryear{{Capuzzo-Dolcetta} \& {Tosta e
  Melo}}{{Capuzzo-Dolcetta} \& {Tosta e
  Melo}}{2017}]{capuzzo-dolcetta_tosta-e-melo2017}
{Capuzzo-Dolcetta} R.,  {Tosta e Melo} I.,  2017, \mn@doi [\mnras]
  {10.1093/mnras/stx2246}, \href
  {https://ui.adsabs.harvard.edu/abs/2017MNRAS.472.4013C} {472, 4013}

\bibitem[\protect\citeauthoryear{{Chon} \& {Omukai}}{{Chon} \&
  {Omukai}}{2020}]{chon_omukai2020}
{Chon} S.,  {Omukai} K.,  2020, \mn@doi [\mnras] {10.1093/mnras/staa863}, \href
  {https://ui.adsabs.harvard.edu/abs/2020MNRAS.494.2851C} {494, 2851}

\bibitem[\protect\citeauthoryear{{Chon}, {Hirano}, {Hosokawa}  \&
  {Yoshida}}{{Chon} et~al.}{2016}]{chon2016}
{Chon} S.,  {Hirano} S.,  {Hosokawa} T.,   {Yoshida} N.,  2016, \mn@doi [\apj]
  {10.3847/0004-637X/832/2/134}, \href
  {https://ui.adsabs.harvard.edu/abs/2016ApJ...832..134C} {832, 134}

\bibitem[\protect\citeauthoryear{{Chon}, {Hosokawa}  \& {Yoshida}}{{Chon}
  et~al.}{2018}]{chon_hosokawa_yoshida2018}
{Chon} S.,  {Hosokawa} T.,   {Yoshida} N.,  2018, \mn@doi [\mnras]
  {10.1093/mnras/sty086}, \href
  {https://ui.adsabs.harvard.edu/abs/2018MNRAS.475.4104C} {475, 4104}

\bibitem[\protect\citeauthoryear{{Cora} et~al.,}{{Cora}
  et~al.}{2018}]{cora2018}
{Cora} S.~A.,  et~al., 2018, \mn@doi [\mnras] {10.1093/mnras/sty1131}, \href
  {https://ui.adsabs.harvard.edu/abs/2018MNRAS.479....2C} {479, 2}

\bibitem[\protect\citeauthoryear{{Croton} et~al.,}{{Croton}
  et~al.}{2006}]{croton2006}
{Croton} D.~J.,  et~al., 2006, \mn@doi [\mnras]
  {10.1111/j.1365-2966.2005.09675.x}, \href
  {https://ui.adsabs.harvard.edu/abs/2006MNRAS.365...11C} {365, 11}

\bibitem[\protect\citeauthoryear{{Das}, {Schleicher}, {Leigh}  \&
  {Boekholt}}{{Das} et~al.}{2021}]{das2021a}
{Das} A.,  {Schleicher} D. R.~G.,  {Leigh} N. W.~C.,   {Boekholt} T. C.~N.,
  2021, \mn@doi [\mnras] {10.1093/mnras/stab402}, \href
  {https://ui.adsabs.harvard.edu/abs/2021MNRAS.503.1051D} {503, 1051}

\bibitem[\protect\citeauthoryear{{De Bennassuti}, {Salvadori}, {Schneider},
  {Valiante}  \& {Omukai}}{{De Bennassuti} et~al.}{2017}]{debennassuti2017}
{De Bennassuti} M.,  {Salvadori} S.,  {Schneider} R.,  {Valiante} R.,
  {Omukai} K.,  2017, \mn@doi [\mnras] {10.1093/mnras/stw2687}, \href
  {https://ui.adsabs.harvard.edu/abs/2017MNRAS.465..926D} {465, 926}

\bibitem[\protect\citeauthoryear{{De Lucia}, {Tornatore}, {Frenk}, {Helmi},
  {Navarro}  \& {White}}{{De Lucia} et~al.}{2014}]{delucia2014}
{De Lucia} G.,  {Tornatore} L.,  {Frenk} C.~S.,  {Helmi} A.,  {Navarro} J.~F.,
   {White} S. D.~M.,  2014, \mn@doi [\mnras] {10.1093/mnras/stu1752}, \href
  {https://ui.adsabs.harvard.edu/abs/2014MNRAS.445..970D} {445, 970}

\bibitem[\protect\citeauthoryear{{DeGraf} \& {Sijacki}}{{DeGraf} \&
  {Sijacki}}{2019}]{degraf_sijacki2019}
{DeGraf} C.,  {Sijacki} D.,  2019, arXiv e-prints, \href
  {https://ui.adsabs.harvard.edu/abs/2019arXiv190611271D} {p. arXiv:1906.11271}

\bibitem[\protect\citeauthoryear{{Delvecchio} et~al.,}{{Delvecchio}
  et~al.}{2020}]{delvecchio2020}
{Delvecchio} I.,  et~al., 2020, \mn@doi [\apj] {10.3847/1538-4357/ab789c},
  \href {https://ui.adsabs.harvard.edu/abs/2020ApJ...892...17D} {892, 17}

\bibitem[\protect\citeauthoryear{{Devecchi} \& {Volonteri}}{{Devecchi} \&
  {Volonteri}}{2009}]{devecchi_volonteri2009}
{Devecchi} B.,  {Volonteri} M.,  2009, \mn@doi [\apj]
  {10.1088/0004-637X/694/1/302}, \href
  {https://ui.adsabs.harvard.edu/abs/2009ApJ...694..302D} {694, 302}

\bibitem[\protect\citeauthoryear{{Di Matteo}, {Croft}, {Feng}, {Waters}  \&
  {Wilkins}}{{Di Matteo} et~al.}{2017}]{dimatteo2017}
{Di Matteo} T.,  {Croft} R. A.~C.,  {Feng} Y.,  {Waters} D.,   {Wilkins} S.,
  2017, \mn@doi [\mnras] {10.1093/mnras/stx319}, \href
  {https://ui.adsabs.harvard.edu/abs/2017MNRAS.467.4243D} {467, 4243}

\bibitem[\protect\citeauthoryear{{Dijkstra}, {Haiman}, {Mesinger}  \&
  {Wyithe}}{{Dijkstra} et~al.}{2008}]{dijkstra2008}
{Dijkstra} M.,  {Haiman} Z.,  {Mesinger} A.,   {Wyithe} J. S.~B.,  2008,
  \mn@doi [\mnras] {10.1111/j.1365-2966.2008.14031.x}, \href
  {https://ui.adsabs.harvard.edu/abs/2008MNRAS.391.1961D} {391, 1961}

\bibitem[\protect\citeauthoryear{{Dijkstra}, {Ferrara}  \&
  {Mesinger}}{{Dijkstra} et~al.}{2014}]{dijkstra2014}
{Dijkstra} M.,  {Ferrara} A.,   {Mesinger} A.,  2014, \mn@doi [\mnras]
  {10.1093/mnras/stu1007}, \href
  {http://adsabs.harvard.edu/abs/2014MNRAS.442.2036D} {442, 2036}

\bibitem[\protect\citeauthoryear{{Dubois} et~al.,}{{Dubois}
  et~al.}{2014}]{dubois2014a}
{Dubois} Y.,  et~al., 2014, \mn@doi [\mnras] {10.1093/mnras/stu1227}, \href
  {https://ui.adsabs.harvard.edu/abs/2014MNRAS.444.1453D} {444, 1453}

\bibitem[\protect\citeauthoryear{{Dunn}, {Bellovary}, {Holley-Bockelmann},
  {Christensen}  \& {Quinn}}{{Dunn} et~al.}{2018}]{dunn2018}
{Dunn} G.,  {Bellovary} J.,  {Holley-Bockelmann} K.,  {Christensen} C.,
  {Quinn} T.,  2018, \mn@doi [\apj] {10.3847/1538-4357/aac7c2}, \href
  {https://ui.adsabs.harvard.edu/abs/2018ApJ...861...39D} {861, 39}

\bibitem[\protect\citeauthoryear{{Dunn}, {Holley-Bockelmann}  \&
  {Bellovary}}{{Dunn} et~al.}{2020}]{dunn2020}
{Dunn} G.,  {Holley-Bockelmann} K.,   {Bellovary} J.,  2020, \mn@doi [\apj]
  {10.3847/1538-4357/ab7cd2}, \href
  {https://ui.adsabs.harvard.edu/abs/2020ApJ...896...72D} {896, 72}

\bibitem[\protect\citeauthoryear{{Dutton} \& {Macci{\`o}}}{{Dutton} \&
  {Macci{\`o}}}{2014}]{dutton_maccio2014}
{Dutton} A.~A.,  {Macci{\`o}} A.~V.,  2014, \mn@doi [\mnras]
  {10.1093/mnras/stu742}, \href
  {https://ui.adsabs.harvard.edu/abs/2014MNRAS.441.3359D} {441, 3359}

\bibitem[\protect\citeauthoryear{{Ebisuzaki} et~al.,}{{Ebisuzaki}
  et~al.}{2001}]{ebisuzaki2001}
{Ebisuzaki} T.,  et~al., 2001, \mn@doi [\apjl] {10.1086/338118}, \href
  {https://ui.adsabs.harvard.edu/abs/2001ApJ...562L..19E} {562, L19}

\bibitem[\protect\citeauthoryear{{Erwin} \& {Gadotti}}{{Erwin} \&
  {Gadotti}}{2012}]{erwin_gadotti2012}
{Erwin} P.,  {Gadotti} D.~A.,  2012, \mn@doi [Advances in Astronomy]
  {10.1155/2012/946368}, \href
  {https://ui.adsabs.harvard.edu/abs/2012AdAst2012E...4E} {2012, 946368}

\bibitem[\protect\citeauthoryear{{Escala}}{{Escala}}{2021}]{escala2021}
{Escala} A.,  2021, \mn@doi [\apj] {10.3847/1538-4357/abd93c}, \href
  {https://ui.adsabs.harvard.edu/abs/2021ApJ...908...57E} {908, 57}

\bibitem[\protect\citeauthoryear{{Fabian}}{{Fabian}}{2012}]{fabian2012}
{Fabian} A.~C.,  2012, \mn@doi [\araa] {10.1146/annurev-astro-081811-125521},
  \href {https://ui.adsabs.harvard.edu/abs/2012ARA&A..50..455F} {50, 455}

\bibitem[\protect\citeauthoryear{{Fanidakis}, {Baugh}, {Benson}, {Bower},
  {Cole}, {Done}  \& {Frenk}}{{Fanidakis} et~al.}{2011}]{fanidakis2011}
{Fanidakis} N.,  {Baugh} C.~M.,  {Benson} A.~J.,  {Bower} R.~G.,  {Cole} S.,
  {Done} C.,   {Frenk} C.~S.,  2011, \mn@doi [\mnras]
  {10.1111/j.1365-2966.2010.17427.x}, \href
  {https://ui.adsabs.harvard.edu/abs/2011MNRAS.410...53F} {410, 53}

\bibitem[\protect\citeauthoryear{{Fernandez}, {Bryan}, {Haiman}  \&
  {Li}}{{Fernandez} et~al.}{2014}]{fernandez2014}
{Fernandez} R.,  {Bryan} G.~L.,  {Haiman} Z.,   {Li} M.,  2014, \mn@doi
  [\mnras] {10.1093/mnras/stu230}, \href
  {https://ui.adsabs.harvard.edu/abs/2014MNRAS.439.3798F} {439, 3798}

\bibitem[\protect\citeauthoryear{{Ferr{\'e}-Mateu}, {Mezcua}  \&
  {Barrows}}{{Ferr{\'e}-Mateu} et~al.}{2021}]{ferre-mateu2021}
{Ferr{\'e}-Mateu} A.,  {Mezcua} M.,   {Barrows} R.~S.,  2021, \mn@doi [\mnras]
  {10.1093/mnras/stab1915}, \href
  {https://ui.adsabs.harvard.edu/abs/2021MNRAS.506.4702F} {506, 4702}

\bibitem[\protect\citeauthoryear{{Fielding}, {Quataert}  \&
  {Martizzi}}{{Fielding} et~al.}{2018}]{fielding2018}
{Fielding} D.,  {Quataert} E.,   {Martizzi} D.,  2018, \mn@doi [\mnras]
  {10.1093/mnras/sty2466}, \href
  {https://ui.adsabs.harvard.edu/abs/2018MNRAS.481.3325F} {481, 3325}

\bibitem[\protect\citeauthoryear{{Gallo}, {Treu}, {Jacob}, {Woo}, {Marshall}
  \& {Antonucci}}{{Gallo} et~al.}{2008}]{gallo2008}
{Gallo} E.,  {Treu} T.,  {Jacob} J.,  {Woo} J.-H.,  {Marshall} P.~J.,
  {Antonucci} R.,  2008, \mn@doi [\apj] {10.1086/588012}, \href
  {https://ui.adsabs.harvard.edu/abs/2008ApJ...680..154G} {680, 154}

\bibitem[\protect\citeauthoryear{{Glebbeek}, {Gaburov}, {de Mink}, {Pols}  \&
  {Portegies Zwart}}{{Glebbeek} et~al.}{2009}]{glebbeek2009}
{Glebbeek} E.,  {Gaburov} E.,  {de Mink} S.~E.,  {Pols} O.~R.,   {Portegies
  Zwart} S.~F.,  2009, \mn@doi [\aap] {10.1051/0004-6361/200810425}, \href
  {https://ui.adsabs.harvard.edu/abs/2009A&A...497..255G} {497, 255}

\bibitem[\protect\citeauthoryear{{Greene}}{{Greene}}{2012}]{greene2012}
{Greene} J.~E.,  2012, \mn@doi [Nature Communications] {10.1038/ncomms2314},
  \href {https://ui.adsabs.harvard.edu/abs/2012NatCo...3.1304G} {3, 1304}

\bibitem[\protect\citeauthoryear{{Greif} \& {Bromm}}{{Greif} \&
  {Bromm}}{2006}]{greif_bromm2006}
{Greif} T.~H.,  {Bromm} V.,  2006, \mn@doi [\mnras]
  {10.1111/j.1365-2966.2006.11017.x}, \href
  {https://ui.adsabs.harvard.edu/abs/2006MNRAS.373..128G} {373, 128}

\bibitem[\protect\citeauthoryear{{Greif}, {Springel}, {White}, {Glover},
  {Clark}, {Smith}, {Klessen}  \& {Bromm}}{{Greif} et~al.}{2011}]{greif2011}
{Greif} T.~H.,  {Springel} V.,  {White} S. D.~M.,  {Glover} S. C.~O.,  {Clark}
  P.~C.,  {Smith} R.~J.,  {Klessen} R.~S.,   {Bromm} V.,  2011, \mn@doi [\apj]
  {10.1088/0004-637X/737/2/75}, \href
  {https://ui.adsabs.harvard.edu/abs/2011ApJ...737...75G} {737, 75}

\bibitem[\protect\citeauthoryear{{Guo} et~al.,}{{Guo} et~al.}{2011}]{guo2011}
{Guo} Q.,  et~al., 2011, \mn@doi [\mnras] {10.1111/j.1365-2966.2010.18114.x},
  \href {http://adsabs.harvard.edu/abs/2011MNRAS.413..101G} {413, 101}

\bibitem[\protect\citeauthoryear{{Habouzit}, {Volonteri}, {Latif}, {Dubois}  \&
  {Peirani}}{{Habouzit} et~al.}{2016}]{habouzit2016}
{Habouzit} M.,  {Volonteri} M.,  {Latif} M.,  {Dubois} Y.,   {Peirani} S.,
  2016, \mn@doi [\mnras] {10.1093/mnras/stw1924}, \href
  {https://ui.adsabs.harvard.edu/abs/2016MNRAS.463..529H} {463, 529}

\bibitem[\protect\citeauthoryear{{Habouzit}, {Volonteri}  \&
  {Dubois}}{{Habouzit} et~al.}{2017}]{habouzit_volonteri_dubois2017}
{Habouzit} M.,  {Volonteri} M.,   {Dubois} Y.,  2017, \mn@doi [\mnras]
  {10.1093/mnras/stx666}, \href
  {https://ui.adsabs.harvard.edu/abs/2017MNRAS.468.3935H} {468, 3935}

\bibitem[\protect\citeauthoryear{{Habouzit} et~al.,}{{Habouzit}
  et~al.}{2021}]{habouzit2021}
{Habouzit} M.,  et~al., 2021, \mn@doi [\mnras] {10.1093/mnras/stab496}, \href
  {https://ui.adsabs.harvard.edu/abs/2021MNRAS.503.1940H} {503, 1940}

\bibitem[\protect\citeauthoryear{{Haemmerl{\'e}}, {Woods}, {Klessen}, {Heger}
  \& {Whalen}}{{Haemmerl{\'e}} et~al.}{2018}]{haemmerle2018}
{Haemmerl{\'e}} L.,  {Woods} T.~E.,  {Klessen} R.~S.,  {Heger} A.,   {Whalen}
  D.~J.,  2018, \mn@doi [\mnras] {10.1093/mnras/stx2919}, \href
  {https://ui.adsabs.harvard.edu/abs/2018MNRAS.474.2757H} {474, 2757}

\bibitem[\protect\citeauthoryear{{Haemmerl{\'e}}, {Klessen}, {Mayer}  \&
  {Zwick}}{{Haemmerl{\'e}} et~al.}{2021}]{haemmerle2021}
{Haemmerl{\'e}} L.,  {Klessen} R.~S.,  {Mayer} L.,   {Zwick} L.,  2021, \mn@doi
  [\aap] {10.1051/0004-6361/202141376}, \href
  {https://ui.adsabs.harvard.edu/abs/2021A&A...652L...7H} {652, L7}

\bibitem[\protect\citeauthoryear{{Hartwig}, {Glover}, {Klessen}, {Latif}  \&
  {Volonteri}}{{Hartwig} et~al.}{2015}]{hartwig2015c}
{Hartwig} T.,  {Glover} S. C.~O.,  {Klessen} R.~S.,  {Latif} M.~A.,
  {Volonteri} M.,  2015, \mn@doi [\mnras] {10.1093/mnras/stv1368}, \href
  {https://ui.adsabs.harvard.edu/abs/2015MNRAS.452.1233H} {452, 1233}

\bibitem[\protect\citeauthoryear{{Heckman} \& {Best}}{{Heckman} \&
  {Best}}{2014}]{heckman_best2014}
{Heckman} T.~M.,  {Best} P.~N.,  2014, \mn@doi [\araa]
  {10.1146/annurev-astro-081913-035722}, \href
  {https://ui.adsabs.harvard.edu/abs/2014ARA&A..52..589H} {52, 589}

\bibitem[\protect\citeauthoryear{{Henriques}, {White}, {Thomas}, {Angulo},
  {Guo}, {Lemson}, {Springel}  \& {Overzier}}{{Henriques}
  et~al.}{2015}]{henriques2015}
{Henriques} B.~M.~B.,  {White} S.~D.~M.,  {Thomas} P.~A.,  {Angulo} R.,  {Guo}
  Q.,  {Lemson} G.,  {Springel} V.,   {Overzier} R.,  2015, \mn@doi [\mnras]
  {10.1093/mnras/stv705}, \href
  {http://adsabs.harvard.edu/abs/2015MNRAS.451.2663H} {451, 2663}

\bibitem[\protect\citeauthoryear{{Holzbauer} \& {Furlanetto}}{{Holzbauer} \&
  {Furlanetto}}{2012}]{holzbauer_furlanetto2012}
{Holzbauer} L.~N.,  {Furlanetto} S.~R.,  2012, \mn@doi [\mnras]
  {10.1111/j.1365-2966.2011.19752.x}, \href
  {https://ui.adsabs.harvard.edu/abs/2012MNRAS.419..718H} {419, 718}

\bibitem[\protect\citeauthoryear{{Inayoshi}, {Visbal}  \&
  {Kashiyama}}{{Inayoshi} et~al.}{2015}]{inayoshi_visbal_kashiyama2015}
{Inayoshi} K.,  {Visbal} E.,   {Kashiyama} K.,  2015, \mn@doi [\mnras]
  {10.1093/mnras/stv1654}, \href
  {https://ui.adsabs.harvard.edu/abs/2015MNRAS.453.1692I} {453, 1692}

\bibitem[\protect\citeauthoryear{{Inayoshi}, {Visbal}  \& {Haiman}}{{Inayoshi}
  et~al.}{2020}]{inayoshi_visbal_haiman2020}
{Inayoshi} K.,  {Visbal} E.,   {Haiman} Z.,  2020, \mn@doi [\araa]
  {10.1146/annurev-astro-120419-014455}, \href
  {https://ui.adsabs.harvard.edu/abs/2020ARA&A..58...27I} {58, 27}

\bibitem[\protect\citeauthoryear{{Izquierdo-Villalba}, {Bonoli}, {Spinoso},
  {Rosas-Guevara}, {Henriques}  \&
  {Hern{\'a}ndez-Monteagudo}}{{Izquierdo-Villalba}
  et~al.}{2019}]{izquierdo-villalba2019}
{Izquierdo-Villalba} D.,  {Bonoli} S.,  {Spinoso} D.,  {Rosas-Guevara} Y.,
  {Henriques} B. M.~B.,   {Hern{\'a}ndez-Monteagudo} C.,  2019, \mn@doi
  [\mnras] {10.1093/mnras/stz1694}, \href
  {https://ui.adsabs.harvard.edu/abs/2019MNRAS.488..609I} {488, 609}

\bibitem[\protect\citeauthoryear{{Izquierdo-Villalba}, {Bonoli}, {Dotti},
  {Sesana}, {Rosas-Guevara}  \& {Spinoso}}{{Izquierdo-Villalba}
  et~al.}{2020}]{izquierdo-villalba2020}
{Izquierdo-Villalba} D.,  {Bonoli} S.,  {Dotti} M.,  {Sesana} A.,
  {Rosas-Guevara} Y.,   {Spinoso} D.,  2020, \mn@doi [\mnras]
  {10.1093/mnras/staa1399}, \href
  {https://ui.adsabs.harvard.edu/abs/2020MNRAS.495.4681I} {495, 4681}

\bibitem[\protect\citeauthoryear{{Izquierdo-Villalba}, {Sesana}, {Bonoli}  \&
  {Colpi}}{{Izquierdo-Villalba} et~al.}{2022}]{izquierdo-villalba2022a}
{Izquierdo-Villalba} D.,  {Sesana} A.,  {Bonoli} S.,   {Colpi} M.,  2022,
  \mn@doi [\mnras] {10.1093/mnras/stab3239}, \href
  {https://ui.adsabs.harvard.edu/abs/2022MNRAS.509.3488I} {509, 3488}

\bibitem[\protect\citeauthoryear{{Katz}, {Sijacki}  \& {Haehnelt}}{{Katz}
  et~al.}{2015}]{katz_sijacki_haehnelt2015}
{Katz} H.,  {Sijacki} D.,   {Haehnelt} M.~G.,  2015, \mn@doi [\mnras]
  {10.1093/mnras/stv1048}, \href
  {https://ui.adsabs.harvard.edu/abs/2015MNRAS.451.2352K} {451, 2352}

\bibitem[\protect\citeauthoryear{{Kauffmann} \& {Haehnelt}}{{Kauffmann} \&
  {Haehnelt}}{2000}]{kauffmann_haenelt2000}
{Kauffmann} G.,  {Haehnelt} M.,  2000, \mn@doi [\mnras]
  {10.1046/j.1365-8711.2000.03077.x}, \href
  {https://ui.adsabs.harvard.edu/abs/2000MNRAS.311..576K} {311, 576}

\bibitem[\protect\citeauthoryear{{Kelly} \& {Shen}}{{Kelly} \&
  {Shen}}{2013}]{kelly_shen2013}
{Kelly} B.~C.,  {Shen} Y.,  2013, \mn@doi [\apj] {10.1088/0004-637X/764/1/45},
  \href {https://ui.adsabs.harvard.edu/abs/2013ApJ...764...45K} {764, 45}

\bibitem[\protect\citeauthoryear{{Kim}, {Ostriker}  \& {Raileanu}}{{Kim}
  et~al.}{2017}]{kim_ostriker_railenau2017}
{Kim} C.-G.,  {Ostriker} E.~C.,   {Raileanu} R.,  2017, \mn@doi [\apj]
  {10.3847/1538-4357/834/1/25}, \href
  {https://ui.adsabs.harvard.edu/abs/2017ApJ...834...25K} {834, 25}

\bibitem[\protect\citeauthoryear{{Kormendy} \& {Ho}}{{Kormendy} \&
  {Ho}}{2013}]{kormendy_ho2013}
{Kormendy} J.,  {Ho} L.~C.,  2013, \mn@doi [\araa]
  {10.1146/annurev-astro-082708-101811}, \href
  {https://ui.adsabs.harvard.edu/abs/2013ARA&A..51..511K} {51, 511}

\bibitem[\protect\citeauthoryear{{Kristensen}, {Pimbblet}, {Gibson}, {Penny}
  \& {Koudmani}}{{Kristensen} et~al.}{2021}]{kristensen2021}
{Kristensen} M.~T.,  {Pimbblet} K.~A.,  {Gibson} B.~K.,  {Penny} S.~J.,
  {Koudmani} S.,  2021, \mn@doi [\apj] {10.3847/1538-4357/ac236d}, \href
  {https://ui.adsabs.harvard.edu/abs/2021ApJ...922..127K} {922, 127}

\bibitem[\protect\citeauthoryear{{Kroupa}, {Subr}, {Jerabkova}  \&
  {Wang}}{{Kroupa} et~al.}{2020}]{kroupa2020}
{Kroupa} P.,  {Subr} L.,  {Jerabkova} T.,   {Wang} L.,  2020, \mn@doi [\mnras]
  {10.1093/mnras/staa2276}, \href
  {https://ui.adsabs.harvard.edu/abs/2020MNRAS.498.5652K} {498, 5652}

\bibitem[\protect\citeauthoryear{{Lacey} \& {Cole}}{{Lacey} \&
  {Cole}}{1993}]{lacey_cole1993}
{Lacey} C.,  {Cole} S.,  1993, \mn@doi [\mnras] {10.1093/mnras/262.3.627},
  \href {https://ui.adsabs.harvard.edu/abs/1993MNRAS.262..627L} {262, 627}

\bibitem[\protect\citeauthoryear{{Lacey} et~al.,}{{Lacey}
  et~al.}{2016}]{lacey2016}
{Lacey} C.~G.,  et~al., 2016, \mn@doi [\mnras] {10.1093/mnras/stw1888}, \href
  {https://ui.adsabs.harvard.edu/abs/2016MNRAS.462.3854L} {462, 3854}

\bibitem[\protect\citeauthoryear{{Lagos}, {Tobar}, {Robotham}, {Obreschkow},
  {Mitchell}, {Power}  \& {Elahi}}{{Lagos} et~al.}{2018}]{lagos2018}
{Lagos} C. d.~P.,  {Tobar} R.~J.,  {Robotham} A. S.~G.,  {Obreschkow} D.,
  {Mitchell} P.~D.,  {Power} C.,   {Elahi} P.~J.,  2018, \mn@doi [\mnras]
  {10.1093/mnras/sty2440}, \href
  {https://ui.adsabs.harvard.edu/abs/2018MNRAS.481.3573L} {481, 3573}

\bibitem[\protect\citeauthoryear{{Latif} \& {Ferrara}}{{Latif} \&
  {Ferrara}}{2016}]{latif_ferrara2016}
{Latif} M.~A.,  {Ferrara} A.,  2016, \mn@doi [\pasa] {10.1017/pasa.2016.41},
  \href {https://ui.adsabs.harvard.edu/abs/2016PASA...33...51L} {33, e051}

\bibitem[\protect\citeauthoryear{{Latif} \& {Khochfar}}{{Latif} \&
  {Khochfar}}{2019}]{latif_khochfar2019}
{Latif} M.~A.,  {Khochfar} S.,  2019, \mn@doi [\mnras] {10.1093/mnras/stz2812},
  \href {https://ui.adsabs.harvard.edu/abs/2019MNRAS.490.2706L} {490, 2706}

\bibitem[\protect\citeauthoryear{{Latif} \& {Volonteri}}{{Latif} \&
  {Volonteri}}{2015}]{latif_volonteri2015}
{Latif} M.~A.,  {Volonteri} M.,  2015, \mn@doi [\mnras]
  {10.1093/mnras/stv1337}, \href
  {https://ui.adsabs.harvard.edu/abs/2015MNRAS.452.1026L} {452, 1026}

\bibitem[\protect\citeauthoryear{{Latif}, {Bovino}, {Van Borm}, {Grassi},
  {Schleicher}  \& {Spaans}}{{Latif} et~al.}{2014a}]{latif2014a}
{Latif} M.~A.,  {Bovino} S.,  {Van Borm} C.,  {Grassi} T.,  {Schleicher}
  D.~R.~G.,   {Spaans} M.,  2014a, \mn@doi [\mnras] {10.1093/mnras/stu1230},
  \href {https://ui.adsabs.harvard.edu/abs/2014MNRAS.443.1979L} {443, 1979}

\bibitem[\protect\citeauthoryear{{Latif}, {Schleicher}, {Bovino}, {Grassi}  \&
  {Spaans}}{{Latif} et~al.}{2014b}]{latif2014b}
{Latif} M.~A.,  {Schleicher} D.~R.~G.,  {Bovino} S.,  {Grassi} T.,   {Spaans}
  M.,  2014b, \mn@doi [\apj] {10.1088/0004-637X/792/1/78}, \href
  {https://ui.adsabs.harvard.edu/abs/2014ApJ...792...78L} {792, 78}

\bibitem[\protect\citeauthoryear{{Latif}, {Bovino}, {Grassi}, {Schleicher}  \&
  {Spaans}}{{Latif} et~al.}{2015}]{latif2015}
{Latif} M.~A.,  {Bovino} S.,  {Grassi} T.,  {Schleicher} D.~R.~G.,   {Spaans}
  M.,  2015, \mn@doi [\mnras] {10.1093/mnras/stu2244}, \href
  {https://ui.adsabs.harvard.edu/abs/2015MNRAS.446.3163L} {446, 3163}

\bibitem[\protect\citeauthoryear{{Latif}, {Volonteri}  \& {Wise}}{{Latif}
  et~al.}{2018}]{latif_volonteri_wise2018}
{Latif} M.~A.,  {Volonteri} M.,   {Wise} J.~H.,  2018, \mn@doi [\mnras]
  {10.1093/mnras/sty622}, \href
  {https://ui.adsabs.harvard.edu/abs/2018MNRAS.476.5016L} {476, 5016}

\bibitem[\protect\citeauthoryear{{Latif}, {Khochfar}, {Schleicher}  \&
  {Whalen}}{{Latif} et~al.}{2021}]{latif2021}
{Latif} M.~A.,  {Khochfar} S.,  {Schleicher} D.,   {Whalen} D.~J.,  2021,
  \mn@doi [\mnras] {10.1093/mnras/stab2708}, \href
  {https://ui.adsabs.harvard.edu/abs/2021MNRAS.508.1756L} {508, 1756}

\bibitem[\protect\citeauthoryear{{Lodato} \& {Natarajan}}{{Lodato} \&
  {Natarajan}}{2006}]{lodato_natarajan2006}
{Lodato} G.,  {Natarajan} P.,  2006, \mn@doi [\mnras]
  {10.1111/j.1365-2966.2006.10801.x}, \href
  {https://ui.adsabs.harvard.edu/abs/2006MNRAS.371.1813L} {371, 1813}

\bibitem[\protect\citeauthoryear{{Lupi}, {Colpi}, {Devecchi}, {Galanti}  \&
  {Volonteri}}{{Lupi} et~al.}{2014}]{lupi2014}
{Lupi} A.,  {Colpi} M.,  {Devecchi} B.,  {Galanti} G.,   {Volonteri} M.,  2014,
  \mn@doi [\mnras] {10.1093/mnras/stu1120}, \href
  {https://ui.adsabs.harvard.edu/abs/2014MNRAS.442.3616L} {442, 3616}

\bibitem[\protect\citeauthoryear{{Lupi}, {Volonteri}, {Decarli}, {Bovino}  \&
  {Silk}}{{Lupi} et~al.}{2021a}]{lupi2021}
{Lupi} A.,  {Volonteri} M.,  {Decarli} R.,  {Bovino} S.,   {Silk} J.,  2021a,
  arXiv e-prints, \href {https://ui.adsabs.harvard.edu/abs/2021arXiv210901679L}
  {p. arXiv:2109.01679}

\bibitem[\protect\citeauthoryear{{Lupi}, {Haiman}  \& {Volonteri}}{{Lupi}
  et~al.}{2021b}]{lupi_haiman_volonteri2021}
{Lupi} A.,  {Haiman} Z.,   {Volonteri} M.,  2021b, \mn@doi [\mnras]
  {10.1093/mnras/stab692}, \href
  {https://ui.adsabs.harvard.edu/abs/2021MNRAS.503.5046L} {503, 5046}

\bibitem[\protect\citeauthoryear{{Madau} \& {Rees}}{{Madau} \&
  {Rees}}{2001}]{madau_rees2001}
{Madau} P.,  {Rees} M.~J.,  2001, \mn@doi [\apjl] {10.1086/319848}, \href
  {http://adsabs.harvard.edu/abs/2001ApJ...551L..27M} {551, L27}

\bibitem[\protect\citeauthoryear{{Madau}, {Ferrara}  \& {Rees}}{{Madau}
  et~al.}{2001}]{madau_ferrara_rees2001}
{Madau} P.,  {Ferrara} A.,   {Rees} M.~J.,  2001, \mn@doi [\apj]
  {10.1086/321474}, \href
  {https://ui.adsabs.harvard.edu/abs/2001ApJ...555...92M} {555, 92}

\bibitem[\protect\citeauthoryear{{Maio}, {Ciardi}, {Dolag}, {Tornatore}  \&
  {Khochfar}}{{Maio} et~al.}{2010}]{maio2010}
{Maio} U.,  {Ciardi} B.,  {Dolag} K.,  {Tornatore} L.,   {Khochfar} S.,  2010,
  \mn@doi [\mnras] {10.1111/j.1365-2966.2010.17003.x}, \href
  {http://adsabs.harvard.edu/abs/2010MNRAS.407.1003M} {407, 1003}

\bibitem[\protect\citeauthoryear{{Maio}, {Borgani}, {Ciardi}  \&
  {Petkova}}{{Maio} et~al.}{2019}]{maio2019}
{Maio} U.,  {Borgani} S.,  {Ciardi} B.,   {Petkova} M.,  2019, \mn@doi [\pasa]
  {10.1017/pasa.2019.10}, \href
  {https://ui.adsabs.harvard.edu/abs/2019PASA...36...20M} {36, e020}

\bibitem[\protect\citeauthoryear{{Malbon}, {Baugh}, {Frenk}  \&
  {Lacey}}{{Malbon} et~al.}{2007}]{malbon2007}
{Malbon} R.~K.,  {Baugh} C.~M.,  {Frenk} C.~S.,   {Lacey} C.~G.,  2007, \mn@doi
  [\mnras] {10.1111/j.1365-2966.2007.12317.x}, \href
  {https://ui.adsabs.harvard.edu/abs/2007MNRAS.382.1394M} {382, 1394}

\bibitem[\protect\citeauthoryear{{Marconi}, {Risaliti}, {Gilli}, {Hunt},
  {Maiolino}  \& {Salvati}}{{Marconi} et~al.}{2004}]{marconi2004}
{Marconi} A.,  {Risaliti} G.,  {Gilli} R.,  {Hunt} L.~K.,  {Maiolino} R.,
  {Salvati} M.,  2004, \mn@doi [\mnras] {10.1111/j.1365-2966.2004.07765.x},
  \href {https://ui.adsabs.harvard.edu/abs/2004MNRAS.351..169M} {351, 169}

\bibitem[\protect\citeauthoryear{{Mart{\'\i}n-Navarro} \&
  {Mezcua}}{{Mart{\'\i}n-Navarro} \&
  {Mezcua}}{2018}]{martin-navarro_mezcua2018}
{Mart{\'\i}n-Navarro} I.,  {Mezcua} M.,  2018, \mn@doi [\apjl]
  {10.3847/2041-8213/aab103}, \href
  {https://ui.adsabs.harvard.edu/abs/2018ApJ...855L..20M} {855, L20}

\bibitem[\protect\citeauthoryear{{Marulli}, {Bonoli}, {Branchini}, {Moscardini}
   \& {Springel}}{{Marulli} et~al.}{2008}]{marulli2008a}
{Marulli} F.,  {Bonoli} S.,  {Branchini} E.,  {Moscardini} L.,   {Springel} V.,
   2008, \mn@doi [\mnras] {10.1111/j.1365-2966.2008.12988.x}, \href
  {https://ui.adsabs.harvard.edu/abs/2008MNRAS.385.1846M} {385, 1846}

\bibitem[\protect\citeauthoryear{{Mayer} \& {Bonoli}}{{Mayer} \&
  {Bonoli}}{2019}]{mayer_bonoli2019}
{Mayer} L.,  {Bonoli} S.,  2019, \mn@doi [Reports on Progress in Physics]
  {10.1088/1361-6633/aad6a5}, \href
  {https://ui.adsabs.harvard.edu/abs/2019RPPh...82a6901M} {82, 016901}

\bibitem[\protect\citeauthoryear{{Mayer}, {Kazantzidis}, {Escala}  \&
  {Callegari}}{{Mayer} et~al.}{2010}]{mayer2010}
{Mayer} L.,  {Kazantzidis} S.,  {Escala} A.,   {Callegari} S.,  2010, \mn@doi
  [\nat] {10.1038/nature09294}, \href
  {https://ui.adsabs.harvard.edu/abs/2010Natur.466.1082M} {466, 1082}

\bibitem[\protect\citeauthoryear{{Mayer}, {Fiacconi}, {Bonoli}, {Quinn},
  {Ro{\v{s}}kar}, {Shen}  \& {Wadsley}}{{Mayer} et~al.}{2015}]{mayer2015}
{Mayer} L.,  {Fiacconi} D.,  {Bonoli} S.,  {Quinn} T.,  {Ro{\v{s}}kar} R.,
  {Shen} S.,   {Wadsley} J.,  2015, \mn@doi [\apj]
  {10.1088/0004-637X/810/1/51}, \href
  {https://ui.adsabs.harvard.edu/abs/2015ApJ...810...51M} {810, 51}

\bibitem[\protect\citeauthoryear{{Mazzucchelli} et~al.,}{{Mazzucchelli}
  et~al.}{2017}]{mazzucchelli2017}
{Mazzucchelli} C.,  et~al., 2017, \mn@doi [\apj] {10.3847/1538-4357/aa9185},
  \href {https://ui.adsabs.harvard.edu/abs/2017ApJ...849...91M} {849, 91}

\bibitem[\protect\citeauthoryear{{Mezcua}}{{Mezcua}}{2017}]{mezcua2017}
{Mezcua} M.,  2017, \mn@doi [International Journal of Modern Physics D]
  {10.1142/S021827181730021X}, \href
  {https://ui.adsabs.harvard.edu/abs/2017IJMPD..2630021M} {26, 1730021}

\bibitem[\protect\citeauthoryear{{Mezcua}}{{Mezcua}}{2019}]{mezcua2019}
{Mezcua} M.,  2019, \mn@doi [Nature Astronomy] {10.1038/s41550-018-0662-2},
  \href {https://ui.adsabs.harvard.edu/abs/2019NatAs...3....6M} {3, 6}

\bibitem[\protect\citeauthoryear{{Mezcua}}{{Mezcua}}{2021}]{mezcua2021}
{Mezcua} M.,  2021, \mn@doi [IAU Symposium] {10.1017/S1743921320002240}, \href
  {https://ui.adsabs.harvard.edu/abs/2021IAUS..359..238M} {359, 238}

\bibitem[\protect\citeauthoryear{{Miller}, {Gallo}, {Greene}, {Kelly}, {Treu},
  {Woo}  \& {Baldassare}}{{Miller} et~al.}{2015}]{miller2015}
{Miller} B.~P.,  {Gallo} E.,  {Greene} J.~E.,  {Kelly} B.~C.,  {Treu} T.,
  {Woo} J.-H.,   {Baldassare} V.,  2015, \mn@doi [\apj]
  {10.1088/0004-637X/799/1/98}, \href
  {https://ui.adsabs.harvard.edu/abs/2015ApJ...799...98M} {799, 98}

\bibitem[\protect\citeauthoryear{{Miyaji} et~al.,}{{Miyaji}
  et~al.}{2015}]{miyaji2015}
{Miyaji} T.,  et~al., 2015, \mn@doi [\apj] {10.1088/0004-637X/804/2/104}, \href
  {https://ui.adsabs.harvard.edu/abs/2015ApJ...804..104M} {804, 104}

\bibitem[\protect\citeauthoryear{{Natarajan}}{{Natarajan}}{2011}]{natarajan2011}
{Natarajan} P.,  2011, arXiv e-prints, \href
  {https://ui.adsabs.harvard.edu/abs/2011arXiv1105.4902N} {p. arXiv:1105.4902}

\bibitem[\protect\citeauthoryear{{Navarro}, {Frenk}  \& {White}}{{Navarro}
  et~al.}{1997}]{navarro_frenk_white1997}
{Navarro} J.~F.,  {Frenk} C.~S.,   {White} S. D.~M.,  1997, \mn@doi [\apj]
  {10.1086/304888}, \href
  {https://ui.adsabs.harvard.edu/abs/1997ApJ...490..493N} {490, 493}

\bibitem[\protect\citeauthoryear{{Nelson} et~al.,}{{Nelson}
  et~al.}{2015}]{nelson2015}
{Nelson} D.,  et~al., 2015, \mn@doi [Astronomy and Computing]
  {10.1016/j.ascom.2015.09.003}, \href
  {https://ui.adsabs.harvard.edu/abs/2015A&C....13...12N} {13, 12}

\bibitem[\protect\citeauthoryear{{Ni}, {Di Matteo}, {Gilli}, {Croft}, {Feng}
  \& {Norman}}{{Ni} et~al.}{2020}]{ni2020}
{Ni} Y.,  {Di Matteo} T.,  {Gilli} R.,  {Croft} R. A.~C.,  {Feng} Y.,
  {Norman} C.,  2020, \mn@doi [\mnras] {10.1093/mnras/staa1313}, \href
  {https://ui.adsabs.harvard.edu/abs/2020MNRAS.495.2135N} {495, 2135}

\bibitem[\protect\citeauthoryear{{O'Shea} \& {Norman}}{{O'Shea} \&
  {Norman}}{2008}]{oshea_norman2008}
{O'Shea} B.~W.,  {Norman} M.~L.,  2008, \mn@doi [\apj] {10.1086/524006}, \href
  {https://ui.adsabs.harvard.edu/abs/2008ApJ...673...14O} {673, 14}

\bibitem[\protect\citeauthoryear{{Omukai}}{{Omukai}}{2001}]{omukai2001a}
{Omukai} K.,  2001, \mn@doi [\apj] {10.1086/318296}, \href
  {https://ui.adsabs.harvard.edu/abs/2001ApJ...546..635O} {546, 635}

\bibitem[\protect\citeauthoryear{{Omukai} \& {Palla}}{{Omukai} \&
  {Palla}}{2001}]{omukai_palla2001}
{Omukai} K.,  {Palla} F.,  2001, \mn@doi [\apjl] {10.1086/324410}, \href
  {https://ui.adsabs.harvard.edu/abs/2001ApJ...561L..55O} {561, L55}

\bibitem[\protect\citeauthoryear{{Omukai}, {Schneider}  \& {Haiman}}{{Omukai}
  et~al.}{2008}]{omukai_schneider_haiman2008}
{Omukai} K.,  {Schneider} R.,   {Haiman} Z.,  2008, \mn@doi [\apj]
  {10.1086/591636}, \href
  {https://ui.adsabs.harvard.edu/abs/2008ApJ...686..801O} {686, 801}

\bibitem[\protect\citeauthoryear{{Pezzulli}, {Volonteri}, {Schneider}  \&
  {Valiante}}{{Pezzulli} et~al.}{2017}]{pezzulli2017}
{Pezzulli} E.,  {Volonteri} M.,  {Schneider} R.,   {Valiante} R.,  2017,
  \mn@doi [\mnras] {10.1093/mnras/stx1640}, \href
  {http://adsabs.harvard.edu/abs/2017MNRAS.471..589P} {471, 589}

\bibitem[\protect\citeauthoryear{{Piana}, {Dayal}, {Volonteri}  \&
  {Choudhury}}{{Piana} et~al.}{2021}]{piana2021}
{Piana} O.,  {Dayal} P.,  {Volonteri} M.,   {Choudhury} T.~R.,  2021, \mn@doi
  [\mnras] {10.1093/mnras/staa3363}, \href
  {https://ui.adsabs.harvard.edu/abs/2021MNRAS.500.2146P} {500, 2146}

\bibitem[\protect\citeauthoryear{{Planck Collaboration} et~al.,}{{Planck
  Collaboration} et~al.}{2016}]{ade2016}
{Planck Collaboration} et~al., 2016, \mn@doi [\aap]
  {10.1051/0004-6361/201525830}, \href
  {https://ui.adsabs.harvard.edu/abs/2016A&A...594A..13P} {594, A13}

\bibitem[\protect\citeauthoryear{{Portegies Zwart} \& {McMillan}}{{Portegies
  Zwart} \& {McMillan}}{2002}]{portegieszwart_mcmillan2002}
{Portegies Zwart} S.~F.,  {McMillan} S. L.~W.,  2002, \mn@doi [\apj]
  {10.1086/341798}, \href
  {https://ui.adsabs.harvard.edu/abs/2002ApJ...576..899P} {576, 899}

\bibitem[\protect\citeauthoryear{{Portegies Zwart}, {Makino}, {McMillan}  \&
  {Hut}}{{Portegies Zwart} et~al.}{1999}]{portegieszwart1999}
{Portegies Zwart} S.~F.,  {Makino} J.,  {McMillan} S.~L.~W.,   {Hut} P.,  1999,
  \aap, \href {https://ui.adsabs.harvard.edu/abs/1999A&A...348..117P} {348,
  117}

\bibitem[\protect\citeauthoryear{{Rasio}, {Freitag}  \& {G{\"u}rkan}}{{Rasio}
  et~al.}{2004}]{rasio_freitag_gurkan2004}
{Rasio} F.~A.,  {Freitag} M.,   {G{\"u}rkan} M.~A.,  2004, in {Ho} L.~C.,  ed.,
  Coevolution of Black Holes and Galaxies. p.~138 (\mn@eprint {arXiv}
  {astro-ph/0304038})

\bibitem[\protect\citeauthoryear{{Rees}, {Begelman}, {Blandford}  \&
  {Phinney}}{{Rees} et~al.}{1982}]{rees1982}
{Rees} M.~J.,  {Begelman} M.~C.,  {Blandford} R.~D.,   {Phinney} E.~S.,  1982,
  \mn@doi [\nat] {10.1038/295017a0}, \href
  {https://ui.adsabs.harvard.edu/abs/1982Natur.295...17R} {295, 17}

\bibitem[\protect\citeauthoryear{{Regan} \& {Downes}}{{Regan} \&
  {Downes}}{2018}]{regan_downes2018}
{Regan} J.~A.,  {Downes} T.~P.,  2018, \mn@doi [\mnras] {10.1093/mnras/sty134},
  \href {https://ui.adsabs.harvard.edu/abs/2018MNRAS.475.4636R} {475, 4636}

\bibitem[\protect\citeauthoryear{{Regan}, {Johansson}  \& {Wise}}{{Regan}
  et~al.}{2014}]{regan2014b}
{Regan} J.~A.,  {Johansson} P.~H.,   {Wise} J.~H.,  2014, \mn@doi [\apj]
  {10.1088/0004-637X/795/2/137}, \href
  {https://ui.adsabs.harvard.edu/abs/2014ApJ...795..137R} {795, 137}

\bibitem[\protect\citeauthoryear{{Regan}, {Johansson}  \& {Wise}}{{Regan}
  et~al.}{2016}]{regan_johansson_wise2016a}
{Regan} J.~A.,  {Johansson} P.~H.,   {Wise} J.~H.,  2016, \mn@doi [\mnras]
  {10.1093/mnras/stw899}, \href
  {https://ui.adsabs.harvard.edu/abs/2016MNRAS.459.3377R} {459, 3377}

\bibitem[\protect\citeauthoryear{{Regan}, {Visbal}, {Wise}, {Haiman},
  {Johansson}  \& {Bryan}}{{Regan} et~al.}{2017}]{regan2017}
{Regan} J.~A.,  {Visbal} E.,  {Wise} J.~H.,  {Haiman} Z.,  {Johansson} P.~H.,
  {Bryan} G.~L.,  2017, \mn@doi [Nature Astronomy] {10.1038/s41550-017-0075},
  \href {https://ui.adsabs.harvard.edu/abs/2017NatAs...1E..75R} {1, 0075}

\bibitem[\protect\citeauthoryear{{Regan}, {Downes}, {Volonteri}, {Beckmann},
  {Lupi}, {Trebitsch}  \& {Dubois}}{{Regan} et~al.}{2019}]{regan2019}
{Regan} J.~A.,  {Downes} T.~P.,  {Volonteri} M.,  {Beckmann} R.,  {Lupi} A.,
  {Trebitsch} M.,   {Dubois} Y.,  2019, \mn@doi [\mnras]
  {10.1093/mnras/stz1045}, \href
  {https://ui.adsabs.harvard.edu/abs/2019MNRAS.486.3892R} {486, 3892}

\bibitem[\protect\citeauthoryear{{Regan}, {Wise}, {Woods}, {Downes}, {O'Shea}
  \& {Norman}}{{Regan} et~al.}{2020}]{regan2020c}
{Regan} J.~A.,  {Wise} J.~H.,  {Woods} T.~E.,  {Downes} T.~P.,  {O'Shea} B.~W.,
    {Norman} M.~L.,  2020, \mn@doi [The Open Journal of Astrophysics]
  {10.21105/astro.2008.08090}, \href
  {https://ui.adsabs.harvard.edu/abs/2020OJAp....3E..15R} {3, 15}

\bibitem[\protect\citeauthoryear{{Reines} \& {Volonteri}}{{Reines} \&
  {Volonteri}}{2015}]{reines_volonteri2015}
{Reines} A.~E.,  {Volonteri} M.,  2015, \mn@doi [\apj]
  {10.1088/0004-637X/813/2/82}, \href
  {https://ui.adsabs.harvard.edu/abs/2015ApJ...813...82R} {813, 82}

\bibitem[\protect\citeauthoryear{{Reinoso}, {Schleicher}, {Fellhauer},
  {Klessen}  \& {Boekholt}}{{Reinoso} et~al.}{2018}]{reinoso2018}
{Reinoso} B.,  {Schleicher} D.~R.~G.,  {Fellhauer} M.,  {Klessen} R.~S.,
  {Boekholt} T.~C.~N.,  2018, \mn@doi [\aap] {10.1051/0004-6361/201732224},
  \href {https://ui.adsabs.harvard.edu/abs/2018A&A...614A..14R} {614, A14}

\bibitem[\protect\citeauthoryear{{Ritter}, {Safranek-Shrader}, {Gnat},
  {Milosavljevi{\'c}}  \& {Bromm}}{{Ritter} et~al.}{2012}]{ritter2012}
{Ritter} J.~S.,  {Safranek-Shrader} C.,  {Gnat} O.,  {Milosavljevi{\'c}} M.,
  {Bromm} V.,  2012, \mn@doi [\apj] {10.1088/0004-637X/761/1/56}, \href
  {https://ui.adsabs.harvard.edu/abs/2012ApJ...761...56R} {761, 56}

\bibitem[\protect\citeauthoryear{{Salvadori}, {Schneider}  \&
  {Ferrara}}{{Salvadori} et~al.}{2007}]{salvadori_schneider_ferrara2007}
{Salvadori} S.,  {Schneider} R.,   {Ferrara} A.,  2007, \mn@doi [\mnras]
  {10.1111/j.1365-2966.2007.12133.x}, \href
  {https://ui.adsabs.harvard.edu/abs/2007MNRAS.381..647S} {381, 647}

\bibitem[\protect\citeauthoryear{{Salvadori}, {Ferrara}  \&
  {Schneider}}{{Salvadori} et~al.}{2008}]{salvadori_ferrara_schneider2008}
{Salvadori} S.,  {Ferrara} A.,   {Schneider} R.,  2008, \mn@doi [\mnras]
  {10.1111/j.1365-2966.2008.13035.x}, \href
  {https://ui.adsabs.harvard.edu/abs/2008MNRAS.386..348S} {386, 348}

\bibitem[\protect\citeauthoryear{{Sassano}, {Schneider}, {Valiante},
  {Inayoshi}, {Chon}, {Omukai}, {Mayer}  \& {Capelo}}{{Sassano}
  et~al.}{2021}]{sassano2021}
{Sassano} F.,  {Schneider} R.,  {Valiante} R.,  {Inayoshi} K.,  {Chon} S.,
  {Omukai} K.,  {Mayer} L.,   {Capelo} P.~R.,  2021, \mn@doi [\mnras]
  {10.1093/mnras/stab1737}, \href
  {https://ui.adsabs.harvard.edu/abs/2021MNRAS.506..613S} {506, 613}

\bibitem[\protect\citeauthoryear{{Schaerer}}{{Schaerer}}{2002a}]{schaerer2002b}
{Schaerer} D.,  2002a, arXiv e-prints, \href
  {https://ui.adsabs.harvard.edu/abs/2002astro.ph..8227S} {pp
  astro--ph/0208227}

\bibitem[\protect\citeauthoryear{{Schaerer}}{{Schaerer}}{2002b}]{schaerer2002a}
{Schaerer} D.,  2002b, \mn@doi [\aap] {10.1051/0004-6361:20011619}, \href
  {https://ui.adsabs.harvard.edu/abs/2002A&A...382...28S} {382, 28}

\bibitem[\protect\citeauthoryear{{Schneider}}{{Schneider}}{2006}]{schneider2006a}
{Schneider} R.,  2006, \mn@doi [\nar] {10.1016/j.newar.2005.11.014}, \href
  {https://ui.adsabs.harvard.edu/abs/2006NewAR..50...64S} {50, 64}

\bibitem[\protect\citeauthoryear{{Schneider}, {Ferrara}, {Natarajan}  \&
  {Omukai}}{{Schneider} et~al.}{2002}]{schneider2002}
{Schneider} R.,  {Ferrara} A.,  {Natarajan} P.,   {Omukai} K.,  2002, \mn@doi
  [\apj] {10.1086/339917}, \href
  {https://ui.adsabs.harvard.edu/abs/2002ApJ...571...30S} {571, 30}

\bibitem[\protect\citeauthoryear{{Schneider}, {Omukai}, {Inoue}  \&
  {Ferrara}}{{Schneider} et~al.}{2006}]{schneider2006b}
{Schneider} R.,  {Omukai} K.,  {Inoue} A.~K.,   {Ferrara} A.,  2006, \mn@doi
  [\mnras] {10.1111/j.1365-2966.2006.10391.x}, \href
  {https://ui.adsabs.harvard.edu/abs/2006MNRAS.369.1437S} {369, 1437}

\bibitem[\protect\citeauthoryear{{Sesana}, {Volonteri}  \& {Haardt}}{{Sesana}
  et~al.}{2007}]{sesana_volonteri_haardt2007}
{Sesana} A.,  {Volonteri} M.,   {Haardt} F.,  2007, \mn@doi [\mnras]
  {10.1111/j.1365-2966.2007.11734.x}, \href
  {https://ui.adsabs.harvard.edu/abs/2007MNRAS.377.1711S} {377, 1711}

\bibitem[\protect\citeauthoryear{{Shakura} \& {Sunyaev}}{{Shakura} \&
  {Sunyaev}}{1973}]{shakura_sunyaev1973}
{Shakura} N.~I.,  {Sunyaev} R.~A.,  1973, \aap, \href
  {https://ui.adsabs.harvard.edu/abs/1973A&A....24..337S} {24, 337}

\bibitem[\protect\citeauthoryear{{Shang}, {Bryan}  \& {Haiman}}{{Shang}
  et~al.}{2010}]{shang_bryan_haiman2010}
{Shang} C.,  {Bryan} G.~L.,   {Haiman} Z.,  2010, \mn@doi [\mnras]
  {10.1111/j.1365-2966.2009.15960.x}, \href
  {https://ui.adsabs.harvard.edu/abs/2010MNRAS.402.1249S} {402, 1249}

\bibitem[\protect\citeauthoryear{{Shankar}}{{Shankar}}{2013}]{shankar2013}
{Shankar} F.,  2013, \mn@doi [Classical and Quantum Gravity]
  {10.1088/0264-9381/30/24/244001}, \href
  {https://ui.adsabs.harvard.edu/abs/2013CQGra..30x4001S} {30, 244001}

\bibitem[\protect\citeauthoryear{{Shankar}, {Salucci}, {Granato}, {De Zotti}
  \& {Danese}}{{Shankar} et~al.}{2004}]{shankar2004}
{Shankar} F.,  {Salucci} P.,  {Granato} G.~L.,  {De Zotti} G.,   {Danese} L.,
  2004, \mn@doi [\mnras] {10.1111/j.1365-2966.2004.08261.x}, \href
  {https://ui.adsabs.harvard.edu/abs/2004MNRAS.354.1020S} {354, 1020}

\bibitem[\protect\citeauthoryear{{Shankar}, {Weinberg}  \&
  {Miralda-Escud{\'e}}}{{Shankar}
  et~al.}{2009}]{shankar_weinberg_miralda-escude2009}
{Shankar} F.,  {Weinberg} D.~H.,   {Miralda-Escud{\'e}} J.,  2009, \mn@doi
  [\apj] {10.1088/0004-637X/690/1/20}, \href
  {https://ui.adsabs.harvard.edu/abs/2009ApJ...690...20S} {690, 20}

\bibitem[\protect\citeauthoryear{{Sharma}, {Roy}, {Nath}  \&
  {Shchekinov}}{{Sharma} et~al.}{2014}]{sharma2014}
{Sharma} P.,  {Roy} A.,  {Nath} B.~B.,   {Shchekinov} Y.,  2014, \mn@doi
  [\mnras] {10.1093/mnras/stu1307}, \href
  {https://ui.adsabs.harvard.edu/abs/2014MNRAS.443.3463S} {443, 3463}

\bibitem[\protect\citeauthoryear{{Sharma}, {Brooks}, {Somerville}, {Tremmel},
  {Bellovary}, {Wright}  \& {Quinn}}{{Sharma} et~al.}{2020}]{sharma2020}
{Sharma} R.~S.,  {Brooks} A.~M.,  {Somerville} R.~S.,  {Tremmel} M.,
  {Bellovary} J.,  {Wright} A.~C.,   {Quinn} T.~R.,  2020, \mn@doi [\apj]
  {10.3847/1538-4357/ab960e}, \href
  {https://ui.adsabs.harvard.edu/abs/2020ApJ...897..103S} {897, 103}

\bibitem[\protect\citeauthoryear{{Shen}, {Hopkins}, {Faucher-Gigu{\`e}re},
  {Alexander}, {Richards}, {Ross}  \& {Hickox}}{{Shen} et~al.}{2020}]{shen2020}
{Shen} X.,  {Hopkins} P.~F.,  {Faucher-Gigu{\`e}re} C.-A.,  {Alexander} D.~M.,
  {Richards} G.~T.,  {Ross} N.~P.,   {Hickox} R.~C.,  2020, \mn@doi [\mnras]
  {10.1093/mnras/staa1381}, \href
  {https://ui.adsabs.harvard.edu/abs/2020MNRAS.495.3252S} {495, 3252}

\bibitem[\protect\citeauthoryear{{Springel} et~al.,}{{Springel}
  et~al.}{2005}]{mill}
{Springel} V.,  et~al., 2005, \mn@doi [\nat] {10.1038/nature03597}, \href
  {http://adsabs.harvard.edu/abs/2005Natur.435..629S} {435, 629}

\bibitem[\protect\citeauthoryear{{Sutherland} \& {Dopita}}{{Sutherland} \&
  {Dopita}}{1993}]{sutherland_dopita1993}
{Sutherland} R.~S.,  {Dopita} M.~A.,  1993, \mn@doi [\apjs] {10.1086/191823},
  \href {https://ui.adsabs.harvard.edu/abs/1993ApJS...88..253S} {88, 253}

\bibitem[\protect\citeauthoryear{{Tanaka} \& {Haiman}}{{Tanaka} \&
  {Haiman}}{2009}]{tanaka_haiman2009}
{Tanaka} T.,  {Haiman} Z.,  2009, \mn@doi [\apj]
  {10.1088/0004-637X/696/2/1798}, \href
  {https://ui.adsabs.harvard.edu/abs/2009ApJ...696.1798T} {696, 1798}

\bibitem[\protect\citeauthoryear{{Tanaka} \& {Li}}{{Tanaka} \&
  {Li}}{2014}]{tanaka_li2014}
{Tanaka} T.~L.,  {Li} M.,  2014, \mn@doi [\mnras] {10.1093/mnras/stu042}, \href
  {https://ui.adsabs.harvard.edu/abs/2014MNRAS.439.1092T} {439, 1092}

\bibitem[\protect\citeauthoryear{{Trinca}, {Schneider}, {Valiante}, {Graziani},
  {Zappacosta}  \& {Shankar}}{{Trinca} et~al.}{2022}]{trinca2022}
{Trinca} A.,  {Schneider} R.,  {Valiante} R.,  {Graziani} L.,  {Zappacosta} L.,
    {Shankar} F.,  2022, \mn@doi [\mnras] {10.1093/mnras/stac062}, \href
  {https://ui.adsabs.harvard.edu/abs/2022MNRAS.tmp...99T} {450, 3545}

\bibitem[\protect\citeauthoryear{{Ueda}, {Akiyama}, {Hasinger}, {Miyaji}  \&
  {Watson}}{{Ueda} et~al.}{2014}]{ueda2014}
{Ueda} Y.,  {Akiyama} M.,  {Hasinger} G.,  {Miyaji} T.,   {Watson} M.~G.,
  2014, \mn@doi [\apj] {10.1088/0004-637X/786/2/104}, \href
  {https://ui.adsabs.harvard.edu/abs/2014ApJ...786..104U} {786, 104}

\bibitem[\protect\citeauthoryear{{Valiante}, {Schneider}, {Salvadori}  \&
  {Bianchi}}{{Valiante} et~al.}{2011}]{valiante2011}
{Valiante} R.,  {Schneider} R.,  {Salvadori} S.,   {Bianchi} S.,  2011, \mn@doi
  [\mnras] {10.1111/j.1365-2966.2011.19168.x}, \href
  {https://ui.adsabs.harvard.edu/abs/2011MNRAS.416.1916V} {416, 1916}

\bibitem[\protect\citeauthoryear{{Valiante}, {Schneider}, {Salvadori}  \&
  {Gallerani}}{{Valiante} et~al.}{2014}]{valiante2014}
{Valiante} R.,  {Schneider} R.,  {Salvadori} S.,   {Gallerani} S.,  2014,
  \mn@doi [\mnras] {10.1093/mnras/stu1613}, \href
  {https://ui.adsabs.harvard.edu/abs/2014MNRAS.444.2442V} {444, 2442}

\bibitem[\protect\citeauthoryear{{Valiante}, {Schneider}, {Volonteri}  \&
  {Omukai}}{{Valiante} et~al.}{2016}]{valiante2016}
{Valiante} R.,  {Schneider} R.,  {Volonteri} M.,   {Omukai} K.,  2016, \mn@doi
  [\mnras] {10.1093/mnras/stw225}, \href
  {https://ui.adsabs.harvard.edu/abs/2016MNRAS.457.3356V} {457, 3356}

\bibitem[\protect\citeauthoryear{{Valiante}, {Agarwal}, {Habouzit}  \&
  {Pezzulli}}{{Valiante} et~al.}{2017}]{valiante2017}
{Valiante} R.,  {Agarwal} B.,  {Habouzit} M.,   {Pezzulli} E.,  2017, \mn@doi
  [\pasa] {10.1017/pasa.2017.25}, \href
  {https://ui.adsabs.harvard.edu/abs/2017PASA...34...31V} {34, e031}

\bibitem[\protect\citeauthoryear{{Valiante} et~al.,}{{Valiante}
  et~al.}{2021}]{valiante2021}
{Valiante} R.,  et~al., 2021, \mn@doi [\mnras] {10.1093/mnras/staa3395}, \href
  {https://ui.adsabs.harvard.edu/abs/2021MNRAS.500.4095V} {500, 4095}

\bibitem[\protect\citeauthoryear{{Van Wassenhove}, {Volonteri}, {Walker}  \&
  {Gair}}{{Van Wassenhove} et~al.}{2010}]{van-wassenhove2010}
{Van Wassenhove} S.,  {Volonteri} M.,  {Walker} M.~G.,   {Gair} J.~R.,  2010,
  \mn@doi [\mnras] {10.1111/j.1365-2966.2010.17189.x}, \href
  {https://ui.adsabs.harvard.edu/abs/2010MNRAS.408.1139V} {408, 1139}

\bibitem[\protect\citeauthoryear{{Visbal}, {Haiman}  \& {Bryan}}{{Visbal}
  et~al.}{2014a}]{visbal_haiman_bryan2014a}
{Visbal} E.,  {Haiman} Z.,   {Bryan} G.~L.,  2014a, \mn@doi [\mnras]
  {10.1093/mnrasl/slu063}, \href
  {https://ui.adsabs.harvard.edu/abs/2014MNRAS.442L.100V} {442, L100}

\bibitem[\protect\citeauthoryear{{Visbal}, {Haiman}  \& {Bryan}}{{Visbal}
  et~al.}{2014b}]{visbal_haiman_bryan2014b}
{Visbal} E.,  {Haiman} Z.,   {Bryan} G.~L.,  2014b, \mn@doi [\mnras]
  {10.1093/mnras/stu1794}, \href
  {https://ui.adsabs.harvard.edu/abs/2014MNRAS.445.1056V} {445, 1056}

\bibitem[\protect\citeauthoryear{{Visbal}, {Bryan}  \& {Haiman}}{{Visbal}
  et~al.}{2020}]{visbal_bryan_haiman2020}
{Visbal} E.,  {Bryan} G.~L.,   {Haiman} Z.,  2020, \mn@doi [\apj]
  {10.3847/1538-4357/ab994e}, \href
  {https://ui.adsabs.harvard.edu/abs/2020ApJ...897...95V} {897, 95}

\bibitem[\protect\citeauthoryear{{Volonteri}}{{Volonteri}}{2007}]{volonteri2007}
{Volonteri} M.,  2007, \mn@doi [\apjl] {10.1086/519525}, \href
  {https://ui.adsabs.harvard.edu/abs/2007ApJ...663L...5V} {663, L5}

\bibitem[\protect\citeauthoryear{{Volonteri}}{{Volonteri}}{2010}]{volonteri2010}
{Volonteri} M.,  2010, \mn@doi [\aapr] {10.1007/s00159-010-0029-x}, \href
  {http://adsabs.harvard.edu/abs/2010A%26ARv..18..279V} {18, 279}

\bibitem[\protect\citeauthoryear{{Volonteri} \& {Bellovary}}{{Volonteri} \&
  {Bellovary}}{2012}]{volonteri_bellovary2012}
{Volonteri} M.,  {Bellovary} J.,  2012, \mn@doi [Reports on Progress in
  Physics] {10.1088/0034-4885/75/12/124901}, \href
  {https://ui.adsabs.harvard.edu/abs/2012RPPh...75l4901V} {75, 124901}

\bibitem[\protect\citeauthoryear{{Volonteri} \& {Rees}}{{Volonteri} \&
  {Rees}}{2005}]{volonteri_rees2005}
{Volonteri} M.,  {Rees} M.~J.,  2005, \mn@doi [\apj] {10.1086/466521}, \href
  {https://ui.adsabs.harvard.edu/abs/2005ApJ...633..624V} {633, 624}

\bibitem[\protect\citeauthoryear{{Volonteri} \& {Reines}}{{Volonteri} \&
  {Reines}}{2016}]{volonteri_reines2016}
{Volonteri} M.,  {Reines} A.~E.,  2016, \mn@doi [\apjl]
  {10.3847/2041-8205/820/1/L6}, \href
  {https://ui.adsabs.harvard.edu/abs/2016ApJ...820L...6V} {820, L6}

\bibitem[\protect\citeauthoryear{{Volonteri}, {Haardt}  \& {Madau}}{{Volonteri}
  et~al.}{2003}]{volonteri_haardt_madau2003}
{Volonteri} M.,  {Haardt} F.,   {Madau} P.,  2003, \mn@doi [\apj]
  {10.1086/344675}, \href
  {https://ui.adsabs.harvard.edu/abs/2003ApJ...582..559V} {582, 559}

\bibitem[\protect\citeauthoryear{{Volonteri}, {Lodato}  \&
  {Natarajan}}{{Volonteri} et~al.}{2008}]{volonteri_lodato_natarajan2008}
{Volonteri} M.,  {Lodato} G.,   {Natarajan} P.,  2008, \mn@doi [\mnras]
  {10.1111/j.1365-2966.2007.12589.x}, \href
  {https://ui.adsabs.harvard.edu/abs/2008MNRAS.383.1079V} {383, 1079}

\bibitem[\protect\citeauthoryear{{Volonteri}, {G{\"u}ltekin}  \&
  {Dotti}}{{Volonteri} et~al.}{2010}]{volonteri_gultekin_dotti2010}
{Volonteri} M.,  {G{\"u}ltekin} K.,   {Dotti} M.,  2010, \mn@doi [\mnras]
  {10.1111/j.1365-2966.2010.16431.x}, \href
  {https://ui.adsabs.harvard.edu/abs/2010MNRAS.404.2143V} {404, 2143}

\bibitem[\protect\citeauthoryear{{Wang} et~al.,}{{Wang}
  et~al.}{2021}]{wang2021}
{Wang} F.,  et~al., 2021, \mn@doi [\apjl] {10.3847/2041-8213/abd8c6}, \href
  {https://ui.adsabs.harvard.edu/abs/2021ApJ...907L...1W} {907, L1}

\bibitem[\protect\citeauthoryear{{White} \& {Frenk}}{{White} \&
  {Frenk}}{1991}]{white_frenk1991}
{White} S. D.~M.,  {Frenk} C.~S.,  1991, \mn@doi [\apj] {10.1086/170483}, \href
  {https://ui.adsabs.harvard.edu/abs/1991ApJ...379...52W} {379, 52}

\bibitem[\protect\citeauthoryear{{White} \& {Rees}}{{White} \&
  {Rees}}{1978}]{white_rees1978}
{White} S.~D.~M.,  {Rees} M.~J.,  1978, \mn@doi [\mnras]
  {10.1093/mnras/183.3.341}, \href
  {http://adsabs.harvard.edu/abs/1978MNRAS.183..341W} {183, 341}

\bibitem[\protect\citeauthoryear{{Wise}, {Demchenko}, {Halicek}, {Norman},
  {Turk}, {Abel}  \& {Smith}}{{Wise} et~al.}{2014}]{wise2014}
{Wise} J.~H.,  {Demchenko} V.~G.,  {Halicek} M.~T.,  {Norman} M.~L.,  {Turk}
  M.~J.,  {Abel} T.,   {Smith} B.~D.,  2014, \mn@doi [\mnras]
  {10.1093/mnras/stu979}, \href
  {https://ui.adsabs.harvard.edu/abs/2014MNRAS.442.2560W} {442, 2560}

\bibitem[\protect\citeauthoryear{{Wolcott-Green} \& {Haiman}}{{Wolcott-Green}
  \& {Haiman}}{2011}]{wolcott-green_haiman2011}
{Wolcott-Green} J.,  {Haiman} Z.,  2011, \mn@doi [\mnras]
  {10.1111/j.1365-2966.2010.18080.x}, \href
  {https://ui.adsabs.harvard.edu/abs/2011MNRAS.412.2603W} {412, 2603}

\bibitem[\protect\citeauthoryear{{Wolcott-Green} \& {Haiman}}{{Wolcott-Green}
  \& {Haiman}}{2019}]{wolcott-green_haiman2019}
{Wolcott-Green} J.,  {Haiman} Z.,  2019, \mn@doi [\mnras]
  {10.1093/mnras/sty3280}, \href
  {https://ui.adsabs.harvard.edu/abs/2019MNRAS.484.2467W} {484, 2467}

\bibitem[\protect\citeauthoryear{{Yadav}, {Mukherjee}, {Sharma}  \&
  {Nath}}{{Yadav} et~al.}{2017}]{yadav2017}
{Yadav} N.,  {Mukherjee} D.,  {Sharma} P.,   {Nath} B.~B.,  2017, \mn@doi
  [\mnras] {10.1093/mnras/stw2522}, \href
  {https://ui.adsabs.harvard.edu/abs/2017MNRAS.465.1720Y} {465, 1720}

\bibitem[\protect\citeauthoryear{{Yates}, {Henriques}, {Thomas}, {Kauffmann},
  {Johansson}  \& {White}}{{Yates} et~al.}{2013}]{yates2013}
{Yates} R.~M.,  {Henriques} B.,  {Thomas} P.~A.,  {Kauffmann} G.,  {Johansson}
  J.,   {White} S. D.~M.,  2013, \mn@doi [\mnras] {10.1093/mnras/stt1542},
  \href {https://ui.adsabs.harvard.edu/abs/2013MNRAS.435.3500Y} {435, 3500}

\bibitem[\protect\citeauthoryear{{Yoshida}, {Abel}, {Hernquist}  \&
  {Sugiyama}}{{Yoshida} et~al.}{2003}]{yoshida2003}
{Yoshida} N.,  {Abel} T.,  {Hernquist} L.,   {Sugiyama} N.,  2003, \mn@doi
  [\apj] {10.1086/375810}, \href
  {https://ui.adsabs.harvard.edu/abs/2003ApJ...592..645Y} {592, 645}

\bibitem[\protect\citeauthoryear{{Yoshida}, {Omukai}  \& {Hernquist}}{{Yoshida}
  et~al.}{2007}]{yoshida_omukai_hernquist2007}
{Yoshida} N.,  {Omukai} K.,   {Hernquist} L.,  2007, \mn@doi [\apjl]
  {10.1086/522202}, \href
  {https://ui.adsabs.harvard.edu/abs/2007ApJ...667L.117Y} {667, L117}

\bibitem[\protect\citeauthoryear{{Yue}, {Ferrara}, {Salvaterra}, {Xu}  \&
  {Chen}}{{Yue} et~al.}{2014}]{yue2014}
{Yue} B.,  {Ferrara} A.,  {Salvaterra} R.,  {Xu} Y.,   {Chen} X.,  2014,
  \mn@doi [\mnras] {10.1093/mnras/stu351}, \href
  {https://ui.adsabs.harvard.edu/abs/2014MNRAS.440.1263Y} {440, 1263}

\makeatother
\end{thebibliography}
}

\appendix
\section{Computation of \texorpdfstring{$\rm J_{PLC}$}{Jplc}}
\label{sec:appendix:jplc_computation}
As discussed in Sect. \ref{sec:model:actual_computation_of_local_metals_and_jlw}, we compute $\rm J_{local}(\overrightarrow{x}_{\rm s})$ by considering the total contribution $\rm J_{PLC}(\overrightarrow{x}_{\rm s})$ provided by LW sources in the past light-cone of a given point of interest $\overrightarrow{x}_{\rm s}$. In order to classify galaxies as LW sources, we perform a preliminary run of \lgal during which, at each SF event of each galaxy, we compute the $\rm J_{LW}$ produced by the newly-formed stars at a distance from the galaxy equal to the $\rm R_{vir}$ of its DM halo host. If this $\rm J_{LW}(R_{vir})$ is greater than $\rm 10\,J_{21}$ (i.e. the $\rm J_{crit,DCBH}$ we impose, see Sect. \ref{sec:model:BH_seeding:RSM_and_DCBH_seeding_in_LGal}), we classify the galaxy \textit{at the current time} as a LW source, and store its properties in an external list. Since the computation of $\rm J_{PLC}$ is only needed before the time at which the last DCBH or RSM seed can form, we stop storing LW sources when $\rm\langle Z_{IGM}\rangle\!>\!\rm Z_{\,crit,RSM}$ (i.e. at $z\!\sim\!6$, see Fig. \ref{fig:dynaranges}). This latter condition helps to keep the total number of LW sources in our catalog relatively low (i.e. few $\sim10^4$ LW sources in the whole \milltwo box).

Nevertheless, our actual computation of $\rm J_{PLC}$ at a given point of interest $\rm\overrightarrow{x}_s$, at a given time $\rm t_s$, employs a loop over all the stored LW sources (i.e. an inefficient $\rm N^2$ operation). We underline that our computation proceeds farther away in space and backwards in time from $\rm(\overrightarrow{x}_s,t_s)$, looking for potential LW sources which are at the correct look-back time and distance to provide LW photons to $\rm(\overrightarrow{x}_s,t_s)$. The further away the potential LW sources are from $\rm\overrightarrow{x}_s$, the fainter is their contribution. This implies that extremely-far sources would negligibly contribute to the $\rm J_{PLC}$ at $\rm\overrightarrow{x}_s$. Therefore, looking ``too far away'' from potential sites for DCBH or RSM seeds formation only makes inefficient the computation of $\rm J_{PLC}$. To avoid this, we stop our loop over the past light cone of $\rm(\overrightarrow{x}_s,t_s)$ when we reach a fixed look-back distance threshold from $\rm(\overrightarrow{x}_s,t_s)$. We conservatively define this threshold as the distance at which, if the brightest source in the whole LW catalog was to be found there, it would produce 1/100th of our $\rm J_{crit}$ threshold.

This technique allows to sensibly reduce the execution time of our $\rm J_{PLC}$ computation by eliminating the need to explore the entire past light-cone of any BH-seeding candidate. We explicitly checked that 1/100th of $\rm J_{crit}$ is a good compromise between the convergence of our $\rm J_{PLC}$ computation and the execution speed of our code. Indeed, by stopping when the brightest source in the LW catalog would produce 1/10th of $\rm J_{crit}$, we start to measure deviations (of the order of $\rm 10^{-3}\,J_{21}$) on the $\rm J_{PLC}$ values we compute with respect to leaving the $\rm J_{PLC}$ computation unconstrained. On the other hand, by stopping at 1/1000th of $\rm J_{crit}$ the execution time of our model is significantly increased without providing appreciable differences on the computed $\rm J_{PLC}$ with respect to the case of stopping at 1/100th of $\rm J_{crit}$.

\section{Grafting of \gqd outputs in \lgal}
\label{sec:appendix:grafting_details}
In order to complement the discussion of Sect. \ref{sec:model:BH_seeding_prescriptions:GQd_Grafting}, here we further detail our grafting procedure. This appendix is based on the working example of a \milltwo{} halo, newly-resolved with virial mass $\rm M_{vir,init}$ at a given initialization-snapshot of the \milltwo (found at redshift $z_{\rm init}$). The \milltwo snapshot immediately preceding $z_{\rm init}$ would be found at $z_{\rm prev}$. In addition, we use $\rm M_{vir,GQd}$ to refer to the virial mass of \gqd{} halos, while $\rm M^{max}_{vir,GQd}(\mathit{z})$ is the maximum of $\rm M_{vir,GQd}$ at any given  $z$.

Normally, for our grafting procedures we draw randomly a \gqd{} halo with $\rm Log10(M_{vir,init})-0.25\leq M_{vir,GQd}\leq Log10(M_{vir,init})+0.25$ extracted from the sample of \gqd{} DM halos with $z_{\rm init}\leq z\leq z_{\rm prev}$. At $z>16$, this procedure remains identical with only one exception: if $\rm M_{vir,init}\!>\! M^{max}_{vir,GQd}(\mathit{z}_{init})$, we draw \gqd{} halos under the condition: $\rm M_{vir,GQd}\geq Log10[M^{max}_{vir,GQd}(\mathit{z}_{init})]-0.25$. In this sense, we use the highest-mass bin of \gqd{} dynamic range (see Fig. \ref{fig:dynaranges}). Therefore, \milltwo{} halos newly-resolved with $\rm M_{vir,init}\!>\!M^{max}_{vir,GQd}$ at $z>16$ have \gqd{} ``progenitors'' which are undermassive with respect to those of \milltwo{} halos newly-resolved at $z<16$ with the same $\rm M_{vir,init}$. We never apply any extra evolution to the properties of \gqd{} halos used for our grafting procedures, since we only use the BH masses and the BH seed-type (i.e. light or heavy) of \gqd{} halos to initialize \lgal{} structures, as pointed out in Sect. \ref{sec:model:BH_seeding_prescriptions:GQd_Grafting}. We checked that the mismatch between the dynamic ranges of \gqd{} and \milltwo{} at $z\!>\!16$ has a negligible effect on fundamental predictions of \lgal{} such as the BH mass function, stellar mass function and the BHAR evolution. This is because the fraction of halos initialized with $\rm M_{vir,init}\!>\!M^{max}_{vir,GQd}$ over all the \milltwo{} halos undergoing grafting procedures is marginal (i.e. only the $0.38\%$).

\bsp	
\label{lastpage}
\end{document}